\documentclass[preprint, 11pt]{aastex}
\usepackage{natbib}
\usepackage{longtable}

\newcommand{\msun}{M$_{\sun}$}

\newcommand{\ldl}{$\lambda/{\Delta}{\lambda}$}
\newcommand{\teff}{T$_{eff}$}

\newcommand{\lii}{\ion{Li}{1}}

\newcommand{\lha}{$\log_{10}{L_{H\alpha}/L_{bol}}$}
\newcommand{\kms}{km~s$^{-1}$}
\newcommand{\masyr}{mas~yr$^{-1}$}
\newcommand{\vsini}{$v\sin{i}$}
\newcommand{\mjup}{$M_{\mathrm{Jup}}$}

\slugcomment{Accepted to ApJS 2015 June 27}

\shorttitle{BDKP V}

\begin{document}

\title{The Brown Dwarf Kinematics Project (BDKP). IV. Radial Velocities of 85 Late-M and L dwarfs with MagE}

\author{
Adam J. Burgasser\altaffilmark{1},
Sarah E. Logsdon\altaffilmark{2}, 
Jonathan Gagn{\'{e}}\altaffilmark{3}, 
John J. Bochanski\altaffilmark{4}, 
Jaqueline K. Faherty\altaffilmark{5,6,7}, 
Andrew A. West\altaffilmark{8}, 
Eric E. Mamajek\altaffilmark{9}, 
Sarah J. Schmidt\altaffilmark{10}, 
and Kelle L. Cruz\altaffilmark{6,11,12}}

\altaffiltext{1}{Center for Astrophysics and Space Science, University of California San Diego, La Jolla, CA, 92093, USA; aburgasser@ucsd.edu}
\altaffiltext{2}{Department of Physics and Astronomy, UCLA, 430 Portola Plaza, Box 951547, Los Angeles, CA 90095-1547, USA}
\altaffiltext{3}{Institute for Research on Exoplanets (iREx), Universit\'e de Montr\'eal, D\'epartement de Physique, C.P.~6128 Succ.\ Centre-ville, Montr\'eal, QC H3C~3J7, Canada}
\altaffiltext{4}{Rider University, 2083 Lawrenceville Road, Lawrenceville, NJ 08648, USA}
\altaffiltext{5}{Department of Terrestrial Magnetism, Carnegie Institution of Washington, Washington, DC 20015, USA}
\altaffiltext{6}{Department of Astrophysics, American Museum of Natural History, Central Park West at 79th Street, New York, NY 10034, USA}
\altaffiltext{7}{Hubble Postdoctoral Fellow}
\altaffiltext{8}{Department of Astronomy, Boston University, 725 Commonwealth Ave Boston, MA 02215 USA}
\altaffiltext{9}{Department of Physics \& Astronomy, University of Rochester, Rochester, NY 14627, USA}
\altaffiltext{10}{Department of Astronomy, Ohio State University, 140 West 18th Avenue, Columbus, OH 43210, USA}
\altaffiltext{11}{Department of Physics and Astronomy, Hunter College, 695 Park Avenue, New York, NY10065, USA}
\altaffiltext{12}{Department of Physics, Graduate Center, City University of New York, 365 Fifth Avenue, New York, NY 10016, USA}

\begin{abstract}
Radial velocity measurements are presented for 85 { late M- and L-type very low mass stars and brown} dwarfs obtained with the Magellan Echellette (MagE) spectrograph. Targets primarily have distances within 20 pc of the Sun, with more distant sources selected for their unusual spectral energy distributions. We achieved precisions of 2--3,{\kms}, and combined these with astrometric and spectrophotometric data to calculate $UVW$ velocities.  Most are members of the thin disk of the Galaxy, and velocity dispersions indicate a mean age of 5.2$\pm$0.2~Gyr for sources within 20 pc. We find signficantly different kinematic ages between late-M dwarfs (4.0$\pm$0.2~Gyr) and L dwarfs (6.5$\pm$0.4~Gyr) { in our sample} that are contrary to predictions from prior simulations. This difference appears to be driven by a dispersed population of unusually blue L dwarfs { which may be more prevalent in our local volume-limited sample than in deeper magnitude-limited surveys.} 
The L dwarfs exhibit an asymmetric $U$ velocity distribution with a net inward flow, similar to gradients recently detected in local stellar samples. Simulations incorporating brown dwarf evolution and Galactic orbital dynamics are unable to reproduce the velocity asymmetry, suggesting non-axisymmetric perturbations or two distinct L dwarf populations.
We also find the L dwarfs to have a kinematic age-activity correlation similar to more massive stars.
We identify several sources with low surface gravities, and two new substellar candidate members of nearby young moving groups: the astrometric binary DENIS~J08230313$-$4912012AB, a low-probability member of the $\beta$ Pictoris Moving Group; and 2MASS~J15104786-2818174, a moderate-probability member of the 30-50~Myr Argus Association.
\end{abstract}


\section{Introduction}

Very low-mass (VLM) stars and brown dwarfs with masses $\lesssim$\,0.1~{\msun} comprise a significant fraction of stars in our Galaxy ($\gtrsim$20\%; \citealt{2003PASP..115..763C,2010AJ....139.2679B}). Their ubiquity and extremely long lifetimes make them an important probe of Galaxy structure, chemical evolution and star formation history \citep{2004ApJS..155..191B,2007AJ....134.2418B,2009ApJ...695.1591P,2013AJ....146...50P}. However, the low luminosities and  temperatures of these M, L, T and Y dwarfs  \citep{2005ARA&A..43..195K}, and the steady cooling of substellar VLM dwarfs over time, has made it difficult to identify and study them in statistically significant numbers.  Fortunately, the completion of wide-field red-optical and infrared sky surveys such as 
the DEep Near-Infrared Survey of the Southern Sky (DENIS; \citealt{1997Msngr..87...27E}), 
the Two Micron All Sky Survey (2MASS; \citealt{2006AJ....131.1163S}),
the Sloan Digital Sky Survey (SDSS; \citealt{2000AJ....120.1579Y}), 
the Canada-France Hawaii Telescope Legacy Survey (CFHTLS; \citealt{2008A&A...484..469D}), and
United Kingdom Infrared Telescope Deep Sky Survey (UKIDSS; \citealt{2007MNRAS.379.1599L}); and
the ongoing
Wide-field Infrared Survey Explorer (WISE; \citealt{2010AJ....140.1868W}),
Panoramic Survey Telescope and Rapid Response System (Pan-STARRS; \citealt{2002SPIE.4836..154K}),
and Visible and Infrared Survey Telescope for Astronomy (VISTA; \citealt{2004Msngr.117...27E})
have uncovered roughly 10,000 VLM dwarfs within 100~pc of the Sun\footnote{Current compilations are maintained by C.\ Gelino at \url{http://dwarfarchives.org} and J.\ Gagne at \url{https://jgagneastro.wordpress.com/list-of-ultracool-dwarfs}.}.
The detector technologies enabling these surveys have also led to advances in spectroscopic instrumentation, allowing detailed characterization of spectral energy distributions and corresponding physical parameters for increasingly larger samples of VLM dwarfs (e.g., \citealt{2008AJ....136.1290R,2010AJ....139.1808S,2011AJ....141...97W,2012ApJ...753..156K}).

Characterizing VLM dwarfs as a population relies on accurate measures of their individual characteristics, including kinematics. While most stars in the disk of the Milky Way form in the mid-plane, with Galactic orbits similar to that of the Sun (low inclination, low eccentricity), dynamical encounters with giant molecular clouds, spiral structure
and other gravitational potential gradients scatter stars stochastically \citep{1953ApJ...118..106S,1964ApJ...139.1217T,2013A&A...550A..91J}.  Individual orbits may be chaotic, but the population as a whole evolves toward greater velocity dispersion over time \citep{1977A&A....60..263W}.
The age-{ velocity} dispersion relation has been used extensively in studies of Galactic star formation history with nearby main sequence stars (e.g., \citealt{1998MNRAS.298..387D,2000MNRAS.318..658B,2009MNRAS.397.1286A}). 
%but few studies have approached this issue using VLM dwarfs due to typically small samples \citep{2009AJ....137....1F,2010ApJ...723..684B}.
Stellar kinematics also segregate large-scale Galactic populations---the thin disk, thick disk and halo---which trace Galactic structure, formation history and chemical enrichment (e.g., \citealt{1983MNRAS.202.1025G,1993ApJ...409..635R,2008Natur.451..216C}).
Young moving groups (YMGs) near the Sun can also be revealed by coherently moving stars with common spectral signatures
of low surface gravity or abundance patterns (e.g., \citealt{2004ARA&A..42..685Z,2006ApJ...643.1160L,2006A&A...460..695T,2014ApJ...783..121G}). Finally, periodicity in motion identifies low-mass companions that can be used to make direct mass measurements and test substellar/exoplanet evolutionary models \citep{1999Sci...283.1718M,2009ApJ...692..729D,2010ApJ...711.1087K,2013A&A...556A.133S}.

Full characterization of the three-dimensional motions of VLM dwarfs requires radial velocities (RVs), and hence high resolution spectroscopy, observations which have proven challenging
for these faint sources. Only a small fraction of the known VLM population has sufficiently precise ($<$5~{\kms}) RV measurements necessary for robust membership assignment or RV variability detection (e.g., \citealt{1998MNRAS.301.1031T,2002AJ....124..519R,2006AJ....132..663B,2007ApJ...666.1198B,2010ApJ...723..684B,2007ApJ...666.1205Z,2009ApJ...705.1416R,2010A&A...512A..37S,2012ApJS..203...10T}). Nevertheless, these studies have identified several remarkable---and in some cases conflicting---kinematic trends among the population.
\citet{2007ApJ...666.1205Z} examined the velocity dispersions for 31 late-M  and 21 L and T dwarfs, finding the latter to be less dispersed and hence marginally younger, 1.2$^{+1.1}_{-0.7}$~Gyr versus 3.8$^{+2.8}_{-1.9}$~Gyr based on the age-dispersion relation of \citet{1977A&A....60..263W}. This age difference is qualitatively consistent with population synthesis simulations that predict that L dwarfs should be on average younger due to brown dwarf cooling  \citep{2004ApJS..155..191B,2005ApJ...625..385A}.  However, a large fraction ($\sim$40\%) of the L and T dwarfs examined in that study were also identified as kinematic members of the 0.4--2~Gyr Hyades Stream, which may have biased their collective ages downward.  
More recent work by \citet{2009ApJ...705.1416R, 2010A&A...512A..37S} and \citet{2010ApJ...723..684B}, based on { precise} RV measurements for roughly 150 M and L dwarfs and more accurate application of the Wielen relations, find mean ages of 3~Gyr and 5~Gyr for late-M and L dwarfs, respectively; i.e., the reverse trend.  Tangential velocity studies by \citet{2009AJ....137....1F} and \citet{2010ApJS..190..100K}, and a larger but lower-precision RV sample by \citet{2010AJ....139.1808S}, find equivalent ages for M, L and T dwarfs, of order 3--8~Gyr, with no statistically significant difference. 

{ \citet{2009AJ....137....1F} and \citet{2010AJ....139.1808S} also report color-dependent trends in  VLM dwarf kinematics, with sources identified as unusually blue in near-infrared colors being more widely dispersed than those with unusually red colors. These differences were attributed to age and/or metallicity effects, in particular increased collision-induced H$_2$ absorption in the atmospheres of sources with high surface gravities and/or subsolar metallicities (cf.\ \citealt{2008ApJ...674..451B,2008MNRAS.385.1771J,2008ApJ...686..528L}). However, \citet{2010ApJS..190..100K} found no difference in dispersions between red and blue outliers among a proper-motion selected sample of L dwarfs, with both sets appearing to be drawn from an older population.}

To address these disagreements in the velocity dispersions and age determinations for the local VLM dwarf population, we report RV measurements for 85 late-type M and L dwarfs based on data obtained with the Magellan Echellete spectrograph (MagE; \citealt{2008SPIE.7014E.169M}).  These include measurements for 30 dwarfs without previously reported RVs.  
In Section~2 we describe our sample, observations, and data reduction methods.   
In Section~3 we discuss our RV measurement procedures in detail, comparing the accuracy and precision of three common methods.
In Section~4 we merge our RV measurements with proper motion and distance determinations to calculate UVW space velocities, and assign membership to Galactic thin and thick disk populations.  We also examine velocity dispersions, probability distributions and kinematic ages for the full sample and subsamples based on spectral class, color and activity.
In Section~5 we focus on the distinct velocity distributions of M and L dwarfs in our sample, and use population synthesis plus orbit simulations in an attempt to reproduce the asymmetries in the $U$ velocity distribution of the latter.
In Section~6 we consider kinematic and spectroscopic evidence for association in 
nearby YMGs, and report two new brown dwarf candidate members.
In Section~7 we highlight additional individual sources of interest in detail.
Results are summarized in Section~8.

\section{Observations}

\subsection{Sample}

\begin{figure}
\centering
\plotone{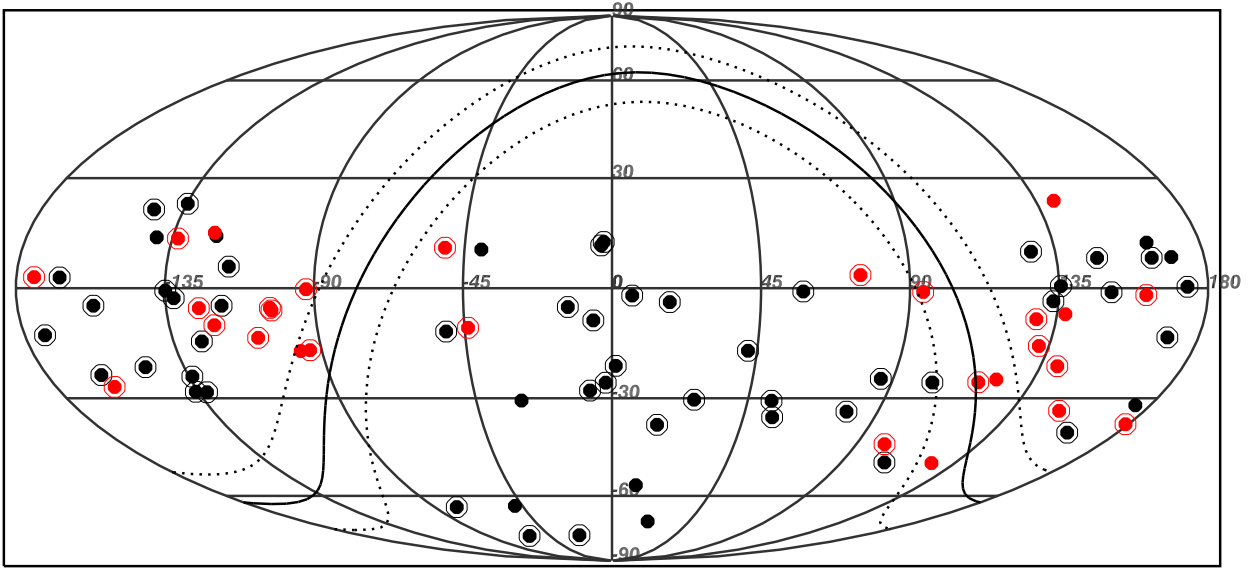}
\caption{Mollweide projection equatorial map of late-M dwarfs (black) and L dwarfs (red) in our sample.
Sources within 20~pc are encircled. 
The Galactic plane and $\pm$10$\degr$ about the plane are indicated by the solid and dotted lines, respectively.
 \label{figure:skymap}}
\end{figure}

\begin{figure}
\centering
\epsscale{1.1}
\plottwo{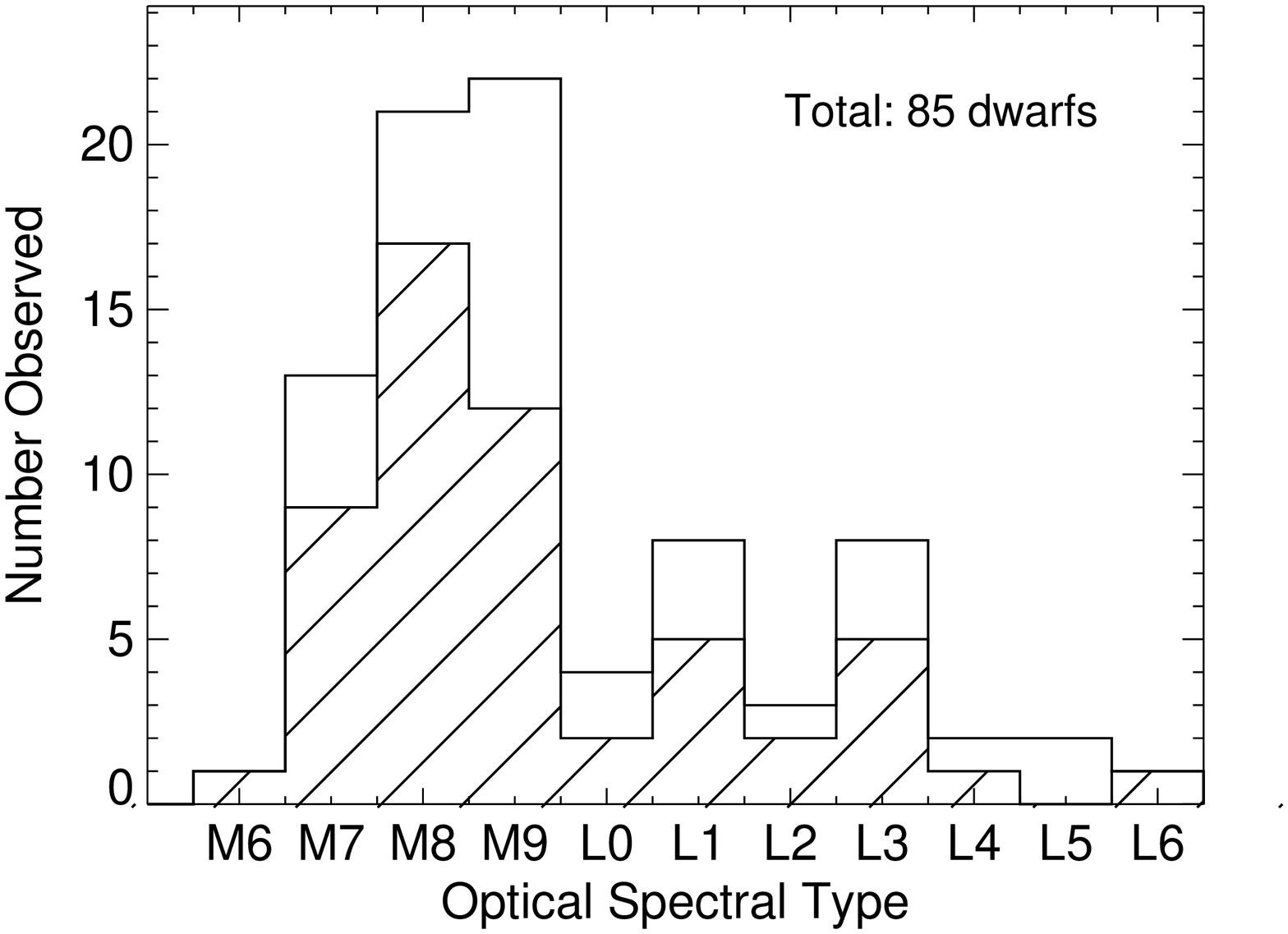}{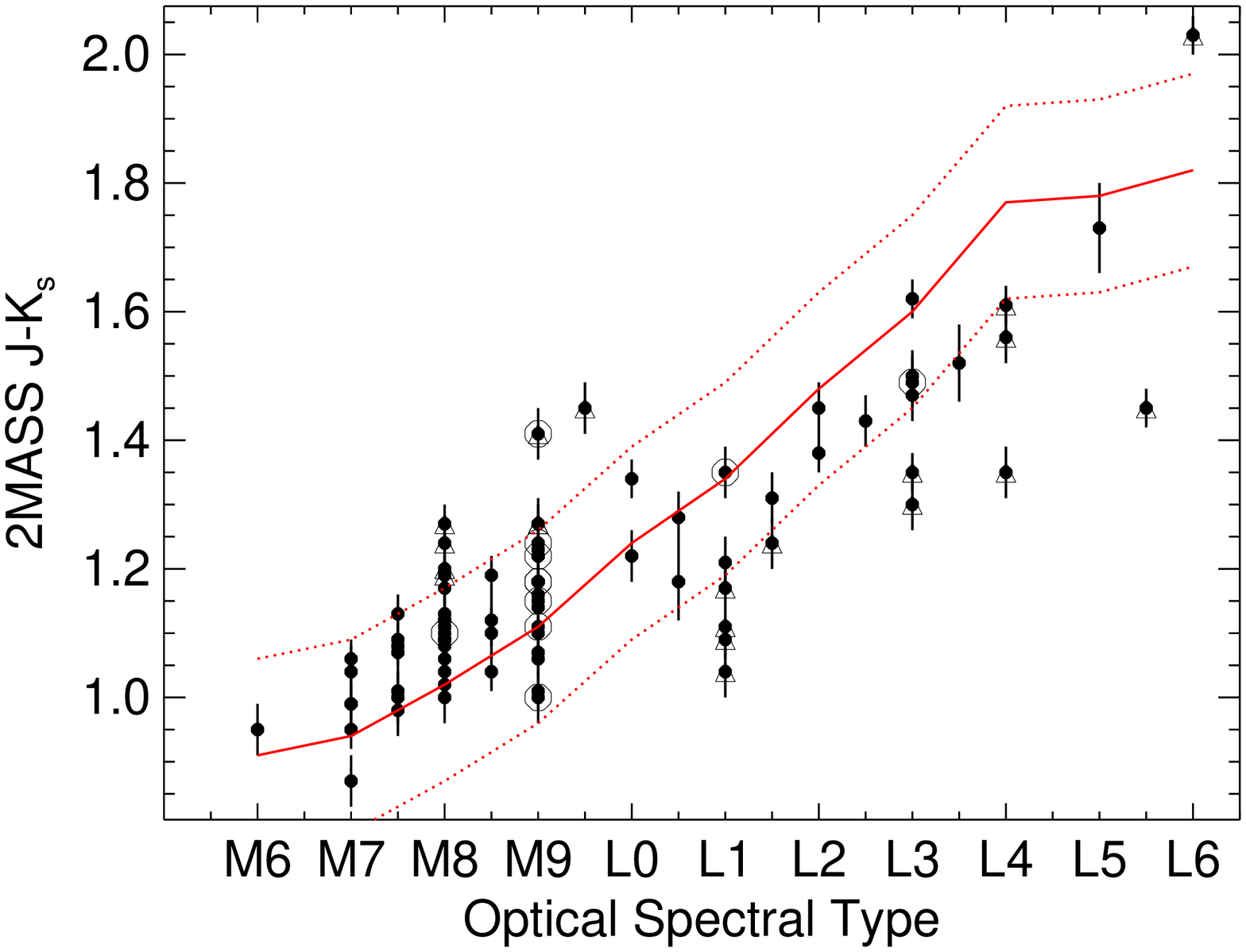}
\caption{(Left): Distribution of optical spectral types for the 85 observed sources, sampled by whole subtype bins. The open histogram refers to all sources observed; the hatched histogram refers to those sources with previously published radial velocities. 
{ (Right): 2MASS $J-K_s$ colors of our targets compared to the mean colors of M6--L6 dwarfs (red line) from \citet{2010AJ....139.1808S} and \citet{2011AJ....141...97W}. Our $\pm$0.15~mag threshold for unusually red and blue dwarfs are indicated by dotted lines, and those sources are highlighted by open triangles.  Young sources in our sample are highlighted with open circles.}
 \label{figure:sptdistribution}}
\end{figure}

Our observational sample is summarized in Table~\ref{table:sample}.  We initially selected 85 late-type dwarfs with published optical spectral types M7 through L5, primarily from the 
Palomar/MSU Survey \citep{1995AJ....110.1838R,1996AJ....112.2799H,2002AJ....124.2721R};
the 2MASS surveys of \citet{1999ApJ...519..802K,2000AJ....120..447K,2002ApJ...575..484G,2003AJ....126.2421C,2007AJ....133..439C} and \citet{2008AJ....136.1290R}; 
the SDSS surveys of \citet{2005AJ....130.1871B,2008AJ....135..785W,2011AJ....141...97W} and \citet{2010AJ....139.1808S}; 
the DENIS surveys of
\citet{1999A&AS..135...41D,1999AJ....118.2466M,2004A&A...416L..17K,2007MNRAS.374..445K} and \citet{2008MNRAS.383..831P}; 
and other individual discoveries.
The majority of our sample were selected to have declinations $\delta \leq 25\degr$ (Figure~\ref{figure:skymap}) and distances $d \lesssim$ 20 pc, the latter based on either astrometric parallax measurements \citep{1995AJ....110.3014T,1995yCat.1174....0V,2002AJ....124.1170D,2005AJ....130..337C,2006AJ....132.1234C,2006AJ....132.2360H,2011AJ....141...54A,2012ApJ...752...56F,2014AJ....147...94D} or spectrophotometric distance estimates based on 2MASS $J$-band photometry and the $M_J$/spectral type relation of \citet{2003AJ....126.2421C}.  
Sources within 20~pc (1$\sigma$) comprise 82\% of our sample. Fifteen sources are at larger distances, most due to previously underestimated distances (e.g., due to multiplicity, misclassification), but several were specifically targetted due to their unusual spectral features indicative of surface gravity, metallicity or cloud effects. There were also several well-resolved binaries selected (e.g., J2200$-$3038AB\footnote{Source identifications in the text are given in shorthand notation based on the sexigesimal right ascension and declination, Jhhmm$\pm$ddmm.  Full source names and coordinates are listed in Table~\ref{table:sample}.}) and sources with unusually large tangential velocities (e.g., J0923+2340).  
We emphasize that while our sample was primarily designed to be volume-limited,  it is not volume-complete.
Space density estimates from \citet{2007AJ....133..439C} predict $\sim$160 late-M dwarfs and $\sim$125 L dwarfs within 20~pc, so the 70 sources in our sample satisfying this distance limit represents $\sim$25\% of the total local population. 
Figure~\ref{figure:sptdistribution} displays a breakdown of our sample by spectral type (based on updated classifications; see below); 56\% are late-type M dwarfs, 44\% are L dwarfs.  
{ We also display the 2MASS $J-K_s$ colors of our targets as compared to the median color relations of  \citet{2010AJ....139.1808S} and \citet{2011AJ....141...97W} based on SDSS sources.  There appears to be a systematic offset in the $J-K_s$ colors of our sample relative to these surveys, with our M dwarfs being on average redder and our L dwarfs being on average bluer. The red offset in our M dwarfs is likely due to our more local sample, as the M dwarfs from \citet{2011AJ....141...97W} extend to $>$100~pc above/below the Galactic disk, and are likely to be on average more metal-poor.  The blue color offset for our L dwarfs, however, is unclear.}
We compiled published proper motions and tangential velocities, or in the absence of a measurement compared multi-epoch astrometry between 2MASS and WISE, yielding typical precisions of 5--15~\masyr; see \citet{2015ApJ...798...73G} for details.

As of March 2015, 43 sources in our sample had RV measurements reported in the literature with precisions $\sigma_{RV}$ $\leq$ 3~{\kms} \citep{1998MNRAS.301.1031T,2003ApJ...583..451M,2009ApJ...705.1416R,2010ApJ...723..684B,2010MNRAS.409..552G,2010A&A...512A..37S,2012ApJS..203...10T}; these values are listed in Table~\ref{table:sample}.  We made use of these prior measurements in our RV measurement analysis (Section~3.3). 

\subsection{MagE Observations}

All sources were observed with the MagE spectrograph, mounted on the Magellan 6.5m Landon Clay Telescope at Las Campanas Observatory.  A complete observing log is given in Table~\ref{table:log}.  Data were obtained in 15 nights over a 2.5-year period (November 2008 through March 2011) in a variety of seeing and weather conditions.  
We used the 0$\farcs$7 slit aligned with the parallactic angle, providing 3200--10050\,{\AA} spectroscopy at an average resolution {\ldl} $\approx$\,4100 ($\Delta$RV = 73~\kms) and dispersion of $\sim$0.5\,{\AA}~pixel$^{-1}$ at 6000\,{\AA}.  Exposure times varied according to source brightness and weather conditions, and ranged from 150--3600~s.  Most sources were observed in a single exposure, although a handful were observed in multiple exposures or over multiple nights to improve data quality.
In addition to the target, we obtained nightly observations of spectrophotometric standards from \citet{1994PASP..106..566H} for flux calibration. ThAr lamps were observed after each source observation for wavelength calibration, and internal quartz and dome flat field lamps were obtained on each night for pixel response calibration.  Data were reduced using the MASE reduction pipeline \citep{2009PASP..121.1409B}, following standard procedures for order tracing, flat field correction, wavelength calibration (including correction to heliocentric motion), optimal source extraction, order stitching, and flux calibration.  We did not perform any correction for telluric absorption on these data, which caused a problem for the flux calibration for several of the spectra around the 7500--7600\,{\AA} O$_2$ telluric band, which resides near an echelle order boundary.  This region is ignored in subequent analyses.
% note scale offset for all spectra < 8288 x 1.1?
% note fix to 1534-1418 spectrum - scaled < 8288 x 1.7 ?
We also note that the spectrum of J0148$-$3024 was affected by poor flux calibration in the 8300--8700\,{\AA} band, and the region is also ignored in the analysis of this source.

\subsection{Spectral Features and Classification}

The red optical components of our spectra (6000--9000\,{\AA}) are shown in Figure~\ref{figure:spectra}. The data are generally of high quality, with signal-to-noise ratios (S/N) of 20--200 at 8600\,{\AA} (mean S/N = 60).
We see all of the characteristic features of late-type dwarfs, including strong TiO, VO, CaH and metal line absorption in the late-M dwarfs. The features fade in the L dwarf spectra and FeH, CrH, and various alkali lines emerge, including the heavily pressure-broadened Na~I and K~I doublets centered at 5500\,{\AA} and 7700\,{\AA}, respectively \citep{1991ApJS...77..417K,1999ApJ...519..802K,1999AJ....118.2466M}.

We revisited the optical classifications for all of our sources by comparing directly to the SDSS M and L dwarf templates from \citet{2007AJ....133..531B} and \citet{2014PASP..126..642S} over the 7100--8800\,{\AA} range (excluding the 7500--7600\,{\AA} O$_2$ band).  The best-match template ($T(\lambda)$) was identified as that with the minimum squared deviation from the observed spectrum ($f(\lambda)$):
\begin{equation}
\sigma^2 = \sum_{\lambda=7100\,{\AA}}^{8800\,{\AA}}\left(f(\lambda)-{\alpha}T(\lambda)\right)^2
\end{equation}
where
\begin{equation}
\alpha = \frac{\sum_{\lambda=7100\,{\AA}}^{8800\,{\AA}}f(\lambda)T(\lambda)}{\sum_{\lambda=7100\,{\AA}}^{8800\,{\AA}}T(\lambda)^2}
\end{equation}
is the optimal scale factor.   For most of our sample, these classifications are within 1 subtype of those reported in the literature. The six discrepant sources (J0041--5621AB, J0123--6921, J0331--3042, J0641--4322, J0751--2530, J0823--4912AB) were re-examined by eye and the revised classification confirmed.
Note that three of these discrepant sources exhibit signatures of low surface gravity, as discussed below.

\subsection{Emission and Absorption Line Features}

For the majority of our sources, we also detect Balmer line emission, including H$\alpha$, H$\beta$ and H$\gamma$ lines. Figure~\ref{figure:halpha} displays the fraction of sources with detectable H$\alpha$ emission as a function of spectral type.  Similar to previous studies (e.g., \citealt{2000AJ....120.1085G,2000AJ....120..447K,2004AJ....128..426W,2008AJ....135..785W, 2007AJ....133.2258S,2014PASP..126..642S}) we find a decline in the frequency of emission between the late-M dwarfs (96\% active) and the L dwarfs (54\% active), although a trend with spectral type is difficult to quantify given the relatively small sample.  

We computed the relative H$\alpha$ luminosities ({\lha}) by two methods. First, we used the sources' 2MASS $J$ magnitudes and mean $i-J$ colors as a function of spectral type from \citet{2010AJ....139.1808S} and \citet{2011AJ....141...97W} to scale the spectra and convert H$\alpha$ emission into apparent flux units. We then used $J$-band bolometric corrections from \citet{2010ApJ...722..311L} to compute apparent bolometric fluxes. { The bolometric corrections are based on the Mauna Kea Observatory (MKO) filter system \citep{2002PASP..114..169S,2002PASP..114..180T}, so we computed a spectral type-dependent correction between 2MASS $J$ and MKO $J$ using spectrophotometry from 533 optically-classified M6--L7 dwarfs in the SpeX Prism Library \citep{2014arXiv1406.4887B}, as described in the Appendix.  The ratio of apparent H$\alpha$ flux to apparent bolometric flux} yields {\lha}. 
{ As a second approach}, we used the { spectral type-dependent} $\chi$-factor defined by \citet{2004PASP..116.1105W}, with updated values from \citet{2014ApJ...795..161D} and \citet{2014PASP..126..642S}.  Both methods gave consistent results, and
values from the first method are listed in Table~\ref{table:lines}.
As in previous studies, we find that the strength of emission monotonically declines from M7 to L5, ultimately falling below our detection limits.  However, two L dwarfs, the L3 J2036+1051 and the L5 J1315$-$2649AB, stand out as being unusually active.  
The latter is a ``hyperactive'' L dwarf observed to have strong H$\alpha$, alkali line and radio emission \citep{2002ApJ...564L..89H, 2002ApJ...580L..77H, 2002ApJ...575..484G, 2008ApJ...689.1295K,2011ApJ...739...49B,2013ApJ...762L...3B}.  We also identify J2037$-$1137 as the only M8 dwarf in our sample to show no sign of H$\alpha$ emission to a stringent limit ({\lha} $< -6.7$).  Both J2036+1051 and J2037$-$1137 are discussed further in Section~7.

Table~\ref{table:lines} also lists equivalent widths (EWs) for alkali lines observed in the MagE data.
Of particular note is Li~I absorption detected in the spectra of nine sources, { all} shown in detail in Figure~\ref{figure:lithium}.  We confirm the detection of Li~I in J0041$-$5621AB, J0123$-$6921, J0823$-$4912, J1139$-$1159, J1411$-$2119 and J2045$-$6332 \citep{1998MNRAS.296L..42T,2003ApJ...593L.109M,2009ApJ...705.1416R,2014MNRAS.439.3890G,sahlmann2015}, and find marginal evidence of absorption previously reported in J0339$-$3525 \citep{2009ApJ...705.1416R}. We report the first identification of Li~I  in the spectra of the M9 dwarfs J0652$-$2534 and J1510$-$2818.  This line is a key age and mass indicator, as it is only present in cool dwarfs less massive than 0.06~{\msun} and/or younger than $\sim$200~Myr \citep{1993ApJ...404L..17M,1997ApJ...482..442B,2004ApJ...604..272B}.  The presence of Li~I absorption in the spectra of these dwarfs therefore indicates that they are likely { to be} young brown dwarfs.

\subsection{Low Surface Gravity Indicators}

As described in previous studies (e.g., \citealt{1999AJ....118.2466M,2008ApJ...689.1295K,2009AJ....137.3345C,2013ApJ...772...79A}), young
brown dwarfs with low surface gravities and low photospheric pressures exhibit enhanced TiO and VO absorption, weak metal hydride bands, and weak alkali lines compared to their equivalently-classified
field dwarf counterparts. Building on work by \citet{2009AJ....137.3345C}, we quantified these features using the Na-a, Na-b, TiO-b, CrH-a, and FeH-a indices defined in \citet{1999ApJ...519..802K}, which sample the 8183\,{\AA} and 8194\,{\AA} Na~I doublets,  8400\,{\AA} TiO band, and 8580\,{\AA} CrH and 8660\,{\AA} FeH bands.  We defined a new index (VO7900) sampling the 7900\,{\AA} VO band as the ratio of integrated flux between 7950--8000\,{\AA} over that between 7825--7875\,{\AA}.
Table~\ref{table:gravindices} lists the mean and standard deviations of these index combinations as a function of spectral type for sources without Li~I absorption.  

We identified sources in our sample for which alkali or metal-hydride bands are consistently weaker, and VO or TiO bands consistently stronger, than these spectral type means. These sources are listed in Table~\ref{table:lowg}. Five show clear signatures of low surface gravity as compared to SDSS templates (Figure~\ref{fig:lowg_strong}). Three of these sources exhibit Li~I absorption and three have been identified as candidate kinematic members of the YMGs TW Hydrae (J1139$-$3159; \citealt{2002ApJ...575..484G}) and $\beta$ Pictoris (J2000$-$7523, J2045$-$6332; \citealt{2010MNRAS.409..552G,2014ApJ...783..121G}).
An additional six sources (Figure~\ref{fig:lowg_weak}) exhibit weak signatures of low surface gravity. All of these have Li~I absorption, and three are previously noted as candidate kinematic members of the Tucana Horologium, Castor and Argus YMGs \citep{2003A&A...400..297R,2010MNRAS.409..552G,2014ApJ...783..121G}.  The low-gravity sources are discussed in further detail below.

\subsection{Metallicity Indicators}

For all of our late-M dwarfs, we computed the metallicity index $\zeta$ defined in \citet{2007ApJ...669.1235L}.
Within our sample, there is very little variation in $\zeta$, with all but one source (the M9.5 J0024$-$0158; $\zeta$ = 0.56$\pm$0.03) falling into the ``subdwarf''  classification.  
Even this source is a poor metal-poor candidate given its late spectral type, although it is unusual in its combination of rapid rotation and sporadic (and occasionally flaring) magnetic emission \citep{1995AJ....109..762B,1999ApJ...527L.105R,2002ApJ...572..503B,2010ApJ...709..332B}. No H$\alpha$ emission { from this source} was detected in our data.

\section{Radial Velocity Measurements}

To obtain the most precise radial velocity measurements for $UVW$ space motion analysis, we examined three different measurement methods.  For interested readers, we detail these methods here; final results are summarized in Table~\ref{table:rvs}.

\subsection{Line Center Measurements}

Our first approach was to measure Doppler shifts for the prominent absorption and emission lines present in the data: K~I, Rb~I, Na~I, Cs~I, and H$\alpha$.  Line centers were  determined by fitting Gaussian profiles to each line using the IRAF\footnote{Image Reduction and Analysis Facility \citep{1986SPIE..627..733T}. IRAF is distributed by the National Optical Astronomy Observatory,
which is operated by the Association of Universities for Research in Astronomy,
Inc., under cooperative agreement with the National Science Foundation} routine {\it splot}, and these were compared to vacuum wavelengths obtained from the National Institute of Standards and Technology (NIST) Atomic Spectra Line database\footnote{\url{http://physics.nist.gov/PhysRefData/ASD/lines\_form.html}}. After rejecting poor line fits, we computed the average and standard deviation of the corresponding velocity shifts (Table~\ref{table:rvs}).

Figure~\ref{figure:litcomp} compares our measured line center RVs to literature values for those sources with previous high precision measurements ($\sigma_{RV} \leq$ 3~{\kms}).
Overall we find good agreement between these measurements, with a mean offset of 1.0$\pm$4.2~{\kms} (line center measurements are slightly more positive).  In only one case, J0517$-$3349, do we find a significant discrepancy between our measurement ($-$36$\pm$7~\kms) and that reported by \citet[31.4~\kms]{2009ApJ...705.1416R}, which we attribute to a sign reversal in that study.  The median measurement uncertainties from this analysis are 7~\kms, roughly one-tenth our nominal resolution but still { unacceptably} large for kinematic analysis.

\subsection{Cross-correlation with SDSS Templates}

To make better use of the full spectrum, we performed a cross-correlation analysis comparing to two sets of RV standards: the zero-velocity, low-resolution M7--L0 SDSS spectral templates of \citet{2007AJ....133..531B} and MagE observations of VLM dwarfs with published RV measurements.  In both cases, we used the IDL\footnote{Interactive Data Language} {\it xcorl} package \citep{2003ApJ...583..451M,2009ApJ...693.1283W} to cross-correlate spectral pairs over five bands sampling distinct spectral features: 
H$\alpha$ emission (6500-6600\,{\AA});
the 7050\,{\AA} TIO absorption band (7150-7250\,{\AA});
K~I and Rb~I doublets (7685-7885\,{\AA});
 the 8183/8195\,{\AA} Na I doublet (8150-8250\,{\AA});
 and TiO, VO and Cs~I absorption (8350-8550\,{\AA}). 
%[POTENTIAL PROBLEM: THERE IS TELLURIC ABSORPTION FROM 7150-7300 (WEAK) AND 7580-7700 (STRONG); MAY NEED TO MODIFY KI TO 7700-7900, ALSO SUGGEST CHANGE TIO TO 7000-7200, AND MAYBE ADD CRH/FEH ] 
%Note that the third region is specifically defined to avoid the telluric O$_2$ A-band at 7580--7700\,{\AA}.
For each source/template pair and spectral region, cross-correlation functions were visually inspected and poor correlations rejected. 
%The final measurement and uncertainty were taken as the average and standard deviation of the remaining correlation velocity offsets.

For the SDSS template sample, we used the non-active templates to compare against our M7--L0 sources.
As the templates have a lower resolution ({\ldl} $\approx$\,1800) than our MagE data, we smoothed the MagE data to an equivalent resolution using a Gaussian kernel and interpolated onto a common wavelength scale.  Each source was compared to the template with the equivalent type; for sources with half-subtype classifications, we compared to both the lower and higher integer subtype template. Velocity shifts from each band (and multiple templates) were then averaged and the standard deviation used as an estimate of the uncertainty.  The latter were typically $\sim$5~\kms, slightly smaller than our line center uncertainties.

Figure~\ref{figure:litcomp} compares our RVs based on this method to literature values.  Unlike the line center measurements, we find a marginally significant offset in our measurements of 6.5$\pm$3.7~{\kms} (excluding J0517$-$3349), with the SDSS  measurements being consistently more positive.  To assess whether this was a consequence of our smoothing function, we performed the same cross-correlation on unsmoothed data and found a comparable offset.  As the SDSS templates are shifted to a zero velocity rest frame in vacuum wavelengths using several of some of the same lines employed in our line centering analysis (K~I, Na~I and H$\alpha$), we are unable to identify the origin of this offset. 
%{ [JOHN: ANY IDEAS? WERE THE LINE WAVELENGTHS USED IN THE 2007 PAPER OR IN OUR LINE ANALYSIS BASED ON AIR NOT VACCUUM WAVELENGTHS?]}

\subsection{Cross-correlation with MagE Radial Velocity Standards}

Our third approach was to use the spectra of 40 sources in our sample with independently measured, high-precision RVs as cross-correlation standards.  We matched sources to standards that had equivalent spectral types to within $\pm$1 subtype (excluding the source itself if it was a standard), shifted the standards to zero velocity, and cross-correlated in the five spectral regions listed above. This resulted in up to 50 measurements per source, more for the late-M dwarfs.  Following the same rejection and averaging procedures above, our reported values are listed in Table~\ref{table:rvs}.  Typical uncertainties are $\sim$1.3~\kms, with over half of the sources having uncertainties below 1~\kms; however, mid-type L dwarfs like J1750$-$0016 have higher uncertainties due to the fewer RV standards available with equivalent spectral types. 

Figure~\ref{figure:litcomp} compares these measurements to the literature values, and we find that they are overall consistent, deviating on average by 0.2$\pm$2.7~{\kms} (again excluding J0517$-$3349).  
However the $\chi^2$ deviation is somewhat high in this case ($\chi^2$ = 123, $N$=55, $p$ = 4$\times$10$^{-7}$), suggesting that our uncertainties are underestimated. We therefore include a 2~{\kms} systematic uncertainty in our reported values, added in quadrature with the standard deviations, { which lowers the $\chi^2$ to a value consistent with no deviation ($p$ = 0.1).} 
The MagE cross-correlation measurements are also in agreement with the line center measurements (average deviation $-$0.8$\pm$3.2~{\kms}) but differ from the SDSS measurements ($-$6.6$\pm$4.5~{\kms}). 
Given the smaller uncertainties and overall fidelity with prior measurements, we adopt the MagE standard cross correlations with an additional systematic uncertainty as our final RV measurements. The only exception is J1750$-$0016, { whose large uncertainty (14~{\kms}) was due to lower S/N data ($\sim$23 at 7400~{\AA}) and the availability of only one RV standard; for this source alone} we adopt the line center value (Table~\ref{table:rvs}).

\subsection{Discrepant Radial Velocities}

In addition to J0517$-$3349, there are three sources whose measured velocities differ by { about 3$\sigma$ or more} from previously published values.
The L dwarfs J0500+0330 and J0835+0819 have radial velocities from \citet{2010ApJ...723..684B} that are more than 10~{\kms} different from our cross-correlation measures, our values being higher and lower, respectively.  For these sources, the lack of mid-L dwarf MagE templates, and noise in some spectral regions used for cross-correlation, resulted in only 3-5 cross-correlation measurements. Hence, we conclude that our measurement uncertainties are likely underestimated. 
For the M8 J2351$-$2537, our value ($-$12$\pm$3~{\kms}) deviates by 2.8$\sigma$ from that of \citet[-3.0$\pm$1.1~{\kms}]{2010A&A...512A..37S}, but is in agreement with that of \citet[$-$10$\pm$3~{\kms}]{2009ApJ...705.1416R}. Hence, either the Seifahrt measure is in error or this source is { binary} RV variable. Additional observations are warranted to test the latter hypothesis.
%J0500+0330 (5.1$\pm$3.1~{\kms} versus 15.94$\pm$0.16~{\kms} from \citealt{2010ApJ...723..684B}) and J0835$-$0819 (40.4$\pm$3.2~{\kms} versus 29.89$\pm$0.06~{\kms} from \citealt{2010ApJ...723..684B} and 26.5$\pm$2.2~{\kms} from \citealt{2010A&A...512A..37S}).

\section{Analysis}

\subsection{$UVW$ Space Motions and Kinematic Populations}

We combined our RV measurements with the proper motions and distances listed in Table~\ref{table:sample} to compute heliocentric $UVW$\footnote{For clarity, all $UVW$ velocities reported in this study are reported in the LSR.} space velocities in the Local Standard of Rest (LSR), following \citet{1987AJ.....93..864J}.  We adopted a right-handed coordinate system centered on the Sun, with $U$ toward the Galactic center, $V$ in the direction of Galactic rotation, and $W$ in the direction of the Galactic north pole.  Velocities were corrected to the LSR
assuming a solar velocity ($U,V,W$)$_{\sun}$ = (11.1, 12.24, 7.25)~{\kms} \citep{2010MNRAS.403.1829S}. Uncertainties were propagated through Monte Carlo sampling, assuming normal distributions for all measurement uncertainties.  We report the means and standard deviations of these calculations in Table~\ref{table:uvw}.

Figure~\ref{figure:UVW} displays the UVW velocities for our M and L dwarfs and compares them to the 2$\sigma$ velocity spheroids of the Galactic thin disk, thick disk and halo populations as tabulated in \citet{2003A&A...410..527B}. Most of our sources cluster around ($U,V,W$) = (0,0,0), with broader dispersions in $U$ and $V$ as compared to $W$.  There is also an asymmetric offset in $V$ which correlates with total LSR velocity; this is attributable to asymmetric drift  \citep{1924ApJ....59..228S}.  There is a noticeable trend between $U$ and $V$ among the L dwarfs in our sample, with $U < 0$ sources tending to also have $V < 0$ and vice versa. { We measure a correlation coefficient of $r$ = 0.43$\pm$0.03; a weaker trend is also seen among the M dwarfs ($r$ = 0.21$\pm$0.03)}. { These trends are} discussed { in detail} below.

We find that the majority of our sources, as expected, fall within the thin disk spheroid, consistent with ages $\lesssim$\,5~Gyr.  However, several sources fall well outside of this volume.  To assess kinematic population membership, we used the method of \citet{2003A&A...410..527B} to calculate relative probabilities, $P$(TD)/$P$(D) of membership between the thick (TD) and thin (D) disks.  Memberships were assigned as thin disk ($P$(TD)/$P$(D) $<$ 0.1), thick disk ($P$(TD)/$P$(D) $>$ 10) or intermediate thin/thick disk (0.1 $<$ $P$(TD)/$P$(D) $<$ 10).  The relative probabilities and population assignments are listed in Table~\ref{table:uvw}.
Only one source, J0707$-$4900, falls fully into the thick disk category, while eleven others are intermediate sources, evenly split between late-M and L dwarfs.  We also calculated relative probabilities for halo to thick disk membership, but none exceeded 0.002.

\subsection{Velocity Dispersions and Kinematic Ages}

As discussed in Section~1, the total velocity dispersion of a stellar population,
\begin{equation}
{\sigma}_v^2 \equiv  {\sigma}_U^2+{\sigma}_V^2+{\sigma}_W^2
\end{equation}
increases over time as dynamical scattering perturbs Galactic orbits. This produces a correlation between group dispersion and average age.
For our analysis, we considered two empirical laws for velocity diffusion. 
The first is the time-dependent decaying relation of \citet{1977A&A....60..263W}: 
\begin{equation}
\tilde{\sigma}_v(\tau)^3 = \tilde{\sigma}_{v,0}^3 +1.5\gamma_{v,p}T_{\gamma}\left(e^{{\tau}/{T_{\gamma}}}-1\right),
\label{equation:age1}
\end{equation}
where $\tilde{\sigma}_{v,0}$ = 10~{\kms}, $\gamma_{v,p}$ = 1.1$\times$10$^4$~({\kms})$^3$~Gyr$^{-1}$, $T_{\gamma}$ = 5~Gyr, $\tilde{\sigma}_v$ is the $|W|$-weighted total velocity dispersion\footnote{\citet{1977A&A....60..263W} defines the $|W|$-weighted velocity dispersion as having components $\tilde{\sigma}_x^2 = \alpha_x\sum_i|{W_i}|(X_i-\bar{X})^2/\sum_i|W_i|$ and $\alpha_x$ =  \{1,1,0.5\} for  $x$ = \{$U,V,W$\}.  As noted in \citet{2009ApJ...705.1416R} and \citet{2010A&A...512A..37S}, the weighted dispersion is required to make proper use of the Wielen relations.}  measured in {\kms}, and $\tau$ is the statistical age measured in Gyr.  The second is 
 a power-law relation, 
%derived a diffusion equation based on  $\sim$15,000 F and G stars with revised {\it Hipparcos} astrometry \citep{2007A&A...474..653V}, radial velocities from the Geneva Copenhagen Survey \citep{2004A&A...418..989N,2007A&A...475..519H,2009A&A...501..941H}, and isochrones from \citet{2008A&A...484..815B}.  This study quantifies the diffusion of the {\em unweighted} total velocity dispersion using the form
\begin{equation}
{\sigma}_v(\tau) = v_{10}\left(\frac{\tau+\tau_1}{10~{\rm Gyr}+\tau_1}\right)^{\beta}
\label{equation:age2}
\end{equation}
(cf.\ \citealt{2008gady.book.....B}), where $\sigma_v$ is the unweighted total velocity dispersion, and we used all six best-fit parameter sets in \citet[Table~2]{2009MNRAS.397.1286A}:
55.179 $\leq$ $v_{10}$ $\leq$ 57.975~{\kms}, 0.148 $\leq$ $\tau_1$ $\leq$ 0.261~Gyr and 0.349 $\leq$ $\beta$ $\leq$ 0.385.

Table~\ref{table:dispersions} lists the mean velocities, unweighted and weighted dispersions, and corresponding diffusion ages for various subsets of our sample.
Uncertainties for all values were propagated forward through Monte Carlo sampling, assuming Gaussian errors for our $UVW$ measurements.
For both the full sample (85 sources) and 20~pc sample (70 sources), we find nearly identical results, with unweighted total velocity dispersions around 44~{\kms} and equivalent
ages of 4.8$\pm$0.2~Gyr and 5.2$\pm$0.2~Gyr, respectively, based on the \citet{2009MNRAS.397.1286A} relation (the \citealt{1977A&A....60..263W} relation gives similar ages).
The derived ages are in good agreement with the radioisotopic age of the Sun, and are generally consistent with mean age estimates for M and L dwarfs from prior studies
(e.g., \citealt{2002AJ....124..519R,2009ApJ...705.1416R, 2010A&A...512A..37S,2010ApJ...723..684B}).

\subsubsection{Spectral Class Variations}

Separating the sample into late-M dwarfs (57 sources) and L dwarfs (28 souces), we find the former to be 1.0--2.5~Gyr younger, depending on the relation used. 
This result is robust even if the 11 low surface gravity sources, which may be members of YMGs, are rejected (see Section~6): the corresponding ages are 5.0$\pm$0.2~Gyr for late-M dwarfs versus 7.1$\pm$0.4~Gyr for L dwarfs based on the Aumer \& Binney relation.
These age differences are statistically significant, and similar to results by \citet{2010A&A...512A..37S} and \citet{2010ApJ...723..684B} who found their L dwarf samples to be kinematically more dispersed than the late-M dwarf samples of \citet{2002AJ....124..519R} and \citet{2009ApJ...705.1416R}.  
Our identical conclusion with a uniformly-analyzed sample confirms this unexpected trend, which we analyze in detail in Section~5.  

We note that the ages inferred between the \citet{1977A&A....60..263W} and \citet{2009MNRAS.397.1286A} relations are discrepant by 2--3$\sigma$ for these two subsamples; the relations produce more significantly discrepant ages for the unusually blue and inactive L dwarf samples described below.
For most subsamples, $\tilde{\sigma} \gtrsim {\sigma}$, as the inclusion of $|W|$-weighting tends to increase the inferred dispersion for that component.  However, for the L dwarfs,
$\tilde{\sigma}_V < {\sigma}_V$, driving down the total dispersion and hence Wielen ages.
\citet{1977A&A....60..263W} included $|W|$-weighting to account for an observed correlation between $V^2$ and $|W|$; i.e., the correlation between Galactic orbital eccentricity and inclination. For our M and L dwarfs, we find correlation coefficients $r$ = 0.40$\pm$0.04 and 0.01$\pm$0.07 between these parameters; i.e., there is no correlation for the L dwarfs, which leads to a biased age assessment using the $|W|$-weighting. For this reason, where the inferred ages between the \citet{1977A&A....60..263W} and \citet{2009MNRAS.397.1286A} relations diverge, we favor the latter.

\subsubsection{Color Deviants}

Late M and L dwarfs are known to exhibit broad variations in near-infrared color within a given subtype, variously attributed to surface gravity, metallicity and cloud effects (e.g., \citealt{2002ApJ...564..466G, 2004AJ....127.3553K, 2008ApJ...674..451B, 2008ApJ...689.1295K, 2008ApJ...686..528L}).  
As noted in Section~1, analysis of the velocity dispersions of color outliers have 
led to conflicting conclusions as to the relative ages of unusually red and blue M and L dwarfs \citep{2009AJ....137....1F,2010AJ....139.1808S,2010ApJS..190..100K}.
%The most extreme outliers on the blue end are metal-poor subdwarfs with halo kinematics, and their colors are the result of enhanced collision-induced {\hh} absorption in a high-pressure photosphere depleted of other molecular absorbers (e.g., \citealt{2000ApJ...535..965L,2003ApJ...592.1186B, 2007ApJ...657..494B}).  At the other extreme, very young VLM dwarfs, including planetary-mass companions to nearby stars, are often found to be red outliers, due to reduced {\hh} absorption and possibly enhanced condensate cloud opacity (e.g., \citealt{2000MNRAS.314..858L,2006ApJ...639.1120K, 2006ApJ...651.1166M, 2011ApJ...729..128C}). In line with these extrema, both \citet{2009AJ....137....1F} and \citet{2010AJ....139.1808S} have found distinct tangential velocity dispersions among unusually red and blue dwarfs, being smaller and larger, respectively, compared to the overall population.  In contrast, \citet{2010ApJS..190..100K} have found higher tangential velocity dispersions for both red and blue outliers identified in a proper-motion selected sample, and argue against any age difference between these color groups.

We defined color deviants in our sample as having $J-K_s$ colors more than 0.15~mag redder or bluer than the median color for their optical spectral type, as delineated in 
the SDSS samples of \citet{2010AJ....139.1808S} and \citet{2011AJ....141...97W}. 
{ The threshhold color offset was chosen because it is comparable to the scatter in median color versus spectral type in these studies, is at least three times the color uncertainties of the vast majority of our sample ($\sigma_{J-K_s}$ = 0.03--0.04~mag), and provides a small but statistically robust sample of outliers (i.e., 10--15\% of the sample).  As $J-K_s$ colors generally become more dispersed toward later spectral types (e.g., \citealt{2008ApJ...689.1295K}), a constant threshold value may probe more ``extreme'' M dwarfs as compared to L dwarfs; however, as shown below, our L dwarf color outliers appear to be more kinematically distinct. 
Due to the systematic
differences in the colors of our sources compared to the \citet{2010AJ....139.1808S} and \citet{2011AJ....141...97W} trends (Figure~\ref{figure:sptdistribution}), all but one of our unusually red dwarfs (9 sources) are M dwarfs (J0835--0819 is the sole red L dwarf) and all of our unusually blue dwarfs are L dwarfs (11 sources).}  
%We also restrict their distances to 20~pc or less.
%Table~\ref{table:dispersions} lists the mean velocities, unweighted and weighted dispersions and ages for these subsamples.

For the unusually red dwarfs, we find a kinematic age of 2.0$\pm$0.2~Gyr for the \citet{2009MNRAS.397.1286A} dispersion relation, considerably younger than the full sample. 
Remarkably, only one of these red sources, J2045$-$6332, exhibits Li~I absorption and is identified as a low surface gravity dwarf. The remaining sources may have thicker photospheric condensate clouds and/or comprise a coherent, and possibly metal-rich, stream. 
The unusually blue L dwarfs, on the other hand, have a large velocity dispersion (61~{\kms}) and dispersion age of 12.4$\pm$0.9~Gyr (for the \citealt{1977A&A....60..263W} relation the dispersion age is a more reasonable 7.0$\pm$0.2~Gyr). 
Roughly 45\% of these sources are also identified as intermediate thin/thick disk stars. 
{ The fact that the kinematically colder red dwarfs are mostly type M and the kinematically warmer blue dwarfs are all of type L provides a possible explanation for the age offsets between these spectral classes.}  Indeed, when color deviants are excluded, the kinematic ages of the late-M dwarfs (4.0$\pm$0.2~Gyr) and L dwarfs (3.4$\pm$0.3) conform to expectations from population simulations (see below). Thus, we have evidence that the age discrepancy between late-M and L dwarfs originates as a color discrepancy.  Section~5.2 discusses this insight in further detail.

\subsubsection{Magnetically Active and Inactive Dwarfs}

Nonthermal magnetic emission is a common metric for low-mass stellar ages, as emission declines in strength as stars spin down through wind-driven angular momentum loss (e.g., \citealt{1963ApJ...138..832W,1972ApJ...171..565S,1993ApJ...409..624S,1995ApJ...450..401F}).  This process appears to be less efficient for late-M and L dwarfs based on the longer timescales inferred for magnetic field decay \citep{2006AJ....132.2507W,2007AJ....133.2258S,2008AJ....135..785W,2008ApJ...684.1390R} and faster rotation rates among older L dwarfs \citep{2003ApJ...583..451M,2006AJ....131.1806R,2010ApJ...710..924R}.  There has also been evidence of a reversal in the standard age-activity relation in the L dwarf regime, with only the older and more massive stellar L dwarfs having sufficient field energy to drive nonthermal emission \citep{2000AJ....120.1085G,2010A&A...522A..13R,2011ApJ...739...49B}.

As discussed in Section~2.4, while the majority of our sample exhibits H$\alpha$ emission, the L dwarfs are roughly evenly split between active and inactive sources.  Comparing these subsets, we find the that active L dwarfs have a significantly smaller overall dispersion (45.9$\pm$1.1~{\kms}) than the non-active L dwarfs (51.5$\pm$0.7~{\kms}) and a correspondingly younger dispersion age (5.6$\pm$0.4~Gyr versus 7.8$\pm$0.5~Gyr for the Aumer \& Binney relation).  This would appear to confirm an underlying age-activity relation that is similar to earlier-type stars; i.e., older L dwarfs are less active.  
%We note that the \citet{1977A&A....60..263W} age-dispersion relation comes to the opposite conclusion (5.6$\pm$0.3~Gyr versus 4.86$\pm$0.12~Gyr), but again the lack of correlation between $V^2$ and $|W|$ for the L dwarfs argues against the $|W|$-weighting used in the Wielen relation.

\subsubsection{Li-bearing and Low Surface Gravity Brown Dwarfs}

The nine late M and L dwarfs exhibiting {\lii} absorption exhibit the smallest velocity dispersions in our sample, with $\sigma_v$ = 10.4$\pm$0.5~{\kms}.  Both age dispersion relations are undefined for this value, yielding effective upper limits of order 100~Myr. This is qualitatively in line with evolutionary model predictions from \citet{2003A&A...402..701B}, as a Li-bearing M7.5 ({ versus} L1.5) dwarf with an estimated\footnote{ Based on the {\teff}/spectral type relation of \citet{2009ApJ...702..154S}.} {\teff} $\approx$\,2600~K ({ versus} 2050~K) and mass below 0.06~{\msun} should have an age of no more than 120~Myr ({ versus} 570~Myr).  Several of these sources show low surface gravity features in their optical spectra (Section~2.5) and are kinematic members of nearby young moving groups with ages of 10-100~Myr (Section~5). Hence, the velocity dispersions are consistent with overall spectral properties. Similarly, the total velocity dispersion for the 11 low surface gravity sources is 11.9$\pm$0.6~{\kms}, again implying ages $\lesssim$\,100~Myr.

\subsection{Velocity Probability Distributions}

The ages estimated in the previous section are contingent on the velocity distributions being Gaussian.  However, both visual examination of the $UVW$ plots in Figure~\ref{figure:UVW} and discrepancies between the age-dispersion relations used argue that non-Gaussian effects are likely present (see also \citealt{2014MNRAS.439.1231B}).  We therefore constructed probability (``probit'') plots 
for our various subsamples for each of the $U$, $V$ and $W$ coordinates, following procedures described in previous studies \citep{1980AJ.....85.1390L,2002AJ....124.2721R,2007AJ....134.2418B,2009ApJ...705.1416R}.
Probit plots are a rank order mapping of velocity to Gaussian probability, generating a straight line for a single Gaussian distribution with a slope equal to the standard deviation. Deviations from Gaussian emerge as variations in the slope.  

Figure~\ref{figure:disperse_20pc}--\ref{figure:disperse_ldwarf} displays $UVW$ probit plots for our 20~pc, late-M dwarf and L dwarf subsamples.  It is clear that only the $U$ distributions for the 20~pc and late-M dwarf samples are single Gaussian distributions; both $V$ and $W$ distributions show significant slope variations beyond $\pm$1$\sigma$, with those in $W$ being more pronounced.  
Following \citet{2007AJ....134.2418B}, we performed a piece-wise fit to these trends, over ``core'' ($|\sigma| \le 1$) and ``wing'' components ($|\sigma| > 1$), sampling the data uncertainties through Monte Carlo analysis.  Table~\ref{table:dispersions} lists the resulting $UVW$ and total velocity dispersions (unweighted) and corresponding ages based on the \citet{2009MNRAS.397.1286A} relation.  In general, core components have dispersions and ages just below those of the full sample analysis above, while the wing components have greater dispersions and older kinematic ages.  The $V$ probit plots exhibit extended tails to negative velocities and curvature at positive velocities, which again can be attributed to asymmetric drift.

Focusing on the M and L dwarf subsamples, we find clear differences in the velocity distributions for all three components, most notably in $U$.  Here, the L dwarfs exhibit both a nonzero mean velocity ($\langle{U}\rangle$ = 14.7$\pm$0.5~{\kms}) and a pronounced asymmetry about this mean.  The offset indicates a net flow of L dwarfs toward the Galactic center, and persists even when color deviants are excluded ($\langle{U}\rangle$ = 10.1$\pm$0.8~{\kms}). The slope change in the probit plot across the mean is not seen in any of the other velocity components or subsamples.  This pattern is remarkable and, along with the correlation between $U$ and $V$ and lack of correlation between $V^2$ and $\mid{W}\mid$, suggests that the kinematics of L dwarfs in our sample are distinct, either intrinsically or though sample bias. We focus on this problem in Section~5.

\subsection{Galactic Orbits}

As a final examination of the statistical properties of our sample, we used the $UVW$ velocities and Galactocentric coordinates to compute Galactic orbits in a static, axisymmetric potential.  We followed the same strategy as described in \citet{2011AJ....142..169B}, converting heliocentric
velocities to cylindrical velocities ($V_R$, $V_{\phi}$, $V_Z$) in the Galactic frame of rest, assuming an LSR azimuthal motion of 240~{\kms} \citep{2014ApJ...783..130R,2015ApJ...800L..32A}.
We computed Galactic spatial coordinates for our sources relative to an assumed
solar position of ($X,Y,Z$)$_{\sun}$ = ($-$8.43, 0, 0.027)~kpc \citep{2001ApJ...553..184C,2014ApJ...783..130R}, with $XYZ$ defined in the same manner as $UVW$.
For the Galactic potential, we adopted static,
axisymmetric oblate Plummer's sphere models for the Galactic halo, bulge and disk, using the forms
described in  \citet{kuzmin1956} and \citet{1975PASJ...27..533M}, with parameters from \citet{1995A&A...300..117D}.  A fourth-order Runge-Kutta integrator was used to calculate the orbit over a period of $\pm$250~Myr about the current epoch in 10~kyr
steps, and both energy and the $Z$-component of angular momentum were conserved to better than one part
in 10$^{-3}$.  To sample measurement uncertainties,
we computed 100 orbits for each source, varying the initial distances and velocities in a Monte Carlo fashion assuming normal distributions scaled to the uncertainties listed in Tables~\ref{table:sample} and~\ref{table:uvw}.

Figure~\ref{figure:orbits} displays the distributions of inferred orbital elements for our sample: minimum and maximum Galactic radius, maximum absolute vertical displacement, eccentricity ($e \equiv [R_{max}-R_{min}]/[R_{max}+R_{min}]$), and maximum inclination ($\tan{i} \equiv Z/\sqrt{X^2+Y^2}$). 
%and specific angular momentum ($\vec{r} \times \vec{v}$; for the Sun this is $-$1.90~kpc$^2$~Myr$^{-1}$). 
As expected for a sample dominated by the thin disk population, the majority of our sources exhibit circular (e $\lesssim$ 0.15) and planar orbits ($i \lesssim 2\degr$), although the core of the eccentricity distribution extends to 0.2.  
Figure~\ref{figure:orb0707} shows the orbits of the two sources with the largest values of $P$(TD)/$P$(D), the M8.5  
J0707$-$4900 and the L1 J0921$-$2104.  Both have fairly eccentric orbits ($e$ = 0.4) which carry them to perigals just outside the spherical bulge (R $\approx$ 4~kpc; \citealt{2008gady.book.....B}), suggesting that they may be scattered bulge stars. 
In contrast, several sources currently near perigal reach Galactic radii of over 14~kpc (e.g., the D/TD dwarfs J1539$-$0520 and J2331$-$2749). Our local VLM dwarf population therefore samples a significant region of the Galaxy, a point returned to in Section~5.  Note that all of the orbits are prograde, consistent with formation in the disk (e.g., \citealt{2008Natur.451..216C}).  

Given the distinct velocity distributions of the M and L dwarfs in our sample, we examined whether their Galactic orbits differed as well. Figure~\ref{fig:orbml} compares the distributions of minimum and maximum Galactic radii for these two subsamples. The L dwarfs have a broader distribution of Galactic radii, with a notable skew to higher maximum radii.  Median values for $R_{max}$ for M and L dwarfs are 9.5~kpc and 10.3~kpc, respectively; while median eccentricities are 0.11 and 0.16. In contrast, the median values of $\mid{Z_{max}}\mid$ and orbital inclination are the same for both groups.  It appears that the L dwarfs in our sample are distributed more broadly in Galactocentric radius than the late-M dwarfs, which is { directly attributable} to their unusual $U$ velocity distribution.

\section{Discussion: Why are Local L Dwarfs Blue and Dispersed?}

The confirmation that nearby L dwarfs appear to be, on average, more dispersed and kinematically older than nearby late-M dwarfs, and that this dispersion is driven by a large fraction of unusually blue L dwarfs, suggests that there is something { unusual} about the local L dwarf population.  
%stands in contradiction to population synthesis simulations (e.g., \citealt{2004ApJS..155..191B,2005ApJ...625..385A}).
%In these simulations, older brown dwarfs evolve out of the L spectral class; at 5~Gyr, a 0.06~{\msun} brown dwarf is a $\sim$1000~K T dwarf \citep{2001RvMP...73..719B}.  This leaves only young, substellar L dwarfs and a small fraction (small mass range) of old, stellar L dwarfs. 
%The evolutionary segregation of brown dwarfs explains the lower
%space densities of L dwarfs as compared to late M and T dwarfs in the Solar Neighborhood \citep{1999ApJ...521..613R,2004ApJS..155..191B,2007AJ....133..439C}, so
%it is remarkable that our velocity data, and the studies of \citet{2010A&A...512A..37S} and \citet{2010ApJ...723..684B}, come to an opposite conclusion about L dwarf dispersion ages.
Most notable is the asymmetric $U$ distribution of this population that remains even when color outliers are rejected,
and indicates the existence of a net radial flow of L dwarfs that is not matched by the late-M dwarfs. 
Asymmetries in the radial motions of local stars have been observed, and are generally
attributed to resonances with Galactic structures; e.g., the Galactic bar and/or 
spiral arm patterns \citep{1998MNRAS.298..387D,2000AJ....119..800D,2002MNRAS.336..785S,2004MNRAS.350..627D,2005AJ....130..576Q,2007A&A...461..957F,Minchev2010,2011MNRAS.417..762Q}.
Recent large-scale 
{ RV} surveys, most notably the RAdial Velocity Experiment (RAVE; \citealt{2006AJ....132.1645S,2011MNRAS.412.2026S,2013MNRAS.436..101W,2014MNRAS.439.1231B,2014ApJ...793...51S})
have measured
statistically significant { RV} gradients of order 3--10 {\kms}~kpc$^{-1}$ directed toward
the Galactic center.  
\citet{2015ApJ...800L..32A} also find that thin disk stars in the RAVE survey ([M/H] $\geq$ -0.1, $|Z|$ $<$ 0.5~kpc) exhibit a correlation between radial and azimuthal motions, with trailing sources ($V_{\phi} < 0$)
streaming outward ($V_{r} < 0$) and leading sources ($V_{\phi} > 0$) streaming inward ($V_{r} > 0$).
They attribute this correlation to the local Outer Lindblad Resonance with the Galactic bar, which also
builds the Hercules stream \citep{1991dodg.conf..323K,2000AJ....119..800D,2014A&A...563A..60A}.
We see precisely this same trend among the L dwarfs in their $U$ and $V$ velocities (Figure~\ref{figure:UVW}), but curiously not among the late-M dwarfs.

Given prior evidence of radial motion gradients among local stars, we hypothesize that the variance
between the late-M and L dwarfs in our sample arises from two possible sources: 
(1) an inherent asymmetry in the ages and velocity distribution of local L dwarfs made manifest by brown dwarf thermal evolution; and (2) a bias among the L dwarfs in our sample or in the local 20~pc population. 
We consider each of these hypotheses in turn.

\subsection{Is There an Inherent Asymmetry in L Dwarf Ages and Kinematics?}

The lowest-order inherent asymmetry in the distribution of stars in the Galactic disk is the radial density distribution, which increases toward the Galactic center. To test whether this spatial asymmetry, coupled with brown dwarf evolution, could drive an asymmetry in the radial motions of L dwarfs, we performed a Monte Carlo population simulation combining brown dwarf evolution, age-dependent velocity dispersions, radial mixing through Galactic orbital motion, and selection biases { inherent to} a local sample. 

A population of stars and brown dwarfs was generated as described in \citet{2004ApJS..155..191B,2007ApJ...659..655B}, assuming an initial mass function
\begin{equation}
\frac{dN}{dM} \propto M^{-\alpha}
\end{equation}
and age distribution (star formation rate)
\begin{equation}
\frac{dN}{d\tau} \propto e^{\beta(\tau-T_0)}.
\end{equation}
Here, $M$ is mass, constrained to 0.01~{\msun} $\leq M \leq$ 0.20~{\msun};
$\tau$ is age, constrained to 0.2~Gyr $\leq \tau \leq$ 8~Gyr;
$N$ is the number density of stars in a given volume;
$\alpha$ = \{$-$0.5, 0.0, 0.5, 1.0\} is the mass function power-law index;
$\beta$ = \{$-$0.5, 0.0, 0.5, 1.0\} is the star formation rate power-law index;
and $T_0$ = 8~Gyr was adopted as the oldest age of stars in the Solar Neighborhood \citep{2013A&A...560A.109H}. 
Note that $\alpha > 0$ yields a population dominated by lower-mass objects ($\alpha$ $\approx$\,0.5 in nearby clusters; see \citealt{2010ARA&A..48..339B} and references therein), while $\beta > 0$ yields a population dominated by older objects ($\beta$ = 0 is a common assumption for Galactic population simulations, $\beta$ = 1 is consistent with the integrated star formation rate
history of field galaxies; see \citealt{1998ApJ...498..106M,2009MNRAS.397.1286A}).  Drawing 10$^5$ values of $M$ and $\tau$ from these distributions, we inferred the present-day effective temperatures of each source using the evolutionary models of \citet{2001RvMP...73..719B}, and converted these to spectral types using the empirical relation of \citet{2009ApJ...702..154S}. We
limited our analysis to those sources with spectral types between M7 and L5, which represents 6-17\% of the original simulation sample depending on $\alpha$ and $\beta$.  We did not consider the role of multiplicity.

Asymmetry in the Galactic stellar distribution and radial mixing were implemented by assigning initial Galactic radii in the range 5.5--11.5~kpc, drawing from a radial exponential distribution
\begin{equation}
\frac{dN}{dR} \propto e^{(R_{\odot}-R)/L}
\end{equation}
where $R$ is the radial coordinate, $R_{\odot}$ = 8.43~kpc is the solar Galactic radius \citep{2014ApJ...783..130R}, and $L$ = 2.1~kpc is the radial scaleheight \citep{2008ApJ...673..864J}. 
We then assigned $UVW$ velocities in the LSR based on normal distributions centered on zero and with standard deviations determined from the assigned age and the \citet{2009MNRAS.397.1286A} age-dispersion relations above.  We also included an asymmetric drift term
\begin{equation}
V_a = \frac{\sigma_R^2}{74~{\rm km/s}} = 23.7\left(\frac{\tau}{10~{\rm Gyr}}\right)^{0.614}~{\rm km/s}
\end{equation}
based on \citet{2009MNRAS.397.1286A}.
The assigned velocities and initial Galactic coordinates (assuming $Y$ = $Z$ = 0) were used as initial conditions to compute orbits over 500~Myr in steps of 1~Myr in an axisymmetric potential as described in Section~4.4. From these orbit calculations we identified all timesteps among all sources for which ${\mid}R-R_{\odot}{\mid} \leq$ 50~pc and ${\mid}Z{\mid} \leq$ 50~pc. These 7,000--30,000 orbital snapshots (the number depending on simulation parameters) comprised our ``local'' sample, and included multiple instances of sources which repeatedly fell within the Galactic Solar torus. 
%We did not include the effects of spiral structure in our orbit calculations, the potential impact of which is discussed below.

%The number of stars scattering inward or outward from one radius $R$ to another radius $R^{\prime}$ scales as
%\begin{equation}
%dN_s(R,R^{\prime},\Delta{R}) \propto f(R,R^{\prime},\Delta{R})\rho(R)RdR
%\end{equation}
%where $f(R,R^{\prime},\Delta{R})$ is the fraction of time a star from radius $R$ is found in the ``local'' annulus $R^{\prime}{\pm}\Delta{R}$ during its Galactic orbit. The local stellar population near the Sun is then
%\begin{equation}
%N(R_{\odot},\Delta{R}) \propto \int_{R_1}^{R_2}f(R,R_{\odot},\Delta{R})\rho(R)RdR 
%\end{equation}
%over the radial annulus $R_1$ to $R_2$. To compute $f(R,R_{\odot},\Delta{R})$, we used the ages of our simulated dwarfs to assign $U,V,W$ velocities from normal distributions centered on the LSR and with standard deviations assigned by the \citet{2009MNRAS.397.1286A} dispersion relations used above.  

Figure~\ref{fig:mfsimsummary} summarizes the results of our baseline simulation with $\alpha$ = 0.5 and $\beta$ = 0.0, comparing distributions between the initial simulation sample to the dynamically-evolved and locally-selected sample.
Independent of selection mechanism, these calculations affirm prior results that the L dwarf population should be
on average 0.2--0.4~Gyr younger than the late-M dwarf population due to the loss (through thermal evolution) of old brown dwarfs.
This confirms the results for our ``normal'' color populations, but not the full sample of late-M and L dwarfs.
Dynamical evolution and local selection results in a uniformly younger ``observed'' population, with both late-M and L dwarfs being 0.3--0.6~Gyr younger than the initial simulation sample. This offset stems from preferential selection of young objects originating near R$_{\odot}$, while most of the older objects scatter outward and are not fully replaced by older objects scattered into R$_{\odot}$.
We note that the mean ages of locally-selected late-M and L dwarfs in this baseline simulation are somewhat younger than the ages
inferred from our velocity analysis; indeed, a value of $\beta$ between 0.0 and 0.5 appears to be more consistent with the data, suggesting a decline in the VLM star/brown dwarf formation rate over the age of the Galaxy.
Nevertheless, we see no evidence of a distinct, highly-dispersed population of L dwarfs for any of the simulation parameters examined.

To produce a sample in which L dwarfs are on average older than late-M dwarfs, we explored cases where the star formation history differed between stars and brown dwarfs.
A divergent formation history could arise from mass-dependence in the Galactic birth rate.
Figure~\ref{fig:mfsimsummary2} shows the results for a simulation assuming $\beta$ = 0.5 for brown dwarfs (M $<$ 0.07~{\msun}) and $\beta$ = 0.0 for stars (M $>$ 0.07~{\msun}); we also considered the opposite $\beta$ assignments.  The simulation with older brown dwarfs does indeed produce a more dispersed and kinematically older population of L dwarfs by 0.4--0.7~Gyr, although the mean dispersion ages are again younger for both M and L dwarfs than those observed in our full sample.  
 
{ One possible resolution to the divergent results between simulated and observed kinematic dispersion ages is error in the evolutionary models used to predict the timescales for brown dwarf cooling in our simulations. If the cooling rates are slower than these models predict, we would expect VLM sources to remain L dwarfs for longer periods of time, bringing the simulations into agreement with our observations. Indeed, recent observations of L-type binaries with orbital mass measurements have shown that these sources are 60-100\% more luminous than models predict \citep{2009ApJ...692..729D,2010ApJ...711.1087K,2014ApJ...790..133D}, so this is a valid concern for our simulations.}

The simulations generated two clear asymmetries in the spatial and velocity properties of late-M and L dwarfs. 
First, the majority of locally-selected dwarfs in all of the simulations originate (or are at least initially placed) within the Solar radius, with an average offset of $\sim$1~kpc. This largely reflects the assumed exponential decline in stellar density with Galactic radius. 
Second, the azimuthal velocity distributions are skewed to negative velocities, reflecting both the initial radial distribution of stars (which lose azimuthal speed as they climb out of the Galactic potential) and our input azimuthal drift term.
However, there are no significant differences between the late-M and L dwarfs for these
two parameters, nor for radial and vertical velocity distributions, which are symmetric about and centered on zero.
This is true even for separate values of $\beta$ between stars and brown dwarfs.   
It appears that we cannot reproduce Galactic radial velocity asymmetries with an axisymmetric potential, as individual stars
passing through the local volume have nearly the same probability of moving radially inward as outward on this scale.  

Hence, while an older population of L dwarfs can be produced with an assumption of different star formation rates { between} stars and brown dwarfs, { or may be reproduced if there are errors in the evolutionary models; an}
inherent asymmetry in the $U$ velocity distribution of L dwarfs appears to require perturbations from a non-axisymmetric source; i.e., the Galactic bar and spiral structure. Analysis of { these hypotheses are left to more detailed simulations in} a future publication, although it is important to assert that any radial mixing induced by Galactic structure must influence M and L dwarfs differently to match our results and those of \citet{2010A&A...512A..37S} and \citet{2010ApJ...723..684B}. 
%In particular, perturbations that vary over the history of the Galaxy may be necessary to reproduce these kinematics.

\subsection{Is There Sample or Cosmic Bias in the Local L Dwarf Population?}

A more mundane explanation { for our results} is that the sample considered here is kinematically biased in its construction.
The sample was drawn primarily from all-sky imaging and proper motion surveys, which continue to be incomplete in the Solar Neighborhood (e.g., \citealt{2013ApJ...767L...1L,2014A&A...561A.113S,2014A&A...567A...6P}).  Incompleteness is particularly an issue along the Galactic plane due to source crowding. For our sample, we also have a declination limit imposed by the observing site ($\delta < 25\degr$; Figure~\ref{figure:skymap}). To assess whether these ``pointing'' asymmetries produce velocity trends, Figure~\ref{fig:srcskydist} displays the distributions of $XYZ$ coordinates for our sample.  The M and L dwarfs have similar distributions in $X$ and $Y$, but in $Z$ the L dwarfs are more centrally concentrated with a slight bias toward positive $Z$ (both subsamples have fewer sources at $Z$ = 0 due to Galactic plane exclusion). 
However, our simulations show no correlation between local $Z$ position and $U$ velocity; simulations by \citet{2014MNRAS.440.2564F} which include spiral perturbations are similarly symmetric about the Galactic plane. It is therefore unclear what role this difference in vertical spatial distribution would play in producing an asymmetric radial velocity distribution.

Another possible bias is the contribution of YMG members in our sample. At least 8 of the sources investigated here are kinematically associated with YMGs, 7 of which are late-M dwarfs (Section~6).  Similarily, 9 of the 11 low surface gravity dwarfs and 8 of the 9 sources exhibiting Li~I { absorption} are late-M dwarfs.  There is clearly a ``youth bias'' between the M and L dwarf subsamples. However, as noted in Section~4.2.1, while rejecting sources which exhibit low surface gravity features slightly increases the velocity dispersion for the late-M dwarfs and brings their kinematic age closer to (but still less than) the L dwarfs, editing out the low gravity sources does not change the underlying $U$ velocity distributions. Late-M dwarfs remain symmetrically distributioned about $U$ = 0, and L dwarfs asymmetric and offset.  Indeed, removal of the two low gravity L dwarfs in our sample increases the mean velocity offset of the remainder.  Hence, contamination by YMGs does not appear to resolve the velocity differences between the M and L dwarfs.  

A third possibility is that the local sample itself has an inherent cosmic bias.  As described in Section~4.2.2, the unusually blue dwarfs, which are all L dwarfs, are far more dispersed than the unusually red dwarfs, which are predominantly late-M dwarfs.  The unusually blue L dwarfs represent 39\% of all the L dwarfs in our sample.   This suggests a selection effect.  However, the identification of color deviants is based on mean near-infrared colors from the optically-selected SDSS surveys, which as discussed in \citet{2010AJ....139.1808S} are less color-biased than 2MASS samples and tend to identify bluer sources.  It is therefore remarkable that a plurality of the L dwarfs examined here, mostly identified in the 2MASS survey, are bluer still.  Evidence that this color skew may actually be a local effect emerges from the fact that 10 of the 11 unusually blue L dwarfs are within 20~pc (the exception is J0923+2340), which is a higher fraction (91\%) than the remainder of the L dwarf sample (65\%). In other words, the unusually blue L dwarfs are more representative of the local volume than the ``normal-color'' L dwarfs { delineated in \citet{2010AJ....139.1808S}}.  It is possible that we are seeing two distinct populations of L dwarfs  in the 20~pc volume: a ``disk'' group whose dispersions conform to simulation expectations, and a ``dispersed'' group drawn from an older, possibly thick disk VLM population. Given the relatively small number of L dwarfs examined in this study,
% ($\sim$25\% of the total estimated in the full 20~pc volume; \citealt{2007AJ....133..439C}), 
this speculative hypothesis must be confirmed through a larger study.

\section{Candidate Kinematic Members of Nearby Moving Groups and Associations}

Several of our sources exhibit Li~I absorption and/or spectral features indicative of low surface gravity, and as such are
potential members of YMGs. To assess which YMGs these sources are affiliated with, and their probability of affiliation, 
we used the Bayesian Analysis for Nearby Young AssociatioNs II tool (BANYAN~II; \citealt{2014ApJ...783..121G}) which uses spatial and velocity coordinates and photometry to assess the probability of membership ($P_M$) and field contamination ($P_C$)
for individual sources. Our YMG sample included the 
TW~Hydrae Association (TWA; \citealt{1997Sci...277...67K}; \citealt{1989ApJ...343L..61D}, \citealt{2004ARA&A..42..685Z}; 10~Myr; \citealt{2013ApJ...762..118W}), 
the $\beta$~Pictoris Moving Group ($\beta$PMG; \citealt{2001ApJ...562L..87Z}; 20 -- 26~Myr; \citealt{2014MNRAS.445.2169M}, \citealt{2014ApJ...792...37M}, \citealt{2014MNRAS.438L..11B}), 
the Tucana-Horologium Association (THA; \citealt{2000AJ....120.1410T}, \citealt{2000ApJ...535..959Z}; 40~Myr; \citealt{2014AJ....147..146K}), 
the Carina association (CAR; 20 -- 40~Myr; \citealt{2008hsf2.book..757T}),
the Columba association (COL; 20 -- 40~Myr; \citealt{2008hsf2.book..757T}),  
the Argus/IC~2391 association (ARG; 30 -- 50~Myr; \citealt{2008hsf2.book..757T}), and 
the AB~Doradus moving group (ABD;  \citealt{2004ApJ...613L..65Z}; 110 -- 130~Myr ; \citealt{2005ApJ...628L..69L}, \citealt{2013ApJ...766....6B}). We adopt the spatial and kinematic models for each of these associations given in \cite{2014ApJ...783..121G}, and include distances in our comparison for those sources with trignometic parallax measurements.

Table~\ref{table:ymg} lists the membership and contamination probabilities of sources with membership probabilities $P_M > 10$\%.  We also list the spatial ($\Delta{D}$) and velocity ($\Delta{V}$) offsets from the { respective centers} of the best-match association.  In the following discussion on individual candidates, we used the effective temperature ({\teff})/spectral type calibration of \citet{2009ApJ...702..154S} and the evolutionary models of \citet{2003A&A...402..701B} to estimate physical parameters.

\subsection{Previously Known Candidate Members}

%[ 2045-6332]

{ J0041--5621AB (M6.5 + M9)}: The combined-light spectrum of this resolved binary displays H$\alpha$ emission, Li~I absorption, and weak low surface gravity features. It was previously identified as a possible member of either THA or $\beta$PMG by \citet{2009ApJ...705.1416R} on the basis of its Li~I absorption, kinematics and evidence of ongoing accretion.
\citet{2014ApJ...783..121G} favored association with THA based on a BANYAN II analysis of the same data, and our revised analysis supports that conclusion, with a membership probability $P_M > 99.9$\% and $P_C < 0.1\%$. While this source does not have a trigonometric parallax, the BANYAN~II tool predicts a statistical distance of $41^{+2}_{-3}$~pc if it is a member of THA, placing it 7~pc and 1.4~{\kms} away from the center of the spatial and kinematic model.
\citet{2010A&A...513L...9R} resolved the system into a 142.8$\pm$0.5~mas binary and estimated component types of M6.5+M9 and masses of 30~{\mjup} and 15~{\mjup} for an age of 10~Myr.  Our combined-light spectrum is considerably later than M6.5 (Figure~\ref{fig:lowg_weak}), suggesting that the primary may be cooler than inferred in that study. 
Adopting effective temperatures of 2500~K and 2400~K based on component types of M8+M9, a THA age of 40~Myr, a distance of 41~pc, and the evolutionary models of \citet{2003A&A...402..701B}, we predict masses of 40~{\mjup} and 35~{\mjup}, a projected separation of 6~AU, and an orbit period of 50~yr.  The last value is less than half that estimated by \citet{2010A&A...513L...9R}, and suggests that orbital motion could be detectable over the coming decade.
 
 { J0123--6921 (M9)} was proposed as a THA member by \cite{2014ApJ...783..121G}, and our reanalysis supports this conclusion ($P_M > 99.9$\% and $P_C < 0.1$\%).  Like J0041--5621AB, this source exhibits Li~I absorption, H$\alpha$ emission and weak signatures of low surface gravity. We again infer a signficantly later spectral type for this source than the M7.5 reported in \citet{2009ApJ...705.1416R}, and we estimate its mass to be 35~{\mjup}.

{ J0339--3525 (LP 944-20, M9)} has been identified by \cite{2003A&A...400..297R} as a candidate member of the controversial Castor ``association'' (CAS; \citealt{1998A&A...339..831B}). In contrast, \cite{2014ApJ...783..121G} identify J0339--3525 was a candidate member of ARG using BANYAN~II, but note that its \emph{XYZUVW} coordinates are much closer to CAS. Our findings show a weak membership probability for ARG ($P_M = 16.7$\%) but low field contamination ($P_C = 0.3$\%), suggesting that it is likely part of a different association.  The spectrum of J0339--3525 shows weak features of low surface gravity, with particularly enhanced VO absorption at 7400\,{\AA} and weak and narrow alkali lines, H$\alpha$ emission, and weak Li~I absorption \citep{1998MNRAS.296L..42T,2009ApJ...705.1416R}. All of these observations point to either a very young ($\lesssim$\,30~Myr) low-mass brown dwarf or, as argued by
\citet{1998MNRAS.296L..42T}, \citet{2003A&A...400..297R} and \citet{2007MNRAS.380.1285P} an ``intermediate''-aged ($\approx$\,300--600~Myr) brown dwarf with partial Li depletion.  It has been recently argued that hypothesized CAS ``members'', which have included Fomalhaut ABC (440$\pm$40~Myr; \citealt{2012ApJ...754L..20M}) and Vega (455--700~Myr; \citealt{2010ApJ...708...71Y,2012ApJ...761L...3M}), do not share a common origin or composition, and is likely a dynamical stream (e.g., \citealt{2013AJ....146..154M,2012ApJ...754L..20M,2012ApJ...761L...3M,2013ApJ...778....5Z}).  Indeed, \citet{2013AJ....146..154M} specifically note that J0339--3525's current proximity to Fomalhaut may be short-lived.  J0339--3525 may therefore be a ``juvenile'' brown dwarf caught up in a Galactic dynamical pattern rather than a member of a bona-fide association.

{ J1139--3159 (TWA~26, M9$\gamma$)} is a previously reported member of TWA \citep{2002ApJ...575..484G}, and our results are consistent with this assignment ($P_M = 99.6$\% and $P_C = 0.1$\%). With strong signatures of low surface gravity, Li~I absorption and a full complement of H$\alpha$, H$\beta$ and H$\gamma$ emission, this source is an unambiguous young brown dwarf.

{ J2000--7523 (M9)} is a previously reported candidate of $\beta$PMG \citep{2014ApJ...783..121G} or CAS \citep{2010MNRAS.409..552G} with pronounced low surface gravity features. Our BANYAN analysis favors the $\beta$PMG assignment ($P_M = 99.0$\%, $P_C = 4.1$\%). Notably, this source does not show Li~I absorption, an effect likely related to weak alkali lines in low surface gravity photospheres as posited by \citet{2008ApJ...689.1295K}. That study suggests an age of $\lesssim$\,30~Myr for such sources, which is consistent with $\beta$PMG membership.

{ J2045--6332 (M9)} was identified in \citet{2010MNRAS.409..552G, 2014MNRAS.439.3890G} as a high-probability member of CAS (note discussion above).  We confirm the presence of Li~I absorption reported in those studies, and also note H$\alpha$ and H$\beta$ in emission and strong low surface gravity spectral features, implying a young source. Like J2000--7523, we find this source to have a better membership match to $\beta$PMG ($P_M = 87$\%, $P_C = 0.2$\%), which is also supported by its strongly suppressed alkali lines, although the peculiar motion is somewhat large (7~{\kms}).

\subsection{New Candidate Members}

{ J0823--4912AB (L3)} is a low-probability candidate member of $\beta$PMG based on our analysis ($P_M = 30$\%,  $P_C = 0.4\%$). This source exhibits both Li~I absorption and spectral features consistent with low/intermediate surface gravity. Its distance and proper motion measurements give it relatively large peculiar motion relative to $\beta$PMG (7~\kms).  \citet{2013A&A...556A.133S,sahlmann2015} have identified this source as an astrometric binary with a mid-to-late L dwarf companion, and it is possible that the orbital motion of this system is skewing the inferred systemic motion in the radial direction (tangential motion is likely averaged out). \citet{sahlmann2015} also report Li~I in absorption and the presence of low-surface gravity features in its near-infrared spectrum. However, based on comparison of the system mass function, component classifications and evolutionary models, { the \citet{sahlmann2015}} study supports an age in the range 70--470~Myr for J0823--4912AB, considerably older than $\beta$PMG.  J0823--4912AB may therefore be an interloping member of a older YMG or stream, perhaps CAS.

{ J1510--2818 (M9)} is identified as a modest-probability candidate member of ARG ($P_M = 60$\% and $P_C = 34$\%). The source exhibits H$\alpha$ emission and strong signatures of low surface gravity in its spectrum, but no Li~I. 
Again, the absence of Li~I may be related to weak surface gravity at an age $\lesssim$\,30~Myr, which is marginally consistent with the 30--50~Myr age estimate for ARG.   A trigonometric distance measurement could confirm or refute ARG membership; BANYAN~II predicts a statistical distance of $27^{+2}_{-3}$~pc for ARG membership, and $35^{+11}_{-7}$~pc for membership in the field.

\subsection{Interlopers}

{ J1456--2809 (LHS 3003, M7)} emerged as a candidate member of ABD in our analysis ($P_B = 94.3$\%, $P_C = 4.3$\%). However, this source displays no signs of low surface gravity in either its optical or near-infrared spectra \citep{2014ApJ...794..143B}, and no Li~I absorption, implying an age $\gtrsim$500~Myr. Its H$\alpha$ emission does not provide a good constraint on its age; chromospheric activity lasts up to 8~Gyr in late-M dwarfs \citep{2011AJ....141...97W}. With its  close proximity to the locus of ABD members (21.2~pc and 2.5~{\kms} from the center of our model), J1456--2809 highlights the importance of identifying signs of youth before assigning YMG membership.

\subsection{Candidate Young Brown Dwarfs Not Assigned to a Young Moving Group}

Three additional sources exhibiting spectral signatures of low surface gravity were not found to be members of the well-characterized young associations listed above.  We performed a second-order check of additional systems based exclusively on $UVW$ velocities: 
the Octans Association ($\approx$\,30--40~Myr; \citealt{2015MNRAS.447.1267M});
the Ursa Major Moving Group ($\approx$\,500~Myr; \citealt{2003AJ....125.1980K});
the Hercules-Lyrae Moving Group ($\approx$\,250~Myr; \citealt{2013A&A...556A..53E});
Carina-Near Moving Group ($\approx$\,200~Myr; \citealt{2006ApJ...649L.115Z});
and various streams listed in \citet{2004ARA&A..42..685Z}, including CAS.

{ J0652--2534 (M9)} has a spectrum similar to J0339--3525, with somewhat weakened alkali absorption (including narrow K~I doublet lines) and enhanced VO absorption at 7900\,{\AA}, both suggesting a modestly low surface gravity. In addition, the spectrum exhibits Li~I absorption, but no H$\alpha$ emission. Given these features, J0652--2534 is likely a few hundred Myr old.   Its closest match in $UVW$ space is Octans, but the source is about 20~{\kms} discrepant from that association's $UVW$ center and is probably too old.  This ``juvenile'' brown dwarf must be an member of an  association or stream not considered here, or may be an unassociated system.  

{ J0909--0658 (L1)} exhibits strong signatures of low surface gravity, most notably enhanced VO absorption at 7400\,{\AA} and 7900\,{\AA}, weak FeH and CrH bands, and somewhat narrower K~I lines. This source does not exhibit Li~I absorption, and the presence of clear Na~I, Rb~I and Cs~I lines argues against low surface gravity being the reason. Assuming a mass greater than 0.06~{\msun} and {\teff} = 2100~K, this implies an age of at least 500~Myr.  Of all the kinematic groups listed above, ARG has the closest association in $UVW$ space, but J0909--0658 is clearly too old for that group. If this source is a young brown dwarf, it must be a member of an older association or stream or solivagant. Alternately, its unusual metal-oxide and metal-hydride features may reflect non-solar (possibly super-solar) composition. Thorough analysis of this source's full spectral energy distribution is warranted.

{ J1411--2119 (M9)}, like J0339--3525 and J0652--2534, exhibits weak spectral signatures of low surface gravity (weak Na~I absorption at 8183/8195\,{\AA}) and Li~I absorption, as well as pronounced H$\alpha$ and H$\beta$ emission.  Its closest match in $UVW$ space is the $\approx$\,500~Myr Ursa Majoris Group, but J1411--2119 remains about 13~{\kms} apart from the cluster motion center, and it is positioned far from other members of the group \citep{2003AJ....125.1980K}. Again, this brown dwarf may be part of an an as-yet unrecognized moving group or unassociated.

\section{Additional Sources of Interest}

%{ J0024$-$0158 (BRI~0021-0214)} is the only source with $\zeta < 0.825$ implying a subdwarf, but is this reliable? Known flaring star and exhibits radio emission.  also unusually red ($\Delta(J-K_s)$ = 0.28). we find a slightly different RV than Reiners \& Basri 2009; possible RV variable? check for other RV measurements. Very young kinematics

{ J0707$-$4900 (M8.5)} is remarkable in its conflicting kinematic, activity and color indicators of age. Originally identified by \citet{1991ApJ...367L..59R} as a brown dwarf candidate in the Hyades moving group, this source has a large radial velocity (113$\pm$2~{\kms}) and its $UVW$ velocity components identify it as an intermediate thin/thick disk star (see also \citealt{2009ApJ...699..168D}). It has the most negative $V$ velocity in our sample, and its prograde Galactic orbit is highly eccentric (0.46$\pm$0.01) but with a modest inclination (2$\fdg$5$\pm$1$\fdg$0).
However, its spectrum exhibits no obvious signatures of metal-deficiency, and its near-infrared color is unusually red for its spectral type.  
The weak H$\alpha$ emission detected in our spectrum ({\lha} = $-$5.33$\pm$0.06) is somewhat lower than prior detections ({\lha} = $-$4.9 by \citealt{1998MNRAS.301.1031T};  {\lha} = $-$4.4 by \citealt{2003ApJ...583..451M}), but consistent with a star whose magnetic dynamo has weakened but persists.  \citet{2003ApJ...583..451M} report {\vsini} = 10~{\kms} for this source, making it a relatively rapid rotator with weak magnetic emission.  That study attributes the anomalous behavior of J0707$-$4900 to a low surface gravity associated with low mass and youth.  Hence, the kinematics and low level of activity suggest an old age for J0707$-$4900, while its color and rapid rotation supports a young age.  It is possible that the red color of this source indicates the presence of a mid-to-late L dwarf companion as yet unresolved.  The stability in its radial and astrometric motion (three measurements by \citealt{1995ApJ...441L..47I,1996MNRAS.281..644T} and \citealt{2004AJ....128.2460H}) argues against the existence of a $\lesssim$\,0.3~AU binary, but this source is clearly anomalous and merits further attention.

{ J2036+1051 (L3)} is one of two unusually active L dwarfs identified in our program.  It was first observed by \citet{2007AJ....133.2258S} and found to have no H$\alpha$ emission to a limit of EW $>$ $-$6.3\,{\AA}. In contrast, we measure EW = $-$11.5$\pm$3.0\,{\AA} (H$\beta$ and H$\gamma$ lines were too weak to be detected). 
It appears that we caught this rapidly rotating dwarf ({\vsini} = 67.1$\pm$1.5~{\kms}; \citealt{2010ApJ...723..684B}) in a flare state.
We note that the alkali lines in this source are a little weak compared to other L3 dwarfs in the sample (with the exception of J0823--4912AB), but the kinematics are consistent with an older disk source.  The alkali lines may have been filled in by continuum emission from the flare.

%{\em J1705$-$0516}: 
%reid et al. 2006 - possible binary companion; andrei et al. 2011 probably background source
%seifahrt et al. 2011 - possible member of UMa - check with new velocities?
%jameson et al. 2008 - not identified as UMa
%blake - 12.19(0.11), seifahrt - 12.2(1.1),  us: 11.2(2.1)
%reiners et al. 2008 - vsini = 26 km/s no halpha to <-7.1 more stringent than our <-6.8
%kendall et al. 2004 - L4 NIR classification; reid et al. 2008 - optical classification of L0.5
%inactive radio source Berger et al. 2006
%koen 2005 - variable source on two timescales 8.6hr and 2.5hr- possible pulsator?

{ J2037$-$1137 (M8)} is the earliest-type source in our sample to exhibit no detectable H$\alpha$ emission, to an EW limit of $>$-0.3\,{\AA} and {\lha} $<$ $-$6.74.  In contrast, prior studies have consistently detected emission in the range $-$5.51 $\lesssim$ {\lha} $\lesssim$ $-$4.48 \citep{2007AJ....133.2258S,2010ApJ...710..924R,2010ApJ...708.1482L}.  Unlike J2036+1051, this slowly rotating dwarf ({\vsini} $\leq$ 3$\pm$2~{\kms}; \citealt{2009ApJ...705.1416R}) may have been observed during a minimum in its magnetic emission cycle, or in an orientation exhibiting a relatively quiescent surface.

%The M dwarfs J1534$-$1418 (M7), J0707$-$4900 (M8) and J0517$-$3349 (M8) have the largest absolute values of $U$ ($-$71.7$\pm$5.1~\kms), $V$ ($-$98.5$\pm$2.4~\kms) and $W$ (51.4$\pm$3.9~\kms) in the sample, respectively; while the thick disk star J0707$-$4900 also has the largest RV in our sample (113.4$\pm$0.8~\kms).  Five sources (J0641$-$4322, J0707$-$4900, J0921$-$2104, J0949+0806 and J2306$-$0502) have absolute $V$ velocities $>$50~\kms, all negative (i.e., trailing the Sun in Galactic rotation).  Among these, the thick disk star J0921$-$2104 has previously been noted as a unusually blue L dwarf \citep{2008ApJ...674..451B,2009AJ....137....1F}. Only the M8 J2306$-$0502 exhibits both a large space velocity and H$\alpha$ emission. 

\section{Summary}

This paper has reported the radial velocities of 85 VLM stars and brown dwarfs of spectral types M6-L6 with MagE with typical precisions of 2--3~{\kms}.  Combining these with previously published proper motions and distances, we computed $UVW$ velocities and examined velocity dispersions as a function of spectral type, color and magnetic activity. We find that unusually blue objects are more dispersed than unusually red objects, in support of color effects being driven by higher surface gravities and/or slightly subsolar metallicities among an older, low-mass stellar population.  We also find that magnetically inactive L dwarfs are more dispersed than magnetically active L dwarfs, following the age-activity relations of earlier-type stars. The most interesting finding is the greater dispersions and $U$ velocity offset (net radial flow) of L dwarfs as compared to late-M dwarfs. The greater dispersions of L dwarfs { affirms} prior results by \citet{2010A&A...512A..37S} and \citet{2010ApJ...723..684B}, { but we speculate that this may be driven by a large fraction of unusually blue L dwarfs in local 20~pc sample as compared to deeper imaging surveys}.  Population simulations incorporating brown dwarf evolution, Galactic dynamics in an axisymmetric potential, and local selection still predict that L dwarfs should be younger than late-M dwarfs, although this can be reversed if brown dwarfs have had a different formation history than stars (or alternately, evolve more quickly than evolutionary models predict). However, these simulations cannot reproduce the distinct $U$ velocity distribution of L dwarfs, which is either driven by non-axisymmetric Galactic structure (e.g., bar, spiral arms) or the existence of a distinct, dispersed population of L dwarfs in the Solar Neighborhood. We also identify 8 kinematic members of nearby YMGs, including new candidates J0823$-$4912AB in the $\beta$ Pictoris Moving Group (although likely part of an older association/stream) and J1510$-$2818 in the Argus Association.  Three additional sources, J0652$-$2534,  J0909$-$0658 and J1411$-$2119 have evidence of low surface gravities but no YMG assignment; these may be members of older moving groups or streams, or simply solivagant $\sim$500~Myr-old brown dwarfs.

The ultimate goal of the Brown Dwarf Kinematics Project is a complete kinematic sampling of the lowest-mass stars and brown dwarfs in the vicinity of the Sun.  With 25\% of the 20~pc late-M and L dwarf sample studied here, we are likely still missing additional members of nearby YMGs, and we may still be subject to (unknown) selection biases due to incompleteness.  Work is underway to complete the RV sampling of the local VLM population. Given preliminary evidence of distinct kinematic populations of L dwarfs in the Solar Neighborhood, examining the kinematics of T dwarfs (all of which are substellar) may provide an important anchor for disentangling the mass function and star formation history of our local mixed stellar/substellar population. 

\section{Acknowledgements}
The authors would like to thank Jorge Araya, Mauricio Martinez, Hernan Nunez, and Geraldo Vallardes for assistance in the observations.
We acknowledge useful discussions with John Gizis during the preparation of the manuscript.
A.A.W acknowledges funding from NSF grants AST-1109273 and AST-1255568 and support from the Research Corporation for Science Advancement?s Cottrell Scholarship; 
E.E.M acknowledges support from NSF grant AST-1313029.
This publication makes use of data products from the Two Micron All Sky Survey, which is a joint project of the University of Massachusetts and the Infrared Processing and Analysis Center/California Institute of Technology, funded by the National Aeronautics and Space Administration and the National Science Foundation. This research has benefitted from the M, L, and T dwarf compendium housed at DwarfArchives.org and maintained by Chris Gelino, Davy Kirkpatrick, and Adam Burgasser. It has also made use of the SIMBAD database and VizieR service, operated at CDS, Strasbourg, France. This article has also made use of data from the Sloan Digital Sky Survey.    Funding for the SDSS and SDSS-II has been provided by the Alfred P. Sloan Foundation, the Participating Institutions, the National Science Foundation, the U.S. Department of Energy, the National Aeronautics and Space Administration, the Japanese Monbukagakusho, the Max Planck Society, and the Higher Education Funding Council for England. The SDSS Web Site is http://www.sdss.org/.
The SDSS is managed by the Astrophysical Research Consortium for the Participating Institutions. The Participating Institutions are the American Museum of Natural History, Astrophysical Institute Potsdam, University of Basel, University of Cambridge, Case Western Reserve University, University of Chicago, Drexel University, Fermilab, the Institute for Advanced Study, the Japan Participation Group, Johns Hopkins University, the Joint Institute for Nuclear Astrophysics, the Kavli Institute for Particle Astrophysics and Cosmology, the Korean Scientist Group, the Chinese Academy of Sciences (LAMOST), Los Alamos National Laboratory, the Max-Planck-Institute for Astronomy (MPIA), the Max-Planck-Institute for Astrophysics (MPA), New Mexico State University, Ohio State University, University of Pittsburgh, University of Portsmouth, Princeton University, the United States Naval Observatory, and the University of Washington.

Facilities: \facility{Magellan:Clay}

\appendix

\section{ Conversion between 2MASS and MKO $JHK_s$ Magnitudes}

{ 
To compute the bolometric fluxes for our sources, we made use of their 2MASS $J$-band photometry and MKO $J$-band bolometric corrections from \citet{2010ApJ...722..311L}.  This required a small correction between filter magnitude systems due to the highly structured spectra of late M and L dwarfs. \citet{2004PASP..116....9S} have previously reported corrections for L and T dwarfs as a funciton of spectral type, but did not include late-M dwarfs in their sample; we therefore computed a new conversion relation spanning types M6 to L7.  We selected 533 low-resolution, near-infrared spectra form the SpeX Prism Library (SPL; \citealt{2014arXiv1406.4887B}) with reported optical classifications between M6 and L7 and median S/N $>$ 100. We computed spectrophotometric magnitudes in both 2MASS and MKO $JHK_s$ systems following the procedures described in \citet{2004PASP..116....9S} and \citet{2005ApJ...623.1115C}. Figure~\ref{fig:mko2m} displays the magnitude differences ($\Delta$ = MKO-2MASS) in $J$, $H$ and $K_s$ as a function of spectral type.  We fit all three filter differences to second-order polynomials (higher orders did not significantly improve the fits), iteratively rejecting 3$\sigma$ outliers.  The polynomial coefficients and dispersions are listed in Table~\ref{table:mko2m}. Both $J$ and $H$ show significant (and opposing) trends, and the polynomial fits produce a residual scatter of 0.005--0.008~mag.
}

\clearpage

\clearpage

%lines vs lit
\begin{figure}
\centering
\epsscale{0.8}
\plotone{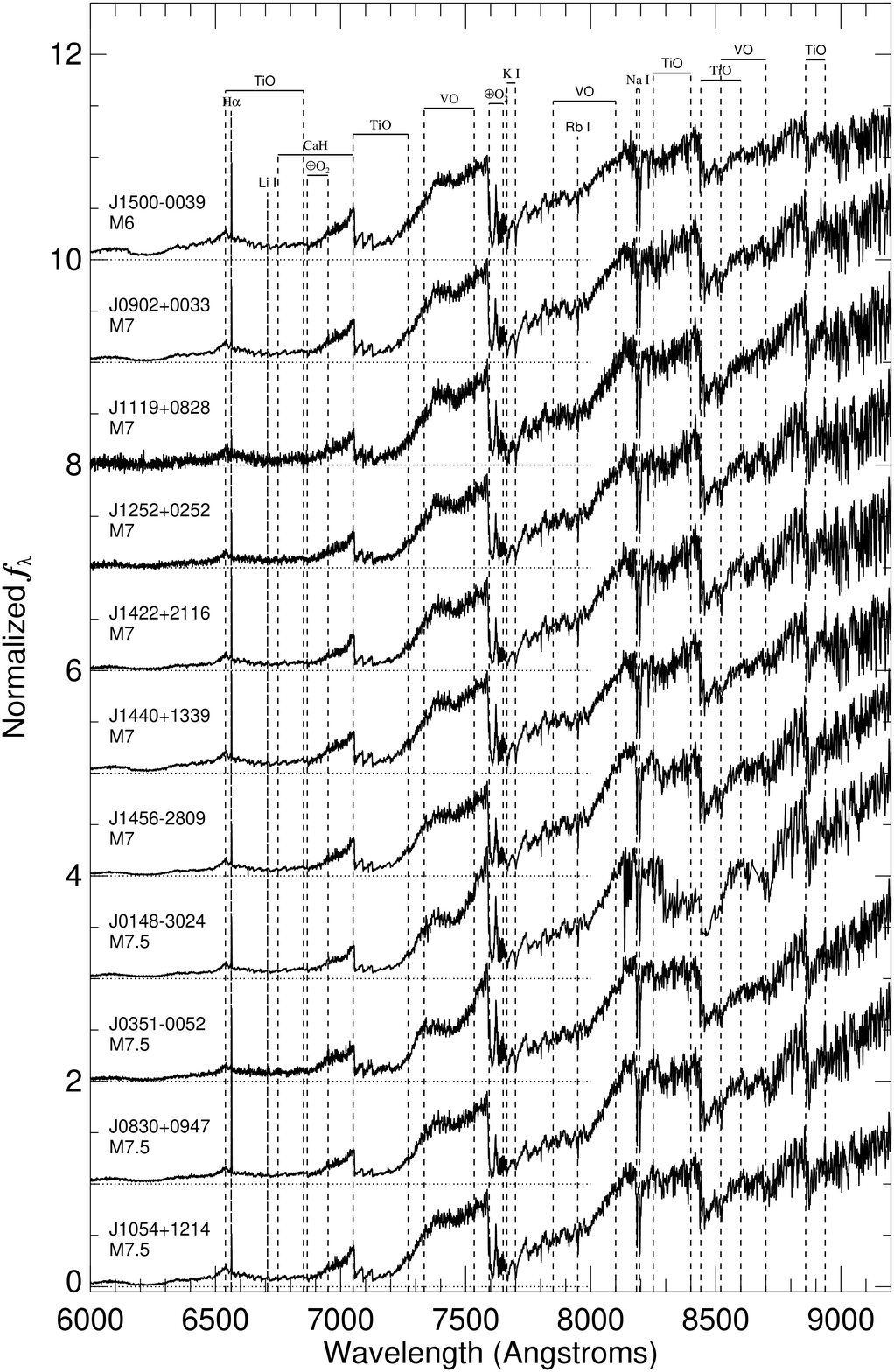}
\caption{Observed MagE spectra of our sample, ordered by spectral type and right ascension.  Data are normalized in the 8200--8600\,{\AA} range and each spectrum offset by a constant for clarity.  Major spectral features from Li~I, K~I, Na~I, Rb~I, Cs~I, TiO, VO, CrH, FeH, CaH and H~I emission are labeled, { as is telluric O$_2$ absorption}. 
\label{figure:spectra}}
\end{figure}

\addtocounter{figure}{-1}
\clearpage

\begin{figure}
\centering
\epsscale{0.8}
\plotone{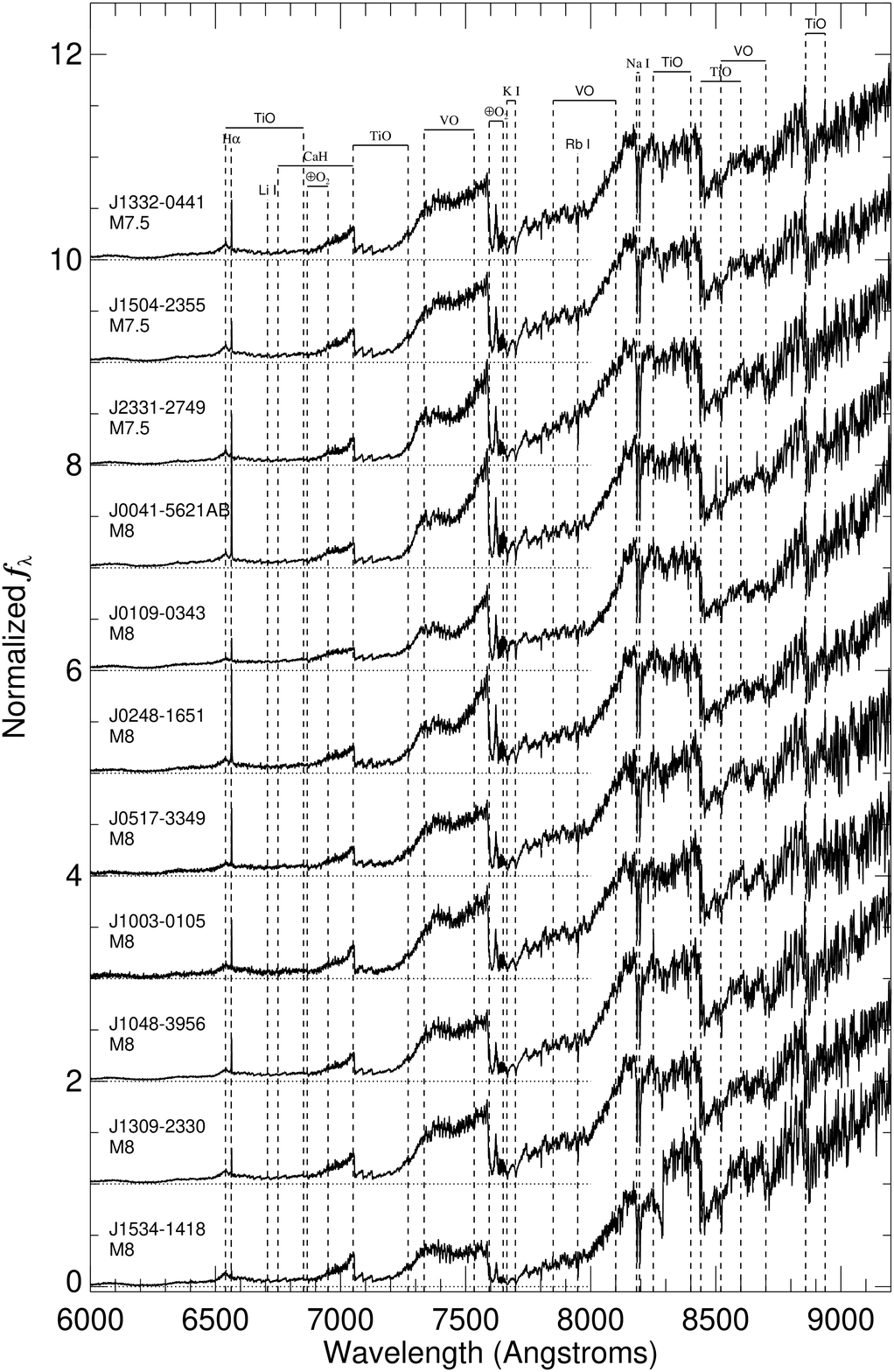}
\caption{Cont.}
\end{figure}

\addtocounter{figure}{-1}
\clearpage

\begin{figure}
\centering
\epsscale{0.8}
\plotone{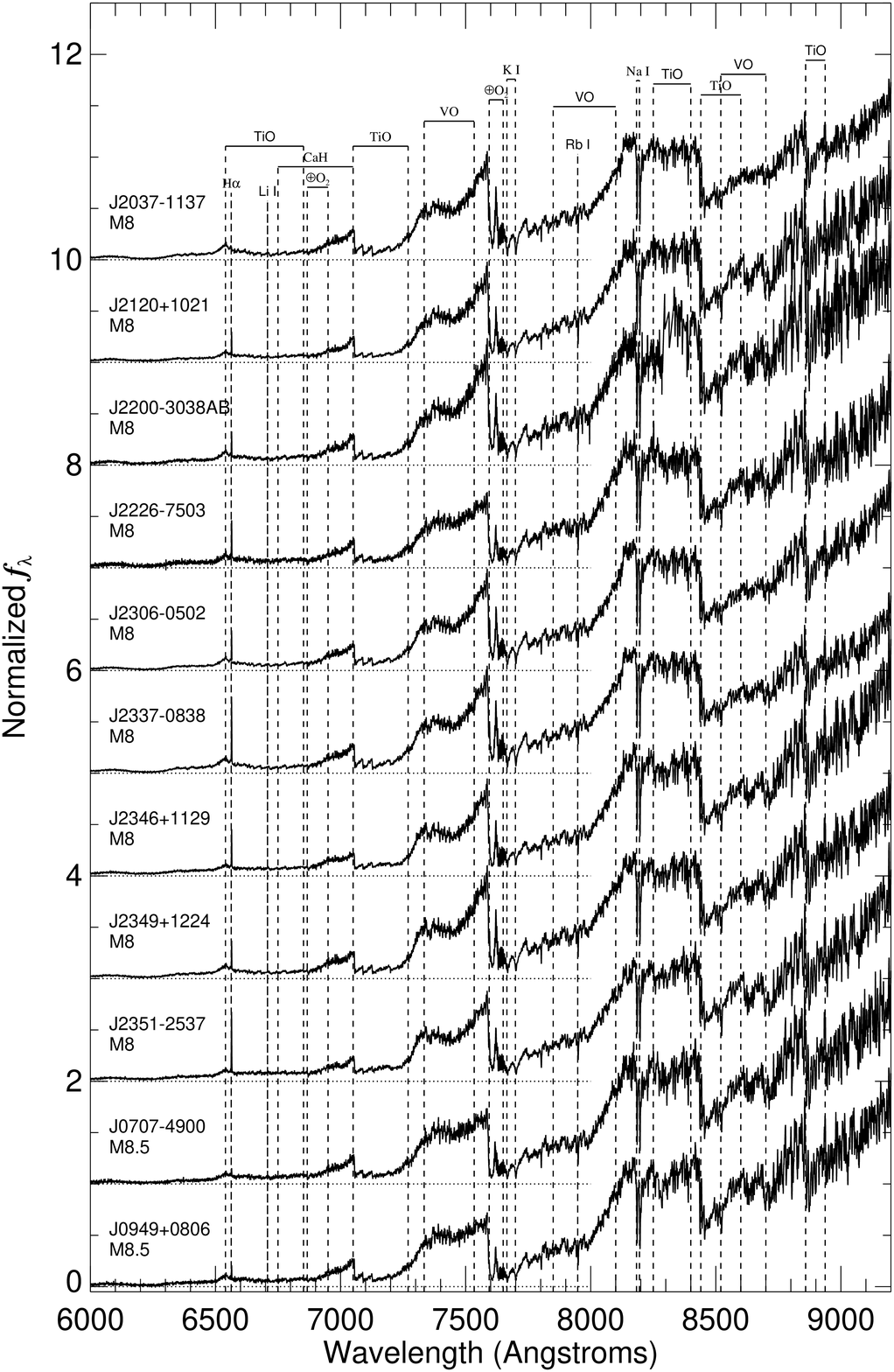}
\caption{Cont.}
\end{figure}

\addtocounter{figure}{-1}
\clearpage

\begin{figure}
\centering
\epsscale{0.8}
\plotone{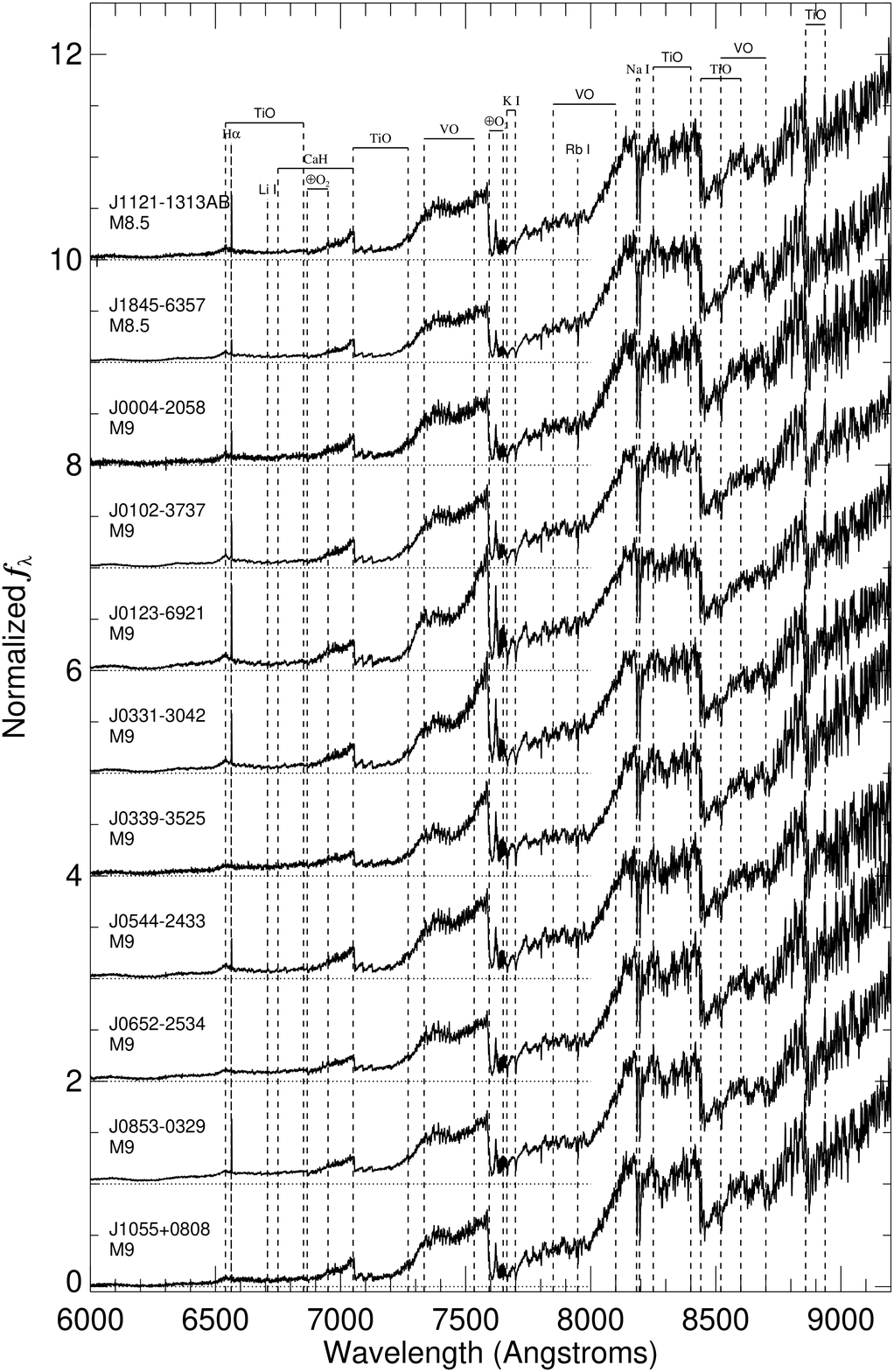}
\caption{Cont.}
\end{figure}

\addtocounter{figure}{-1}
\clearpage

\begin{figure}
\centering
\epsscale{0.8}
\plotone{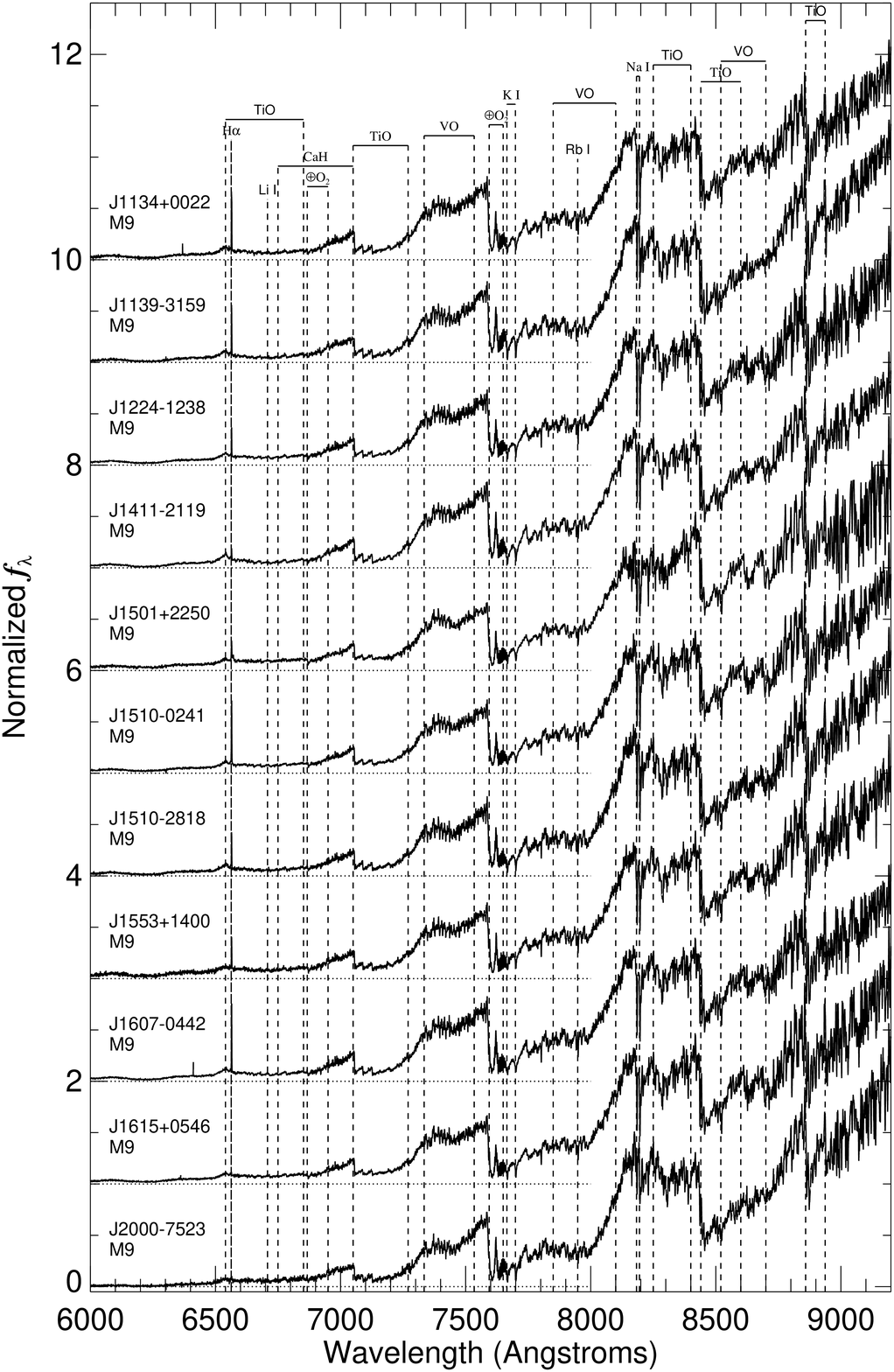}
\caption{Cont.}
\end{figure}

\addtocounter{figure}{-1}
\clearpage

\begin{figure}
\centering
\epsscale{0.8}
\plotone{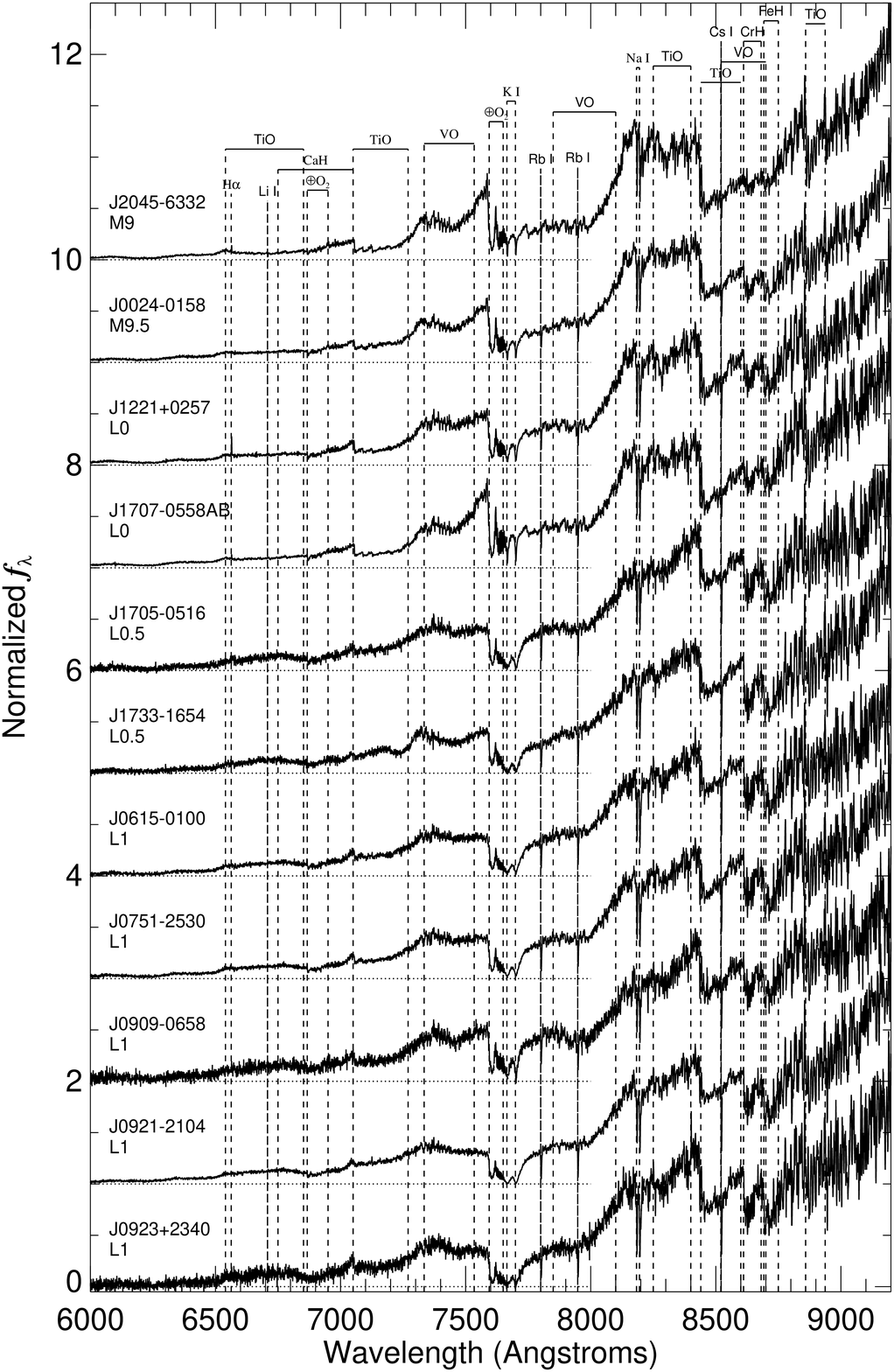}
\caption{Cont.}
\end{figure}

\addtocounter{figure}{-1}
\clearpage

\begin{figure}
\centering
\epsscale{0.8}
\plotone{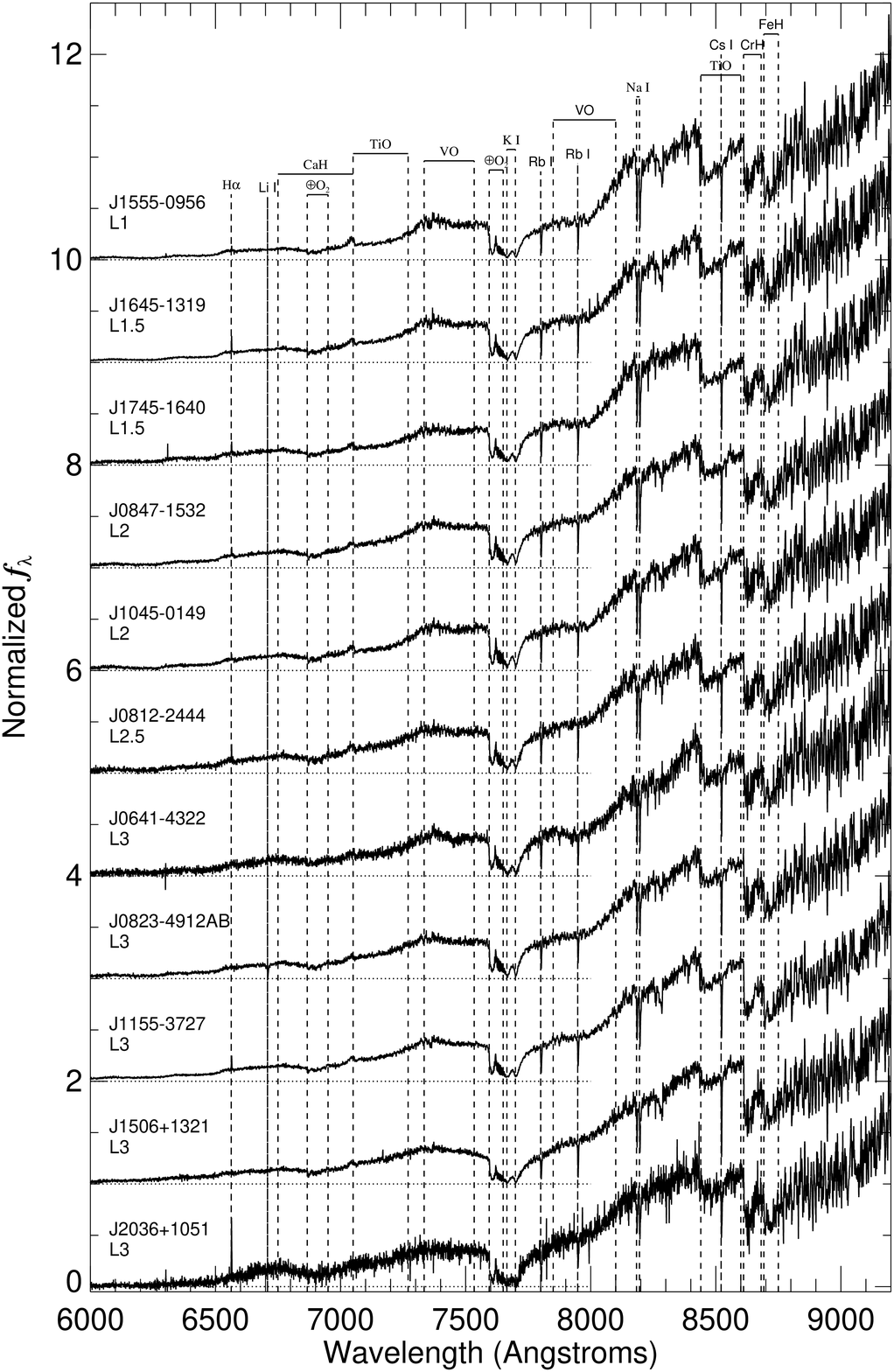}
\caption{Cont.}
\end{figure}

\addtocounter{figure}{-1}
\clearpage

\begin{figure}
\centering
\epsscale{0.8}
\plotone{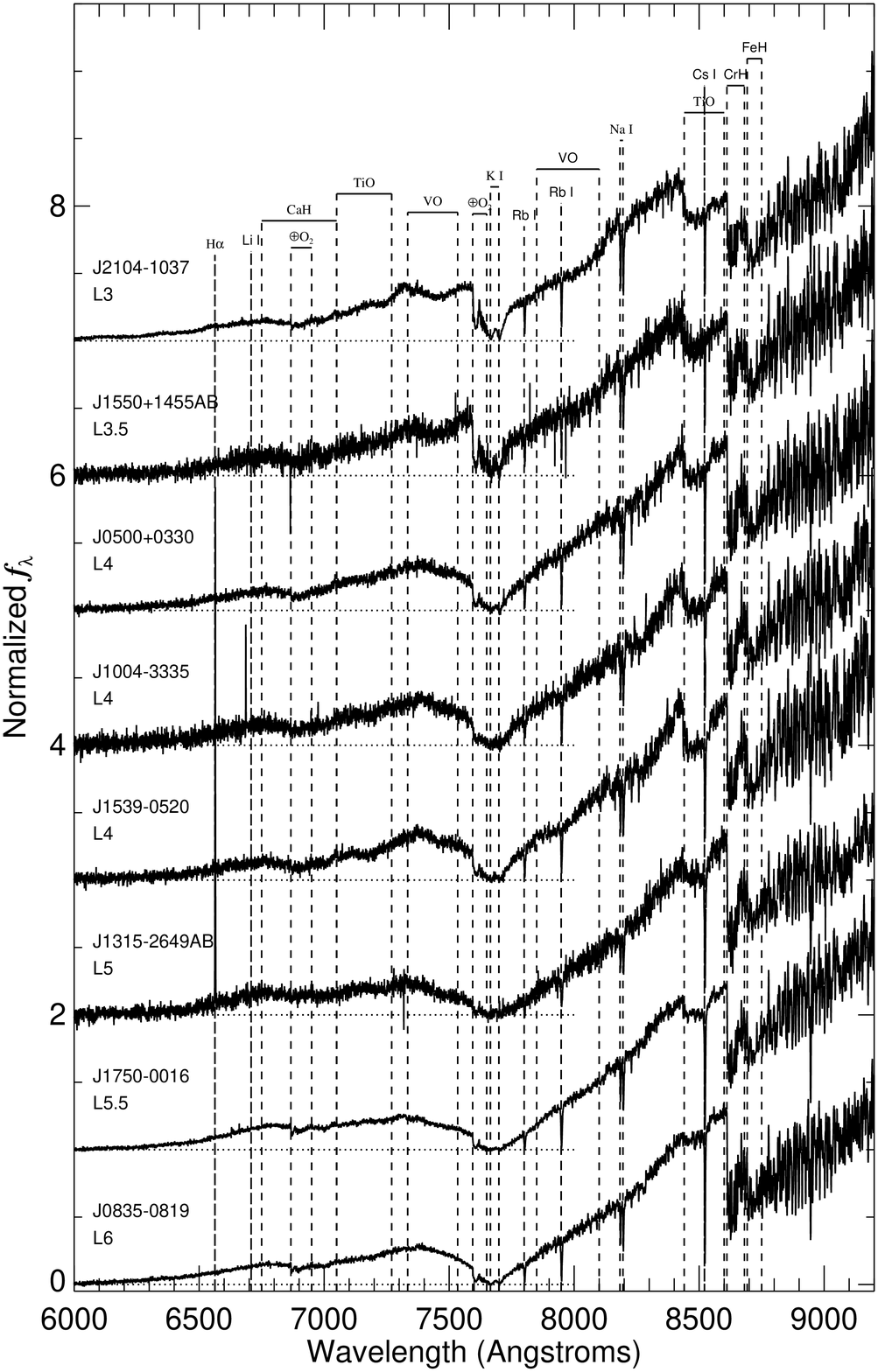}
\caption{Cont.}
\end{figure}

\clearpage

\begin{figure}
\centering
\plottwo{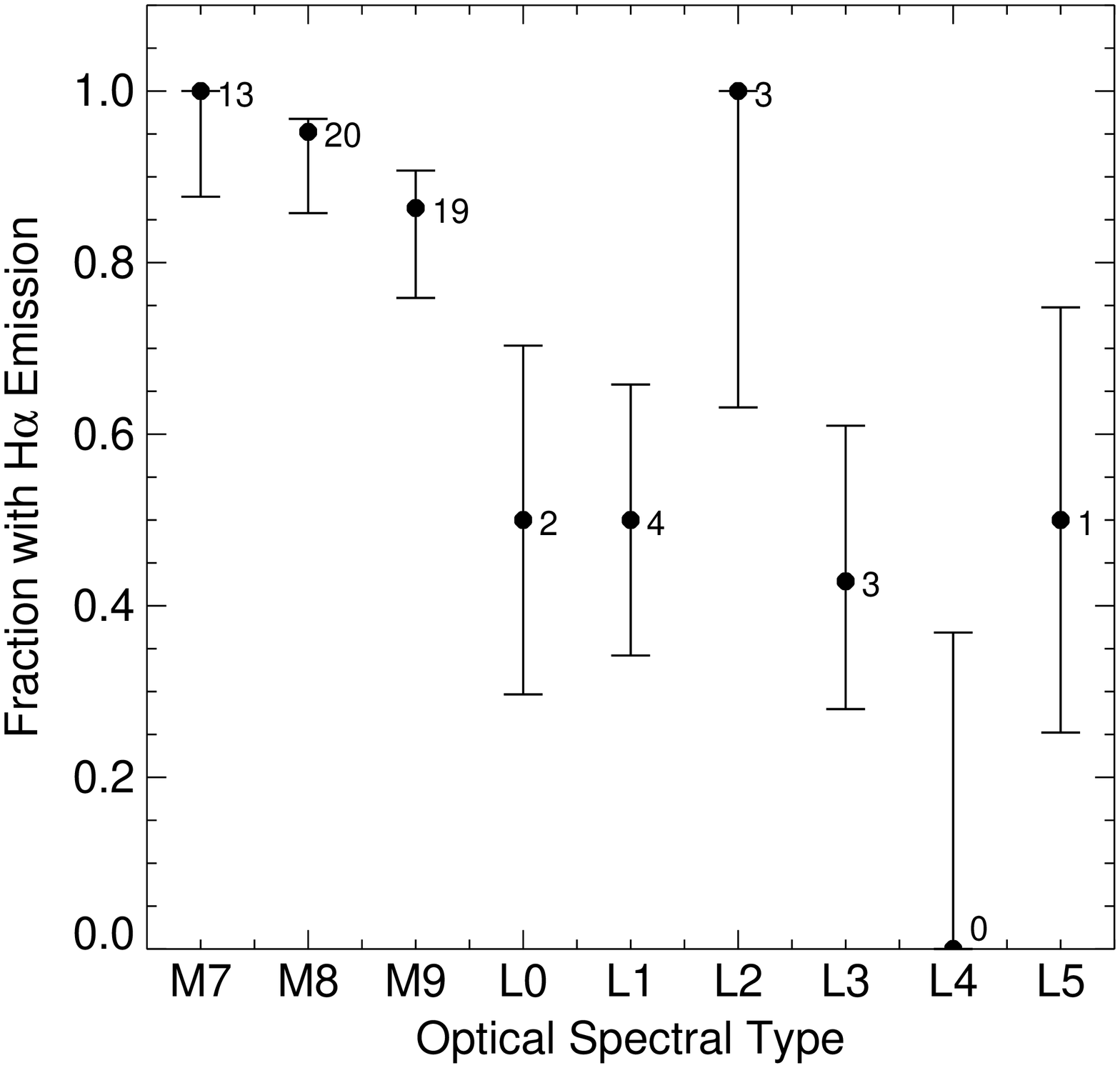}{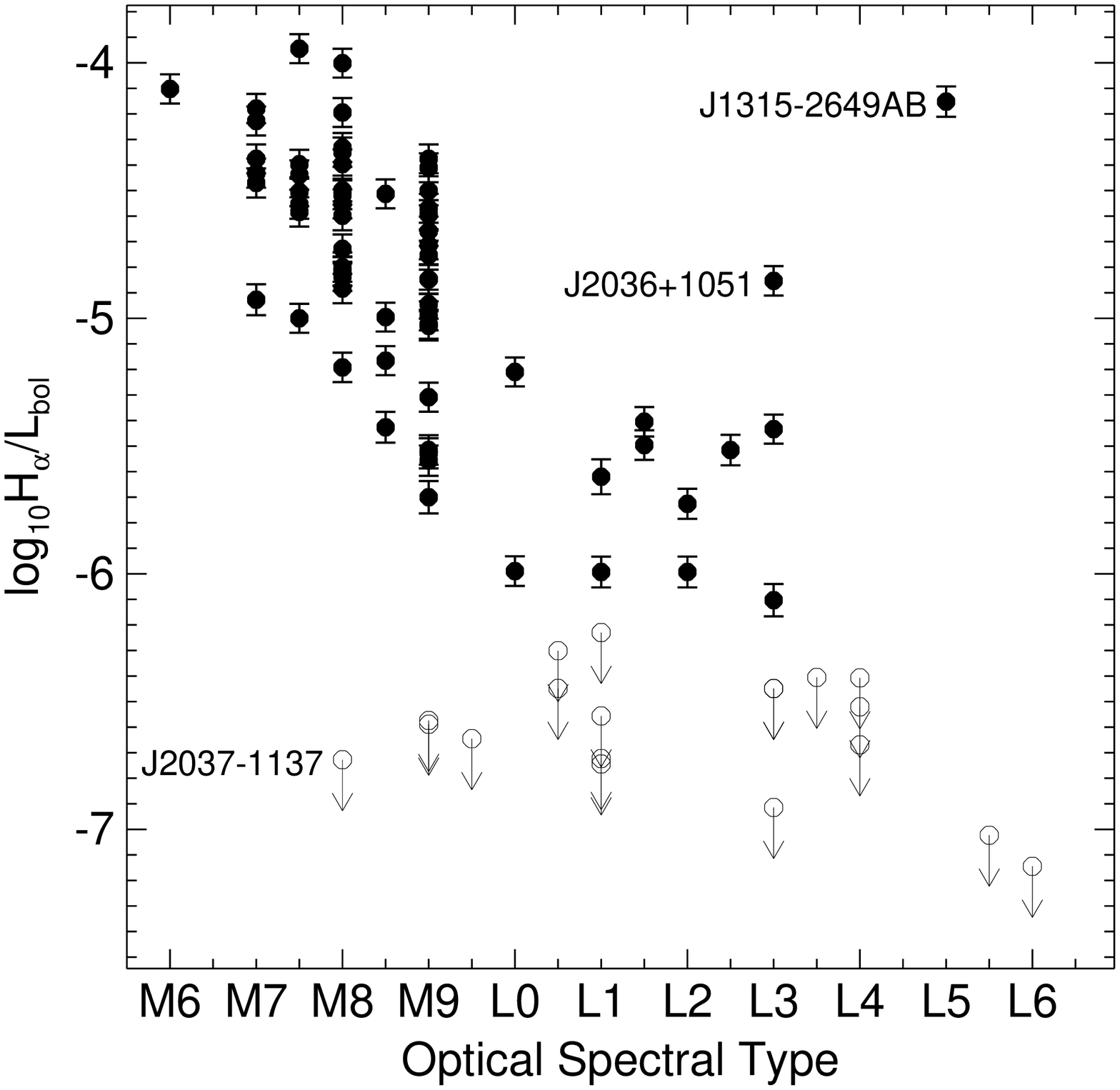}
\caption{(Left): Fraction of sources in our sample exhibiting H$\alpha$ emission as a function of spectral type.  Uncertainties are based on binomial statistics, with the numbers next to each point indicating the total number of objects exhibiting H$\alpha$ in that subtype bin.  
(Right): {\lha} as a function of spectral type; upper limits are indicated by downward arrows.  The two unusually active L dwarfs J2036+1051 and J1315$-$2649 are labeled, as is the unusually { inactive} M8 dwarf J2037--1137.
 \label{figure:halpha}}
\end{figure}

\begin{figure}
\centering
\includegraphics[width=0.19\textwidth]{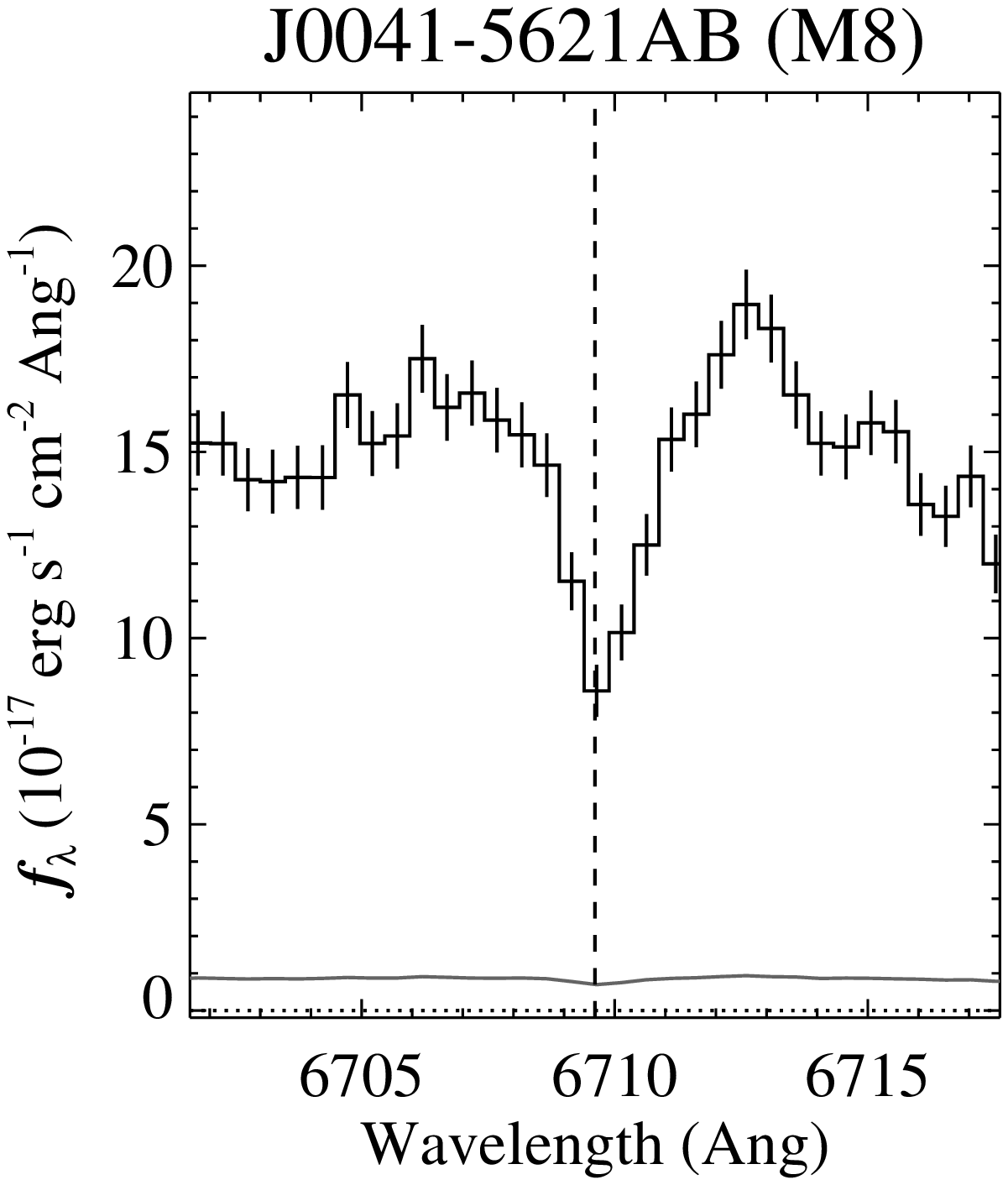}
\includegraphics[width=0.19\textwidth]{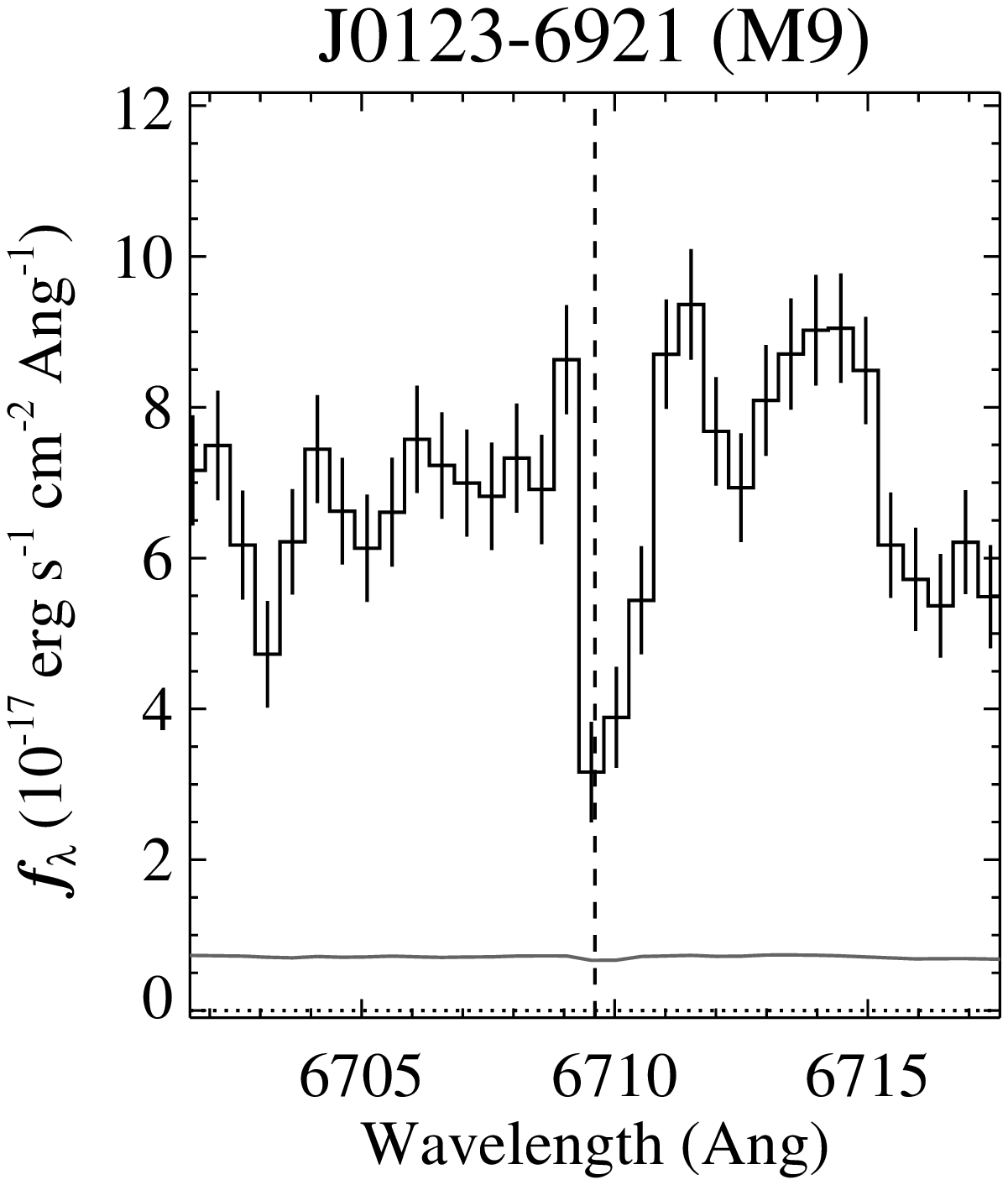}
\includegraphics[width=0.19\textwidth]{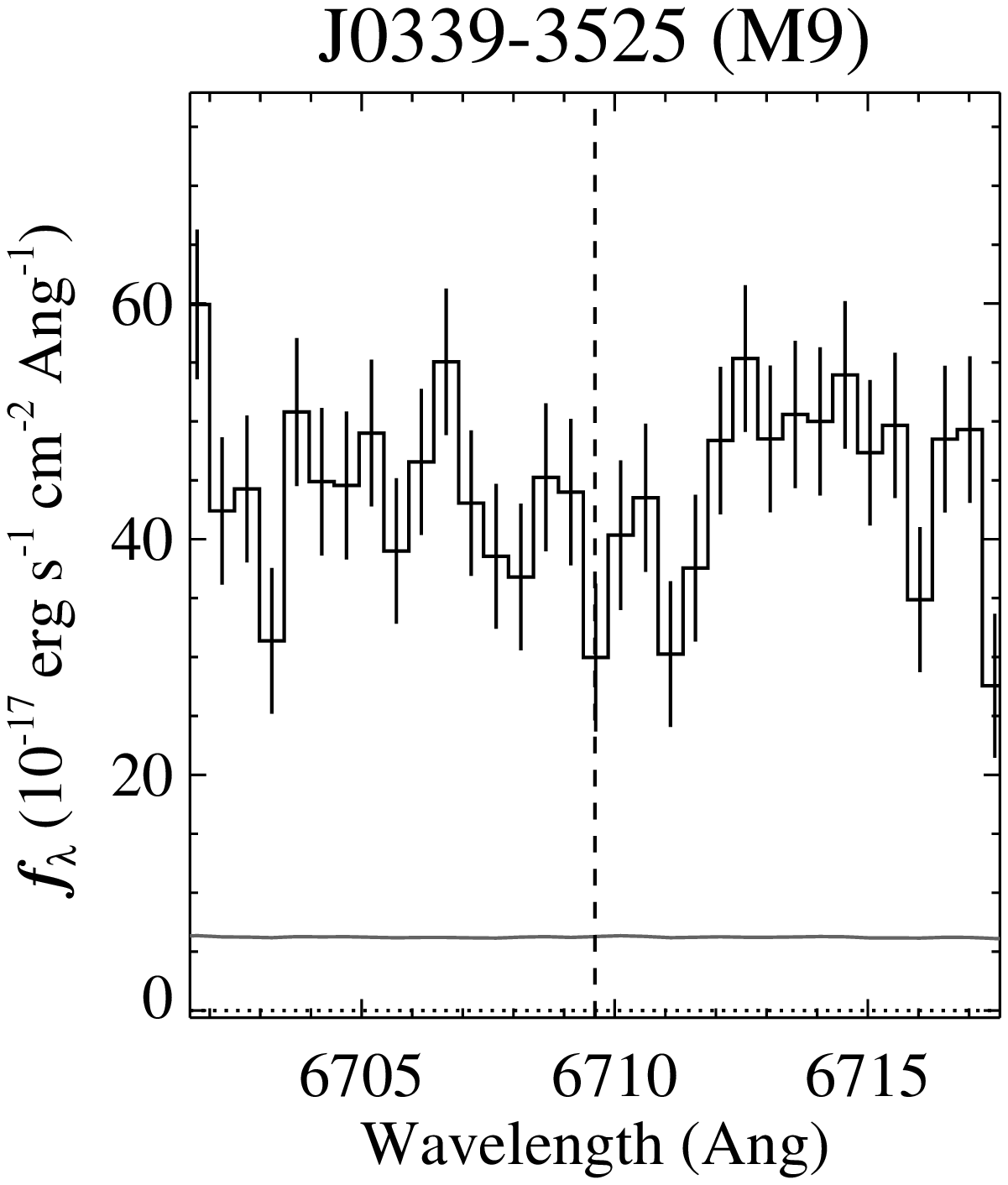}
\includegraphics[width=0.19\textwidth]{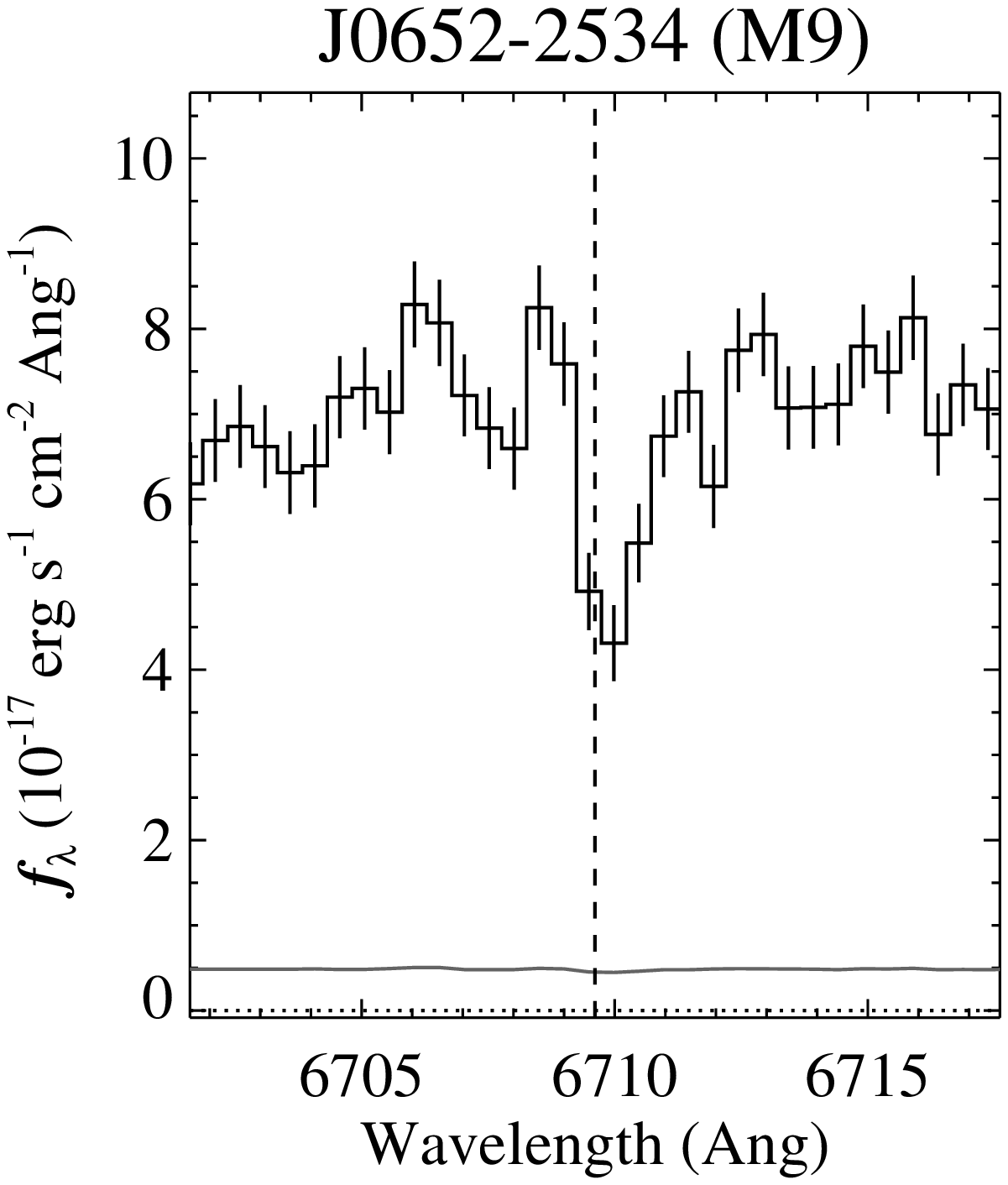}
\includegraphics[width=0.19\textwidth]{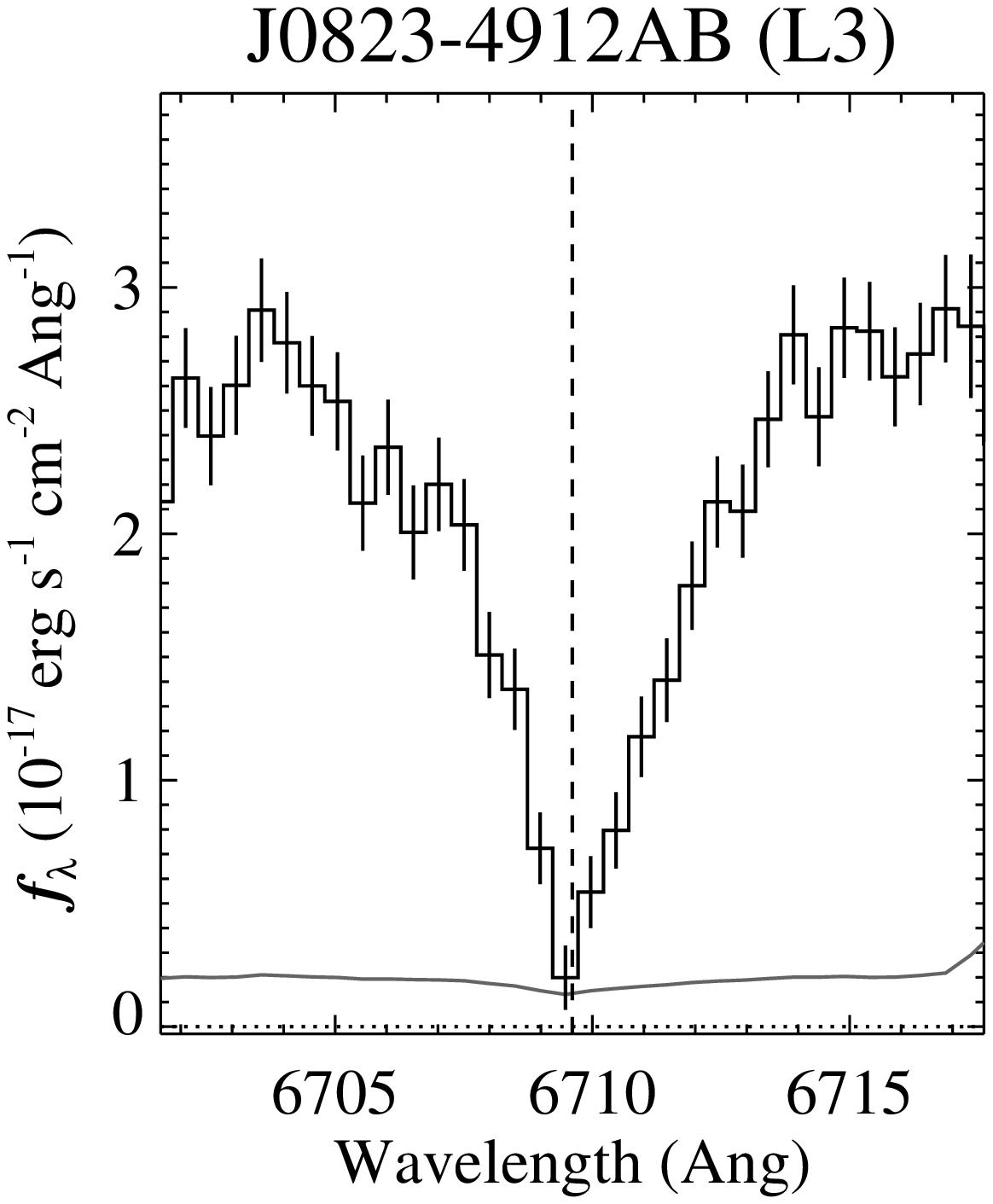}
\includegraphics[width=0.19\textwidth]{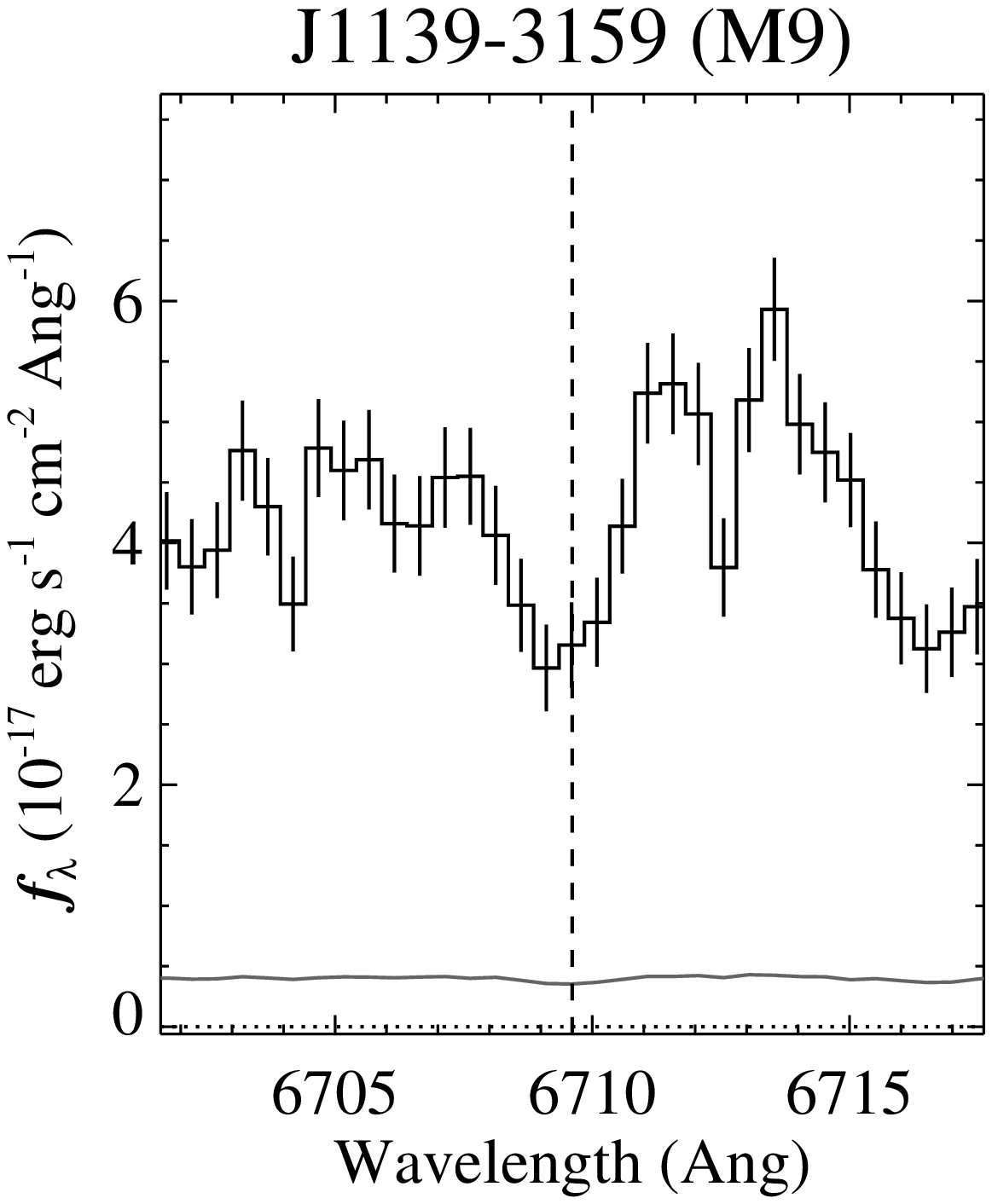}
\includegraphics[width=0.19\textwidth]{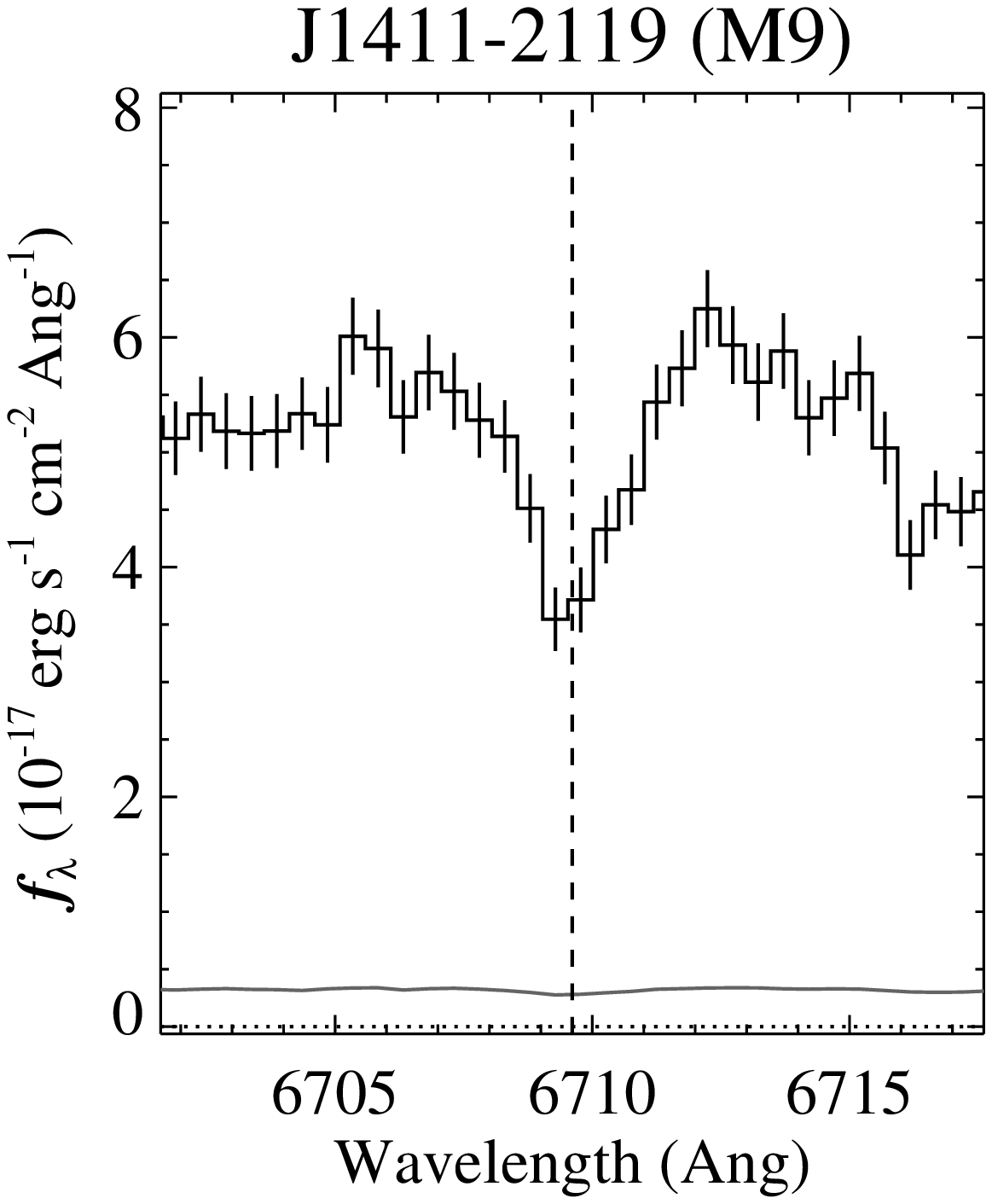}
\includegraphics[width=0.19\textwidth]{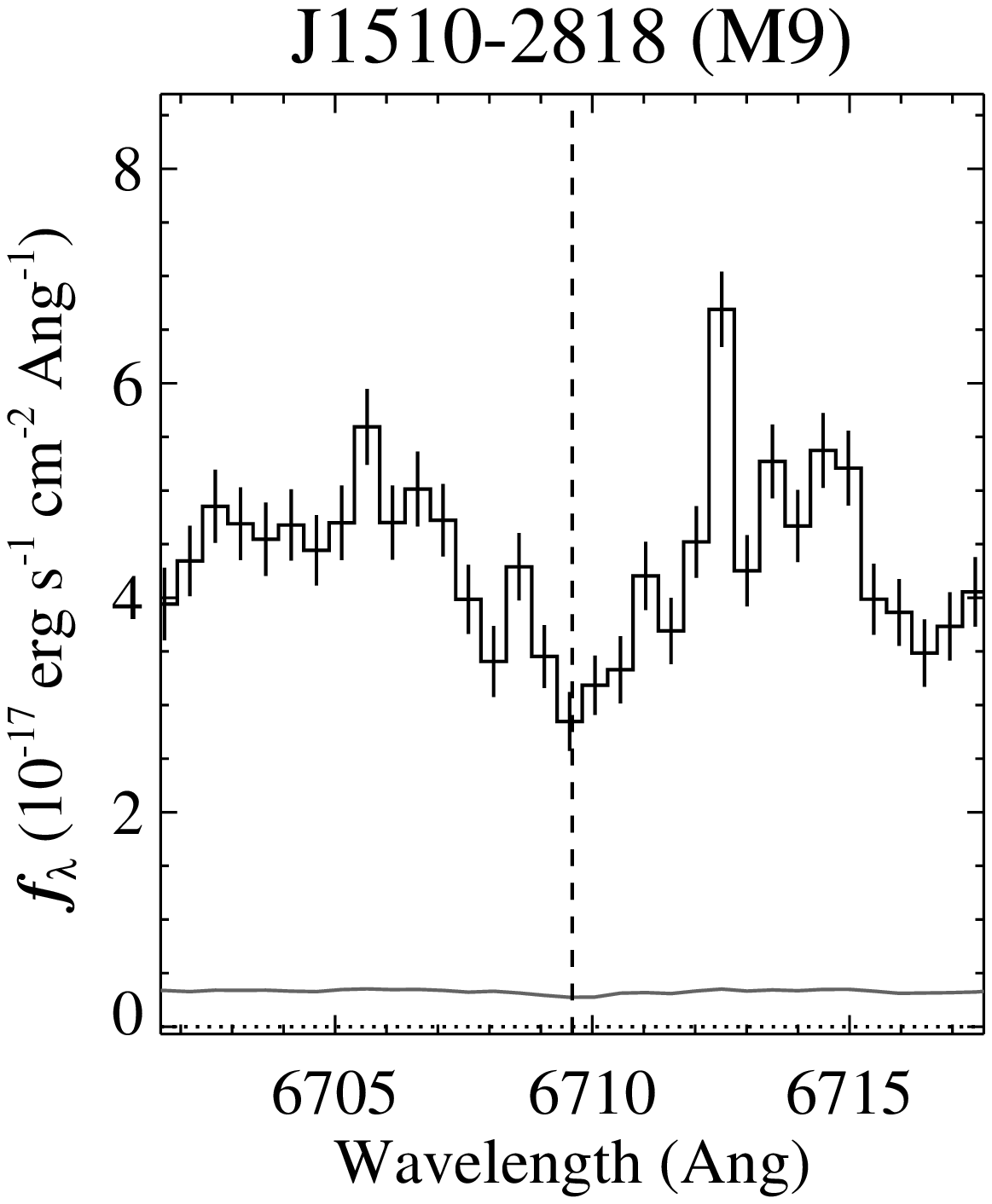}
\includegraphics[width=0.19\textwidth]{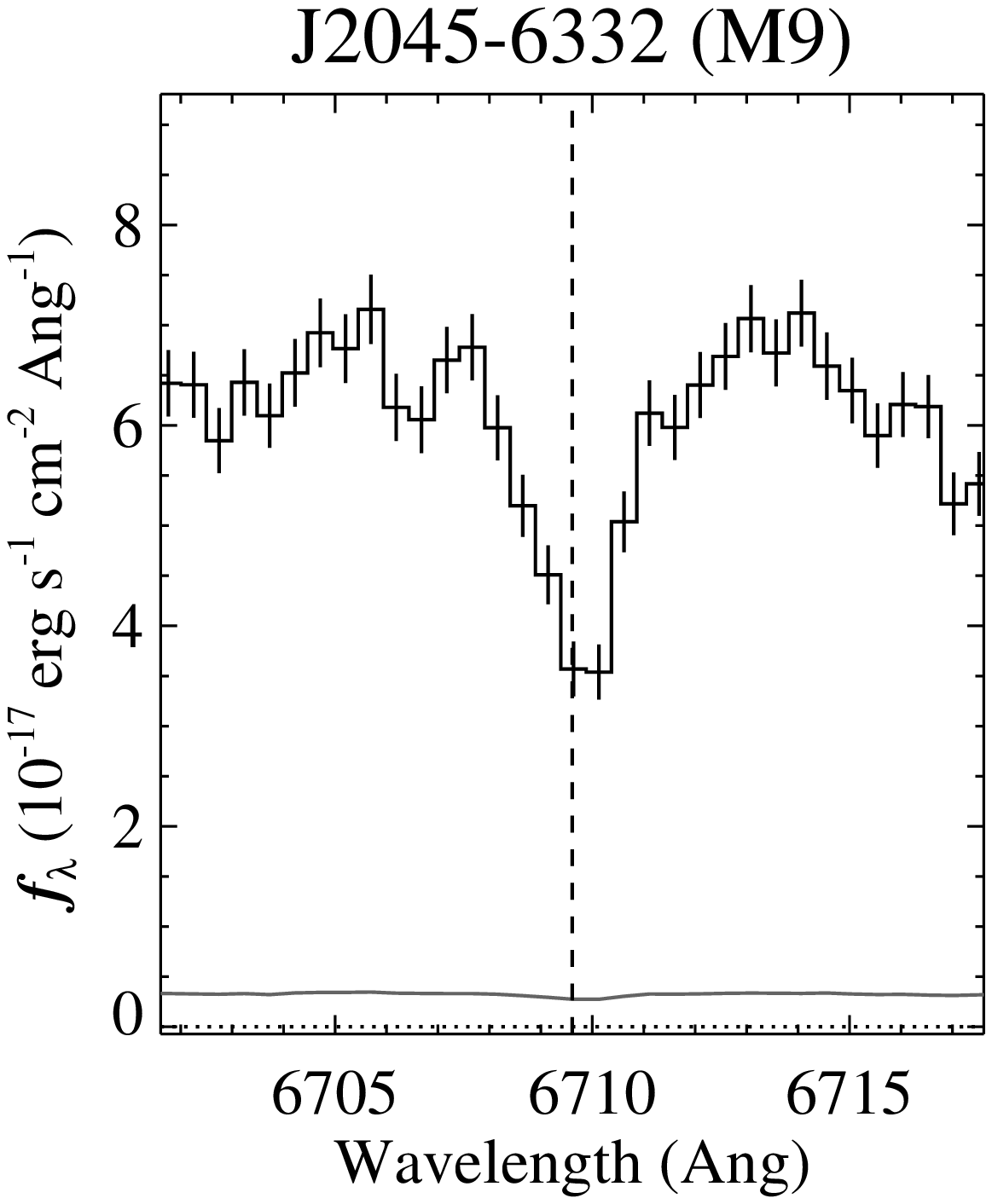}
\caption{6700--6720\,{\AA} spectra of nine sources in our sample that show significant or marginal absorption from the 6710\,{\AA} Li~I line, indicating that they are young brown dwarfs.  
 \label{figure:lithium}}
\end{figure}

\clearpage

\begin{figure}
\centering
\includegraphics[width=0.45\textwidth]{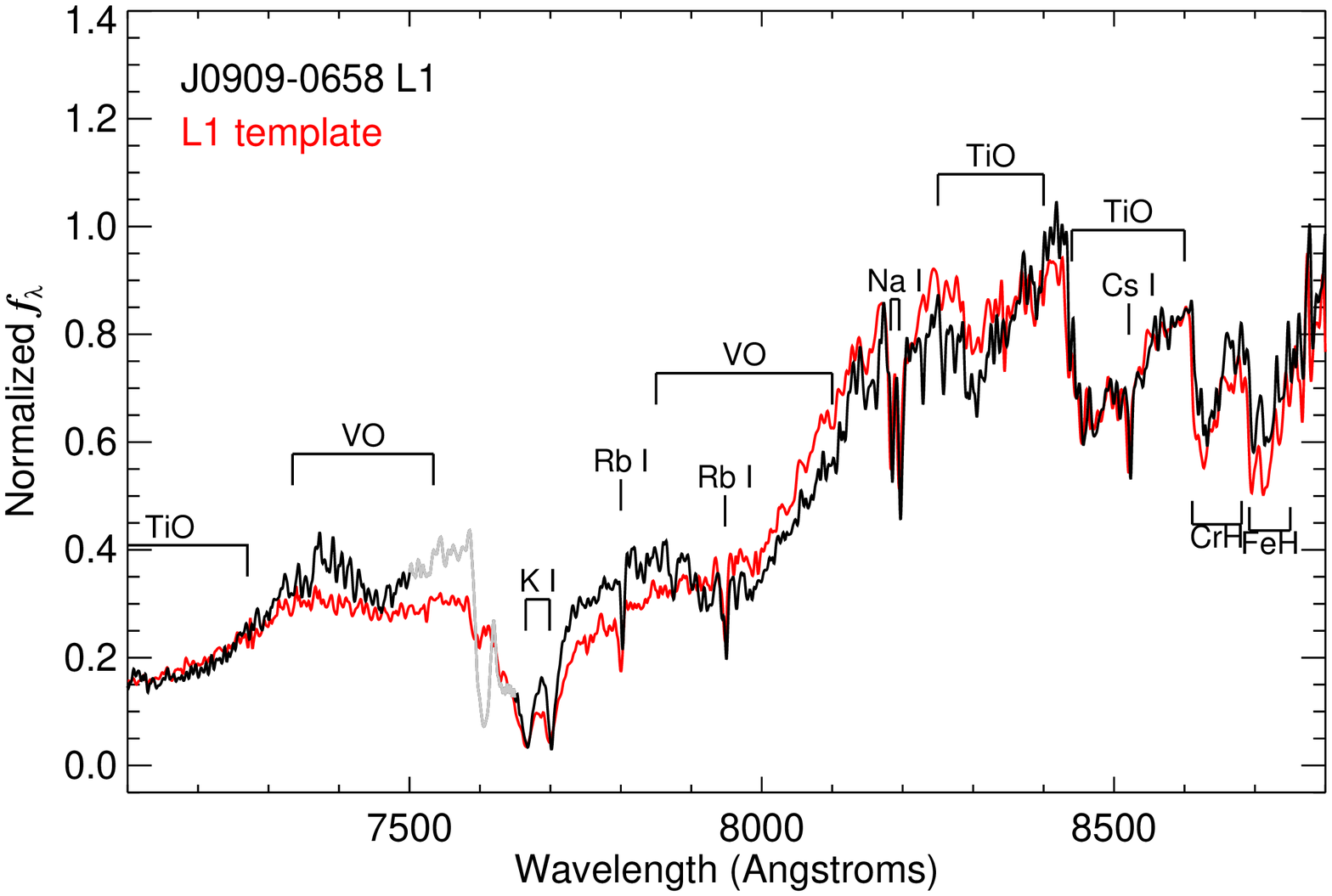}
\includegraphics[width=0.45\textwidth]{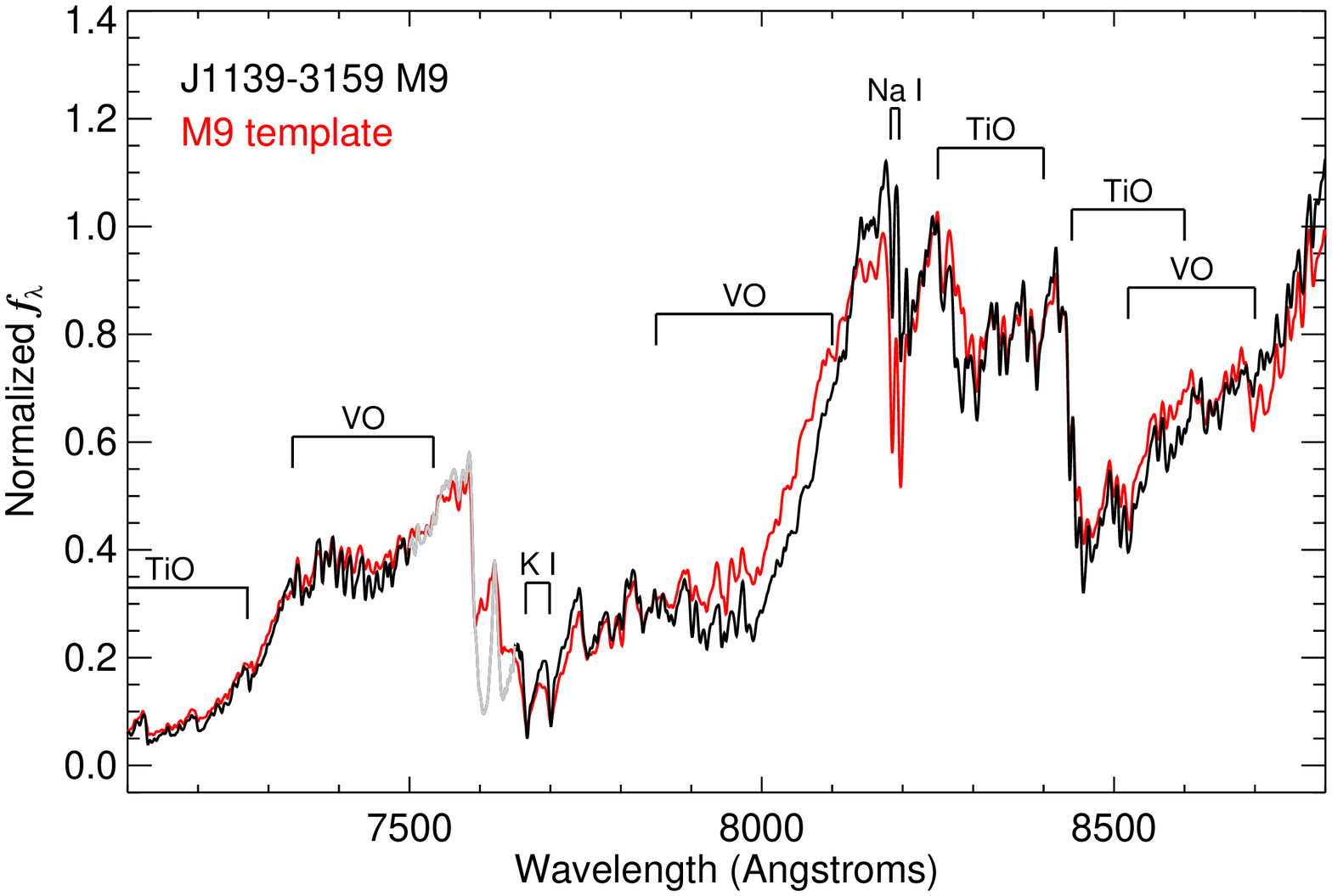} \\
\includegraphics[width=0.45\textwidth]{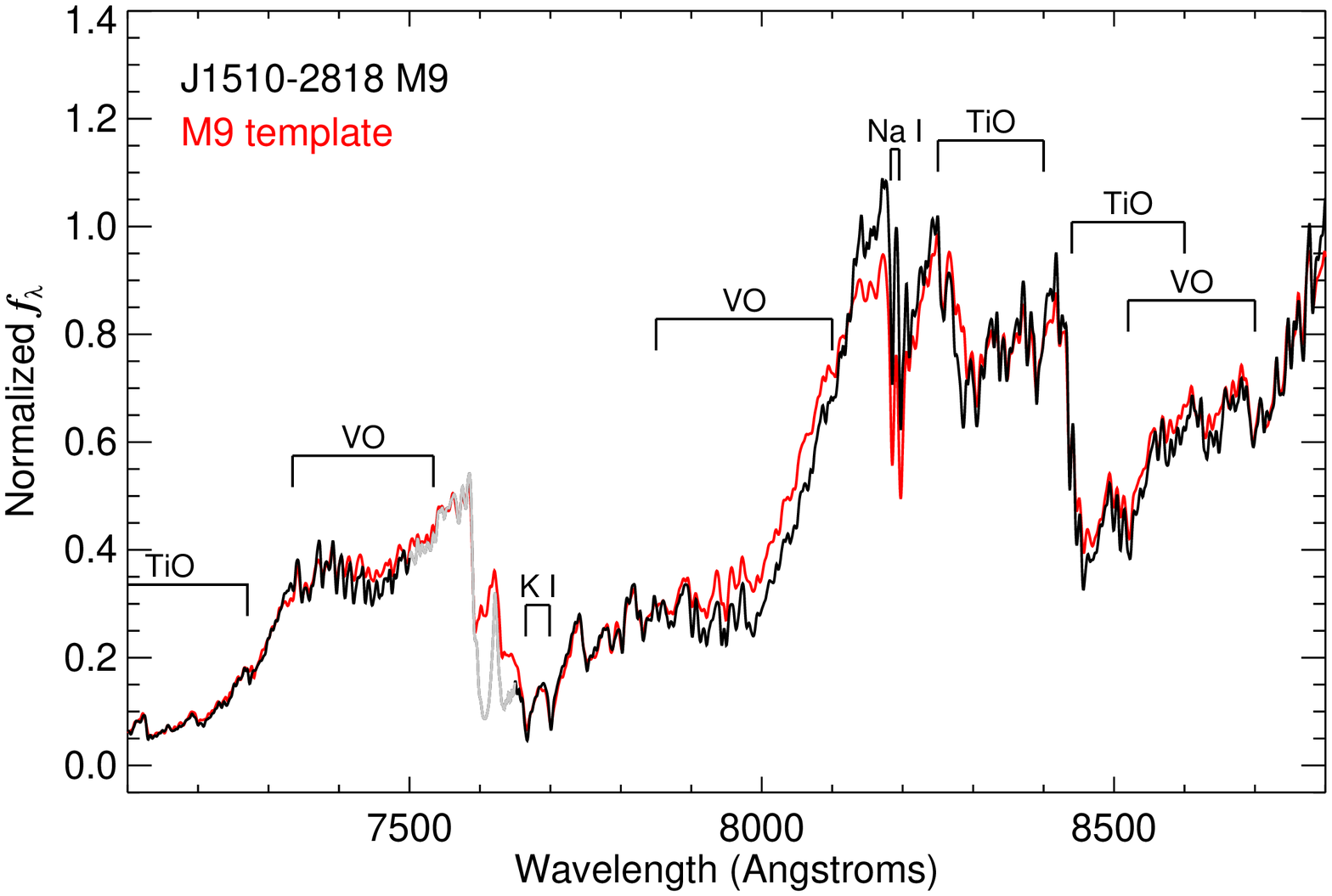} 
\includegraphics[width=0.45\textwidth]{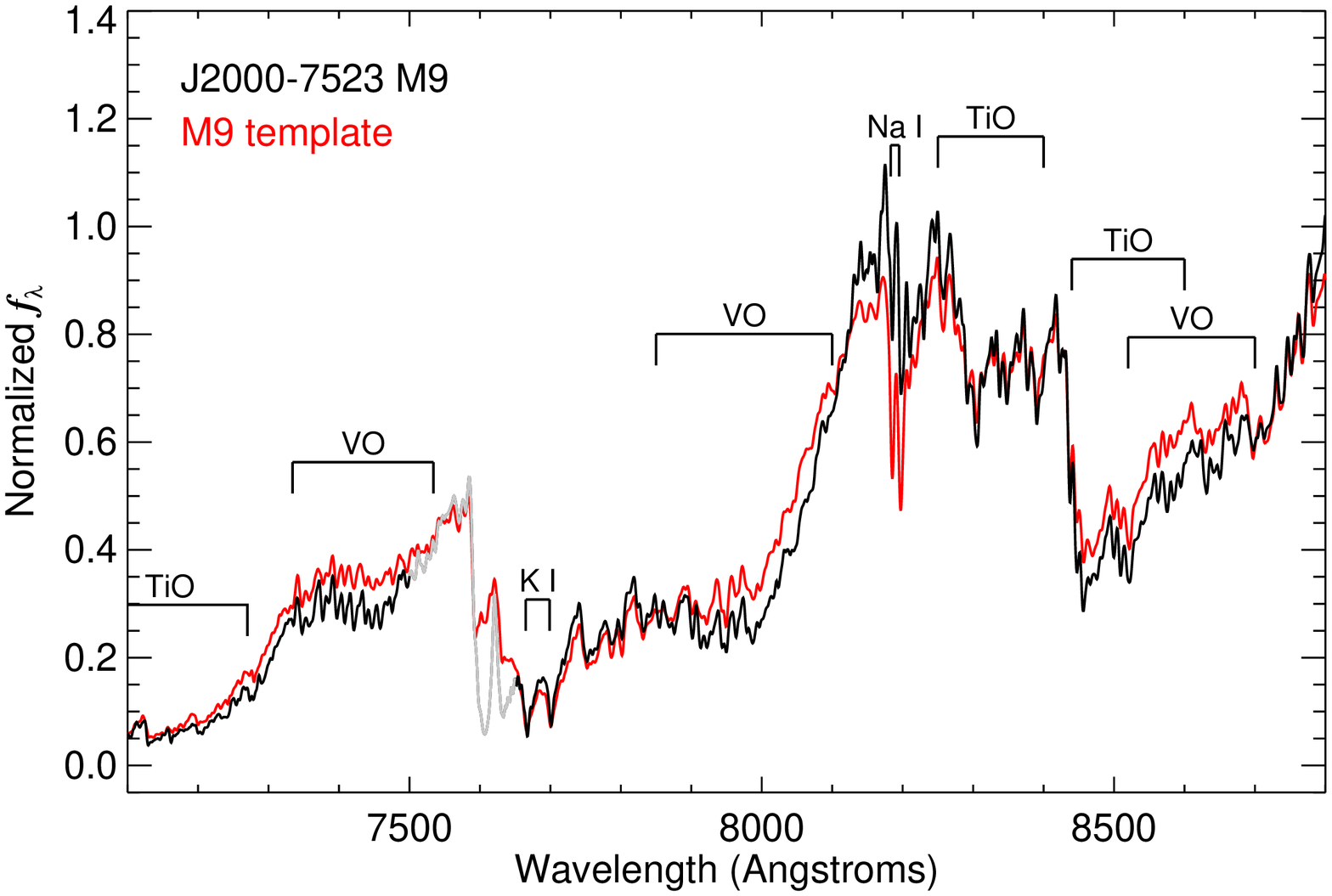} \\
\includegraphics[width=0.45\textwidth]{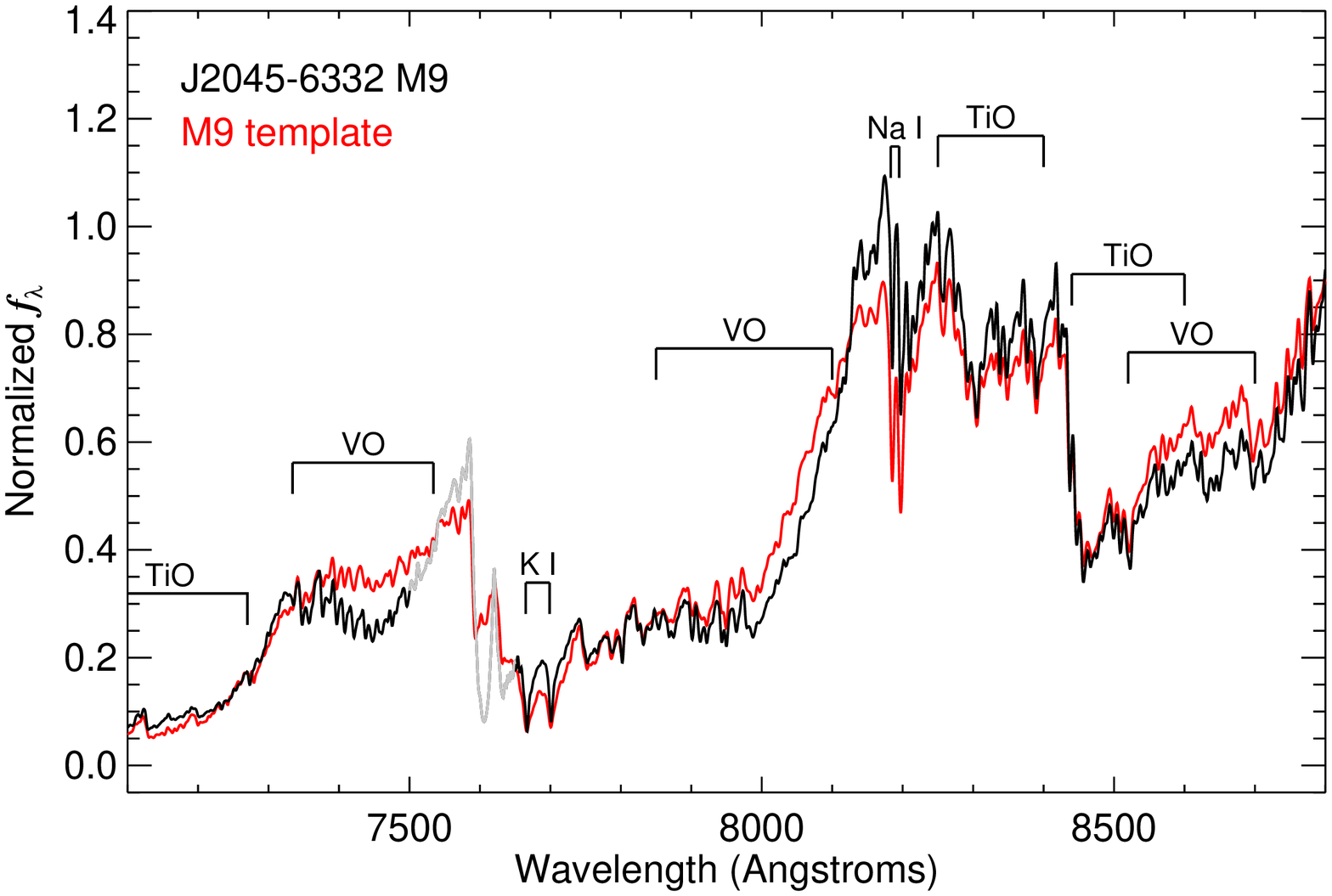}
\caption{7100--8800\,{\AA} spectra of five sources (black) exhibiting strong features of low surface gravity, compared to equivalently-classified SDSS templates (red).  
%Low surface gravity featues include enhanced VO absorption at 7400\,{\AA} and 7900\,{\AA}, weakened alkali lines (especially Na~I at 8183/8195\,{\AA} and weak FeH/CrH absorption at 8600\,{\AA}. 
Spectra are normalized at 8200\,{\AA}; the 7500--7600\,{\AA} region affected by poor flux calibration is indicated in gray and should be ignored.
 \label{fig:lowg_strong}}
\end{figure}

\clearpage

\begin{figure}
\centering
\includegraphics[width=0.45\textwidth]{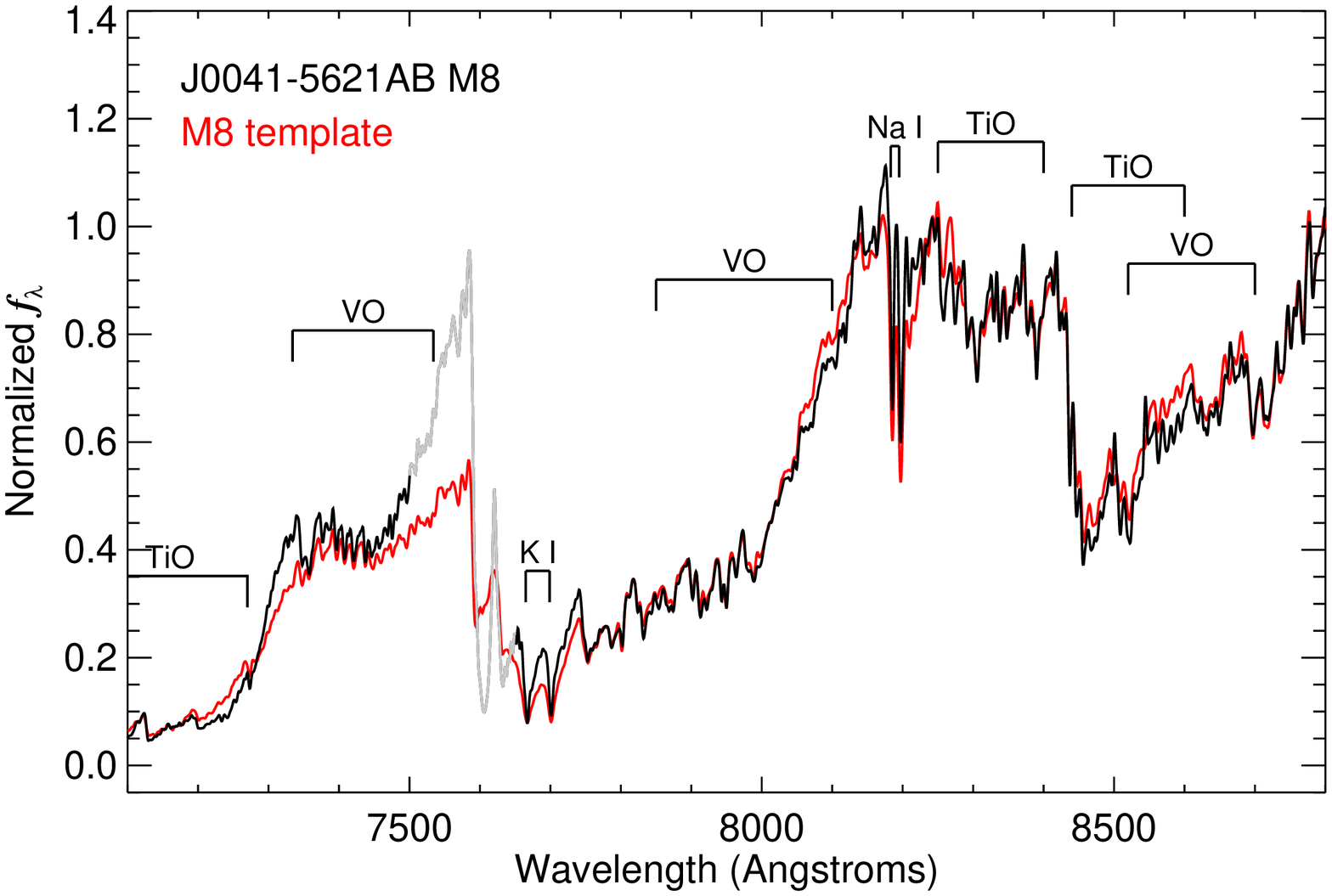}
\includegraphics[width=0.45\textwidth]{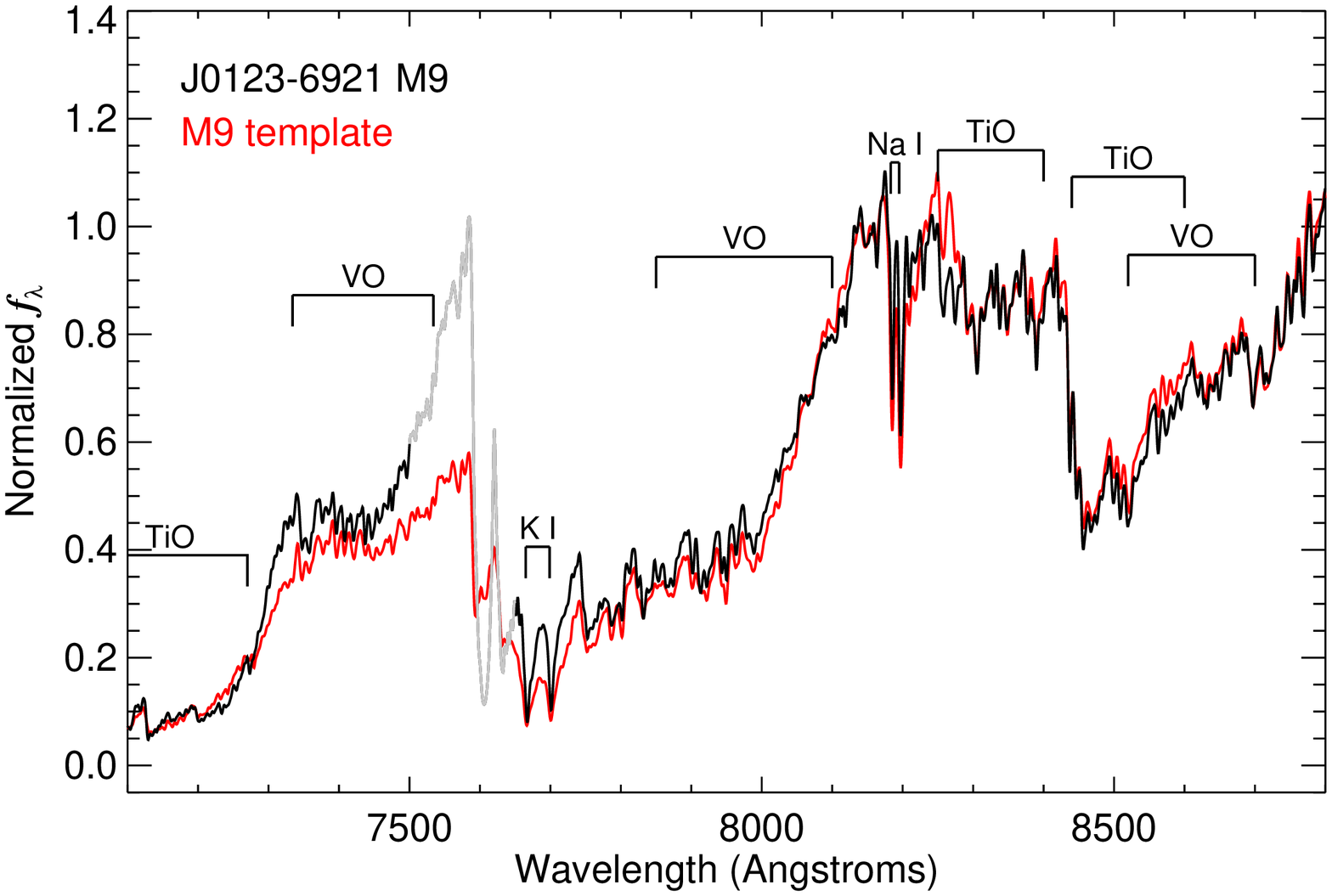} \\
\includegraphics[width=0.45\textwidth]{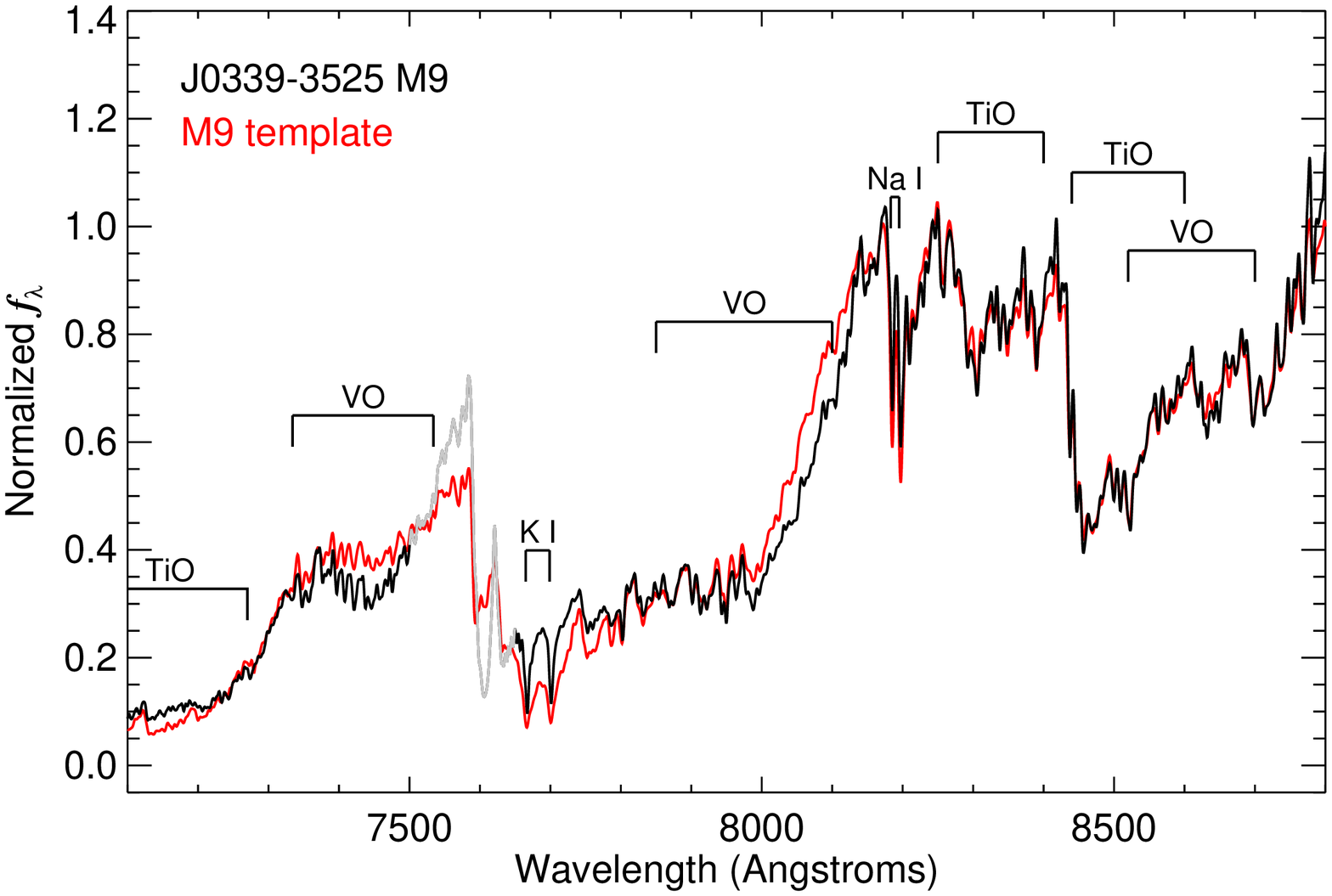} 
\includegraphics[width=0.45\textwidth]{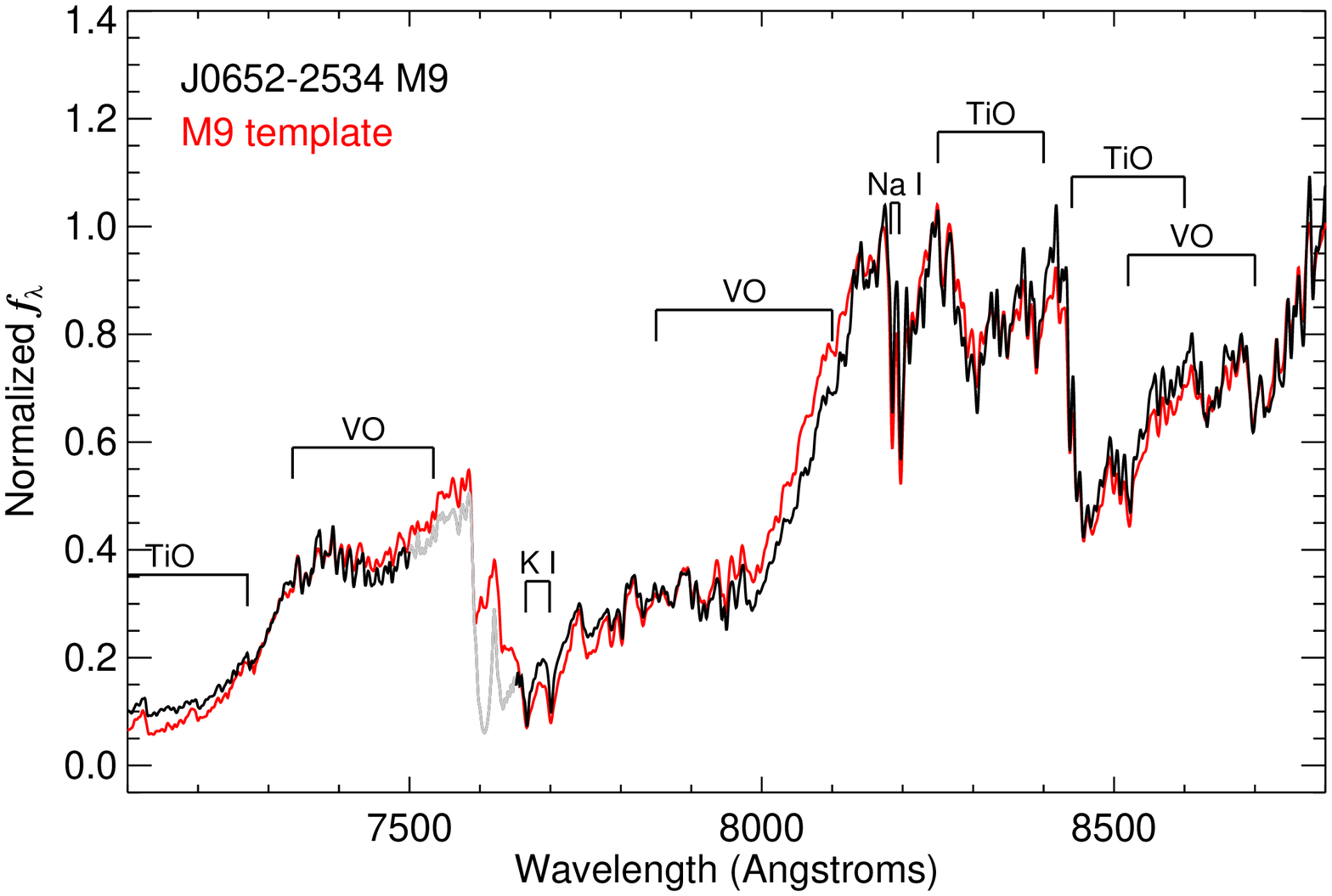} \\
\includegraphics[width=0.45\textwidth]{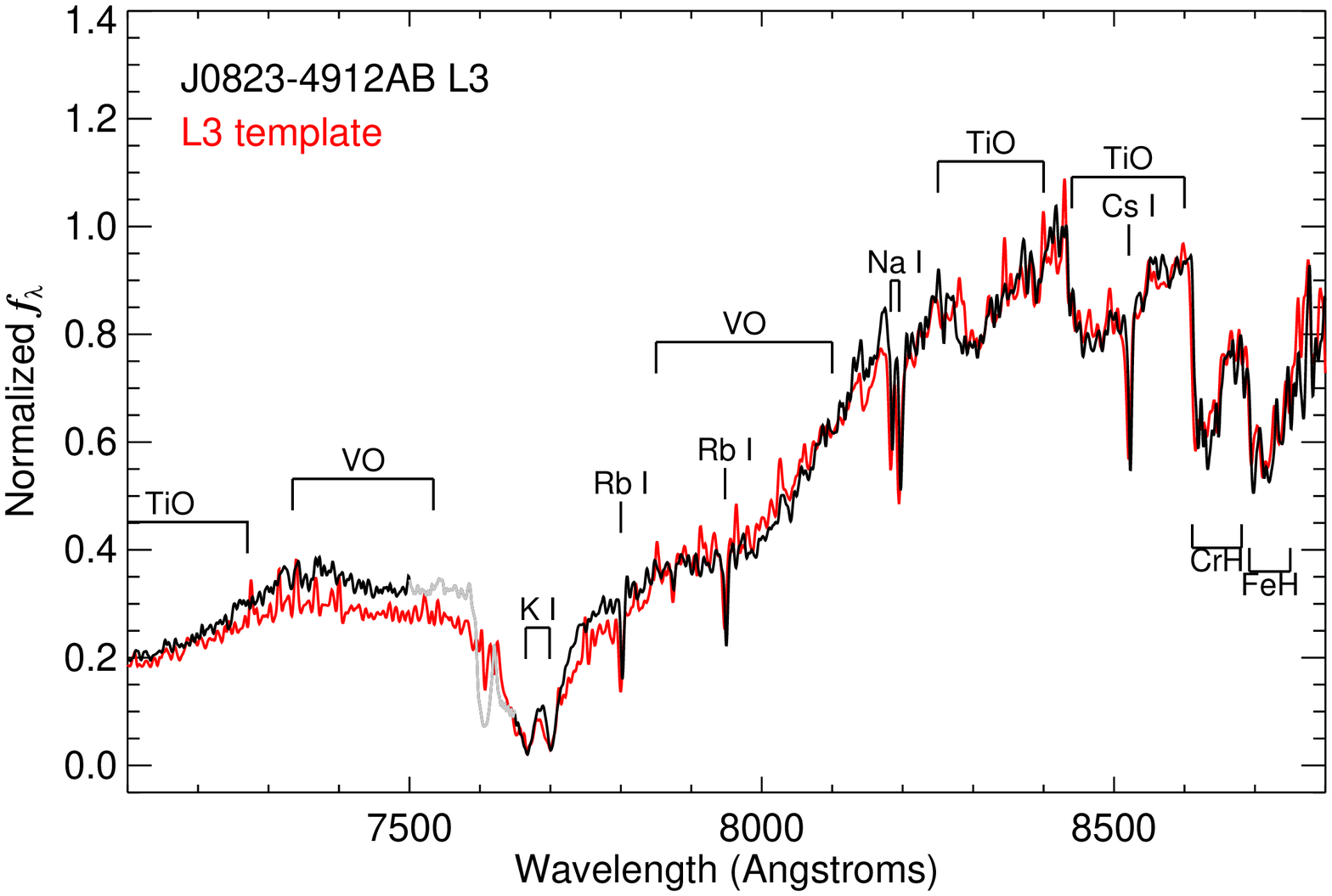}
\includegraphics[width=0.45\textwidth]{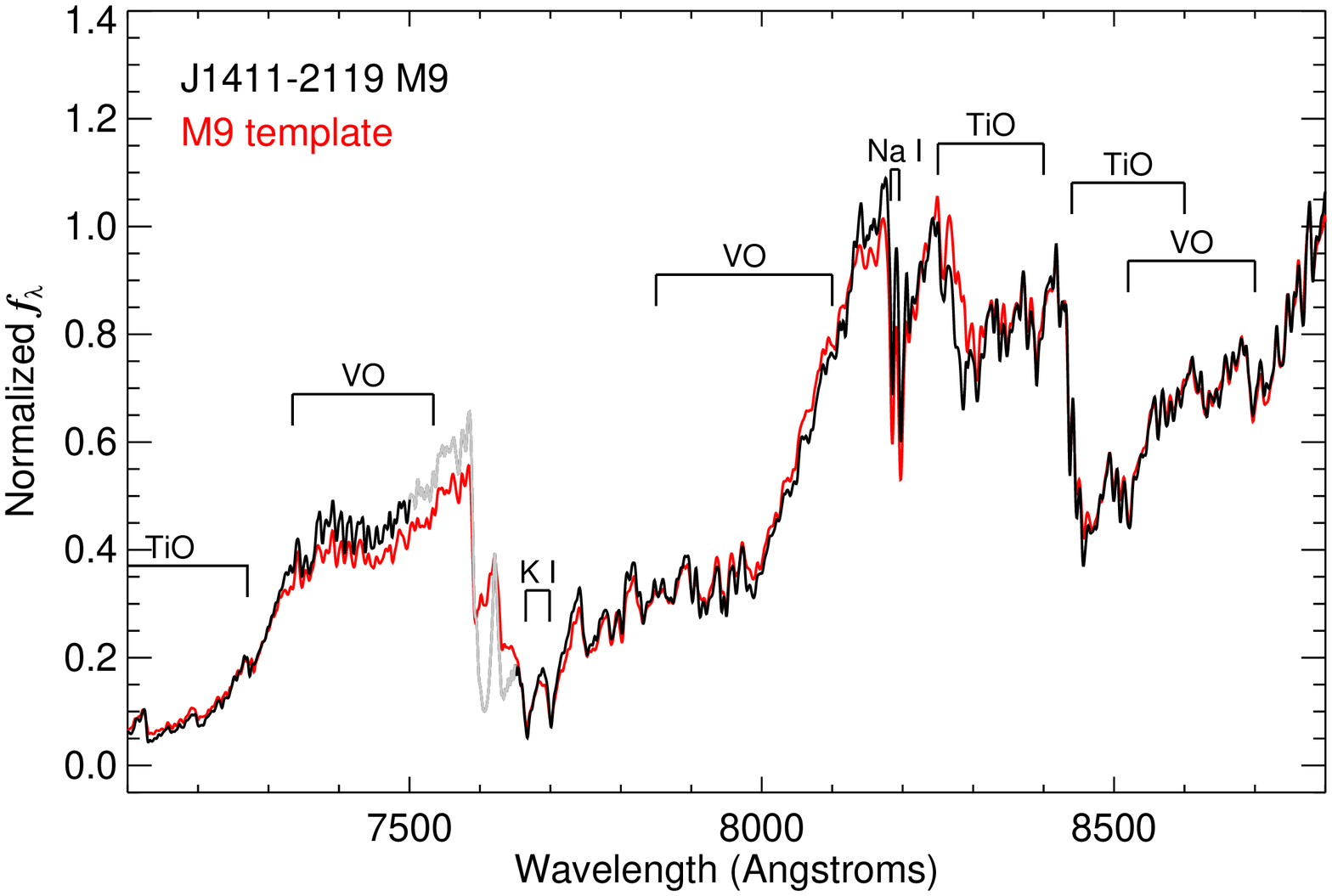}
\caption{Same as Figure~\ref{fig:lowg_strong}, showing sources exhibiting { weak signatures of low surface gravity.}  
 \label{fig:lowg_weak}}
 \end{figure}

\clearpage

\begin{figure}
\centering
\includegraphics[scale=0.45]{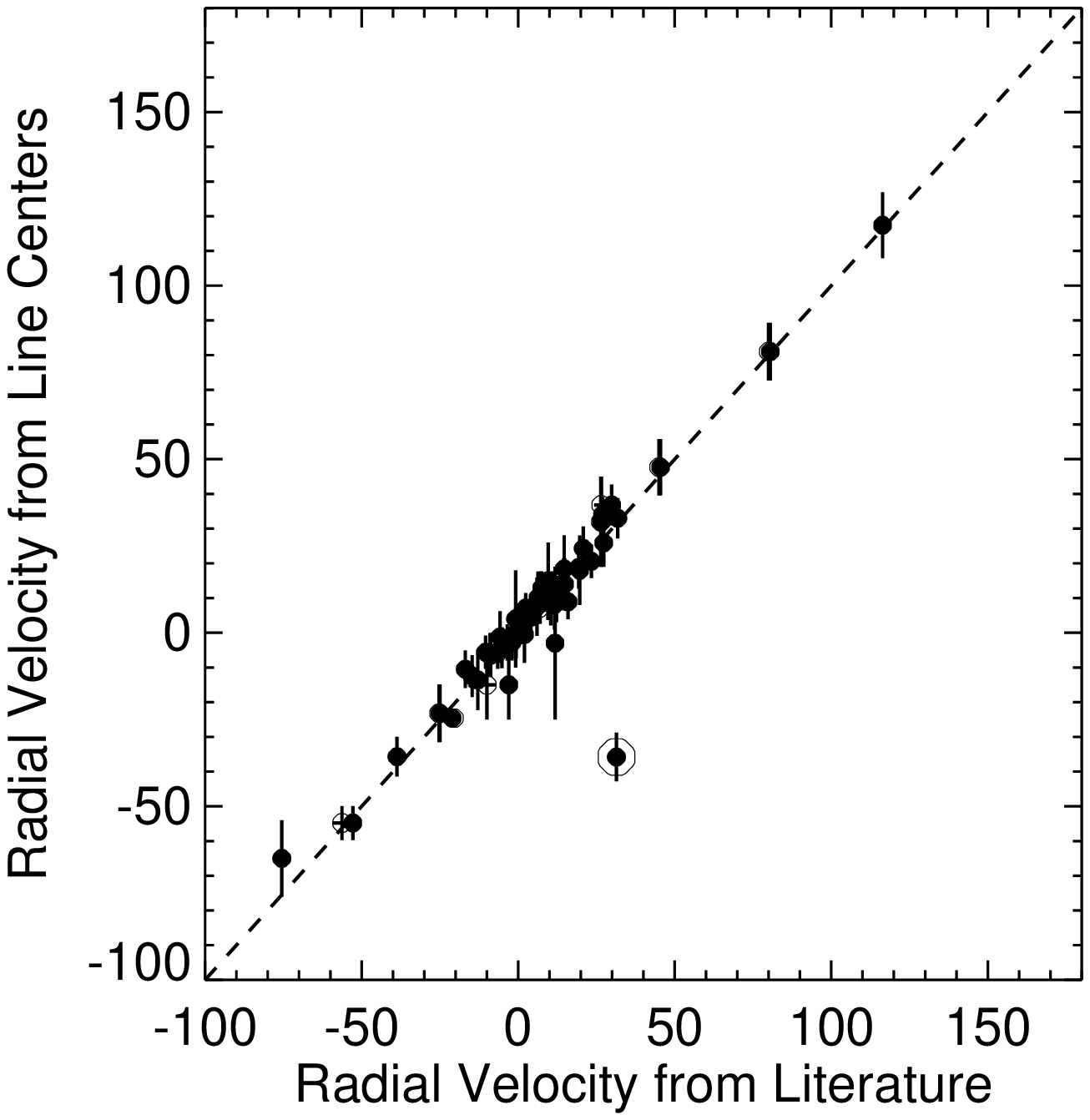}
\includegraphics[scale=0.45]{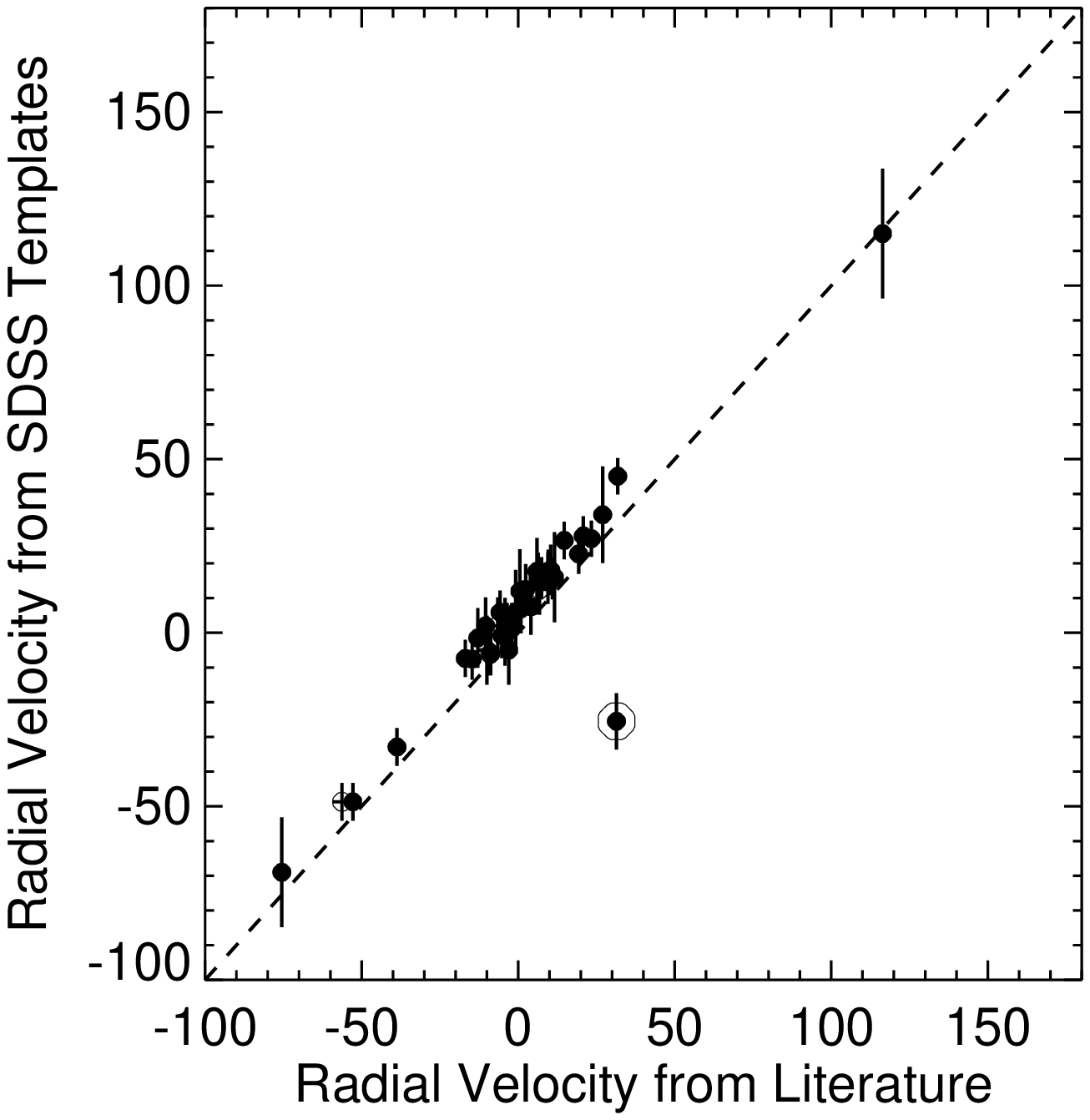}
\includegraphics[scale=0.45]{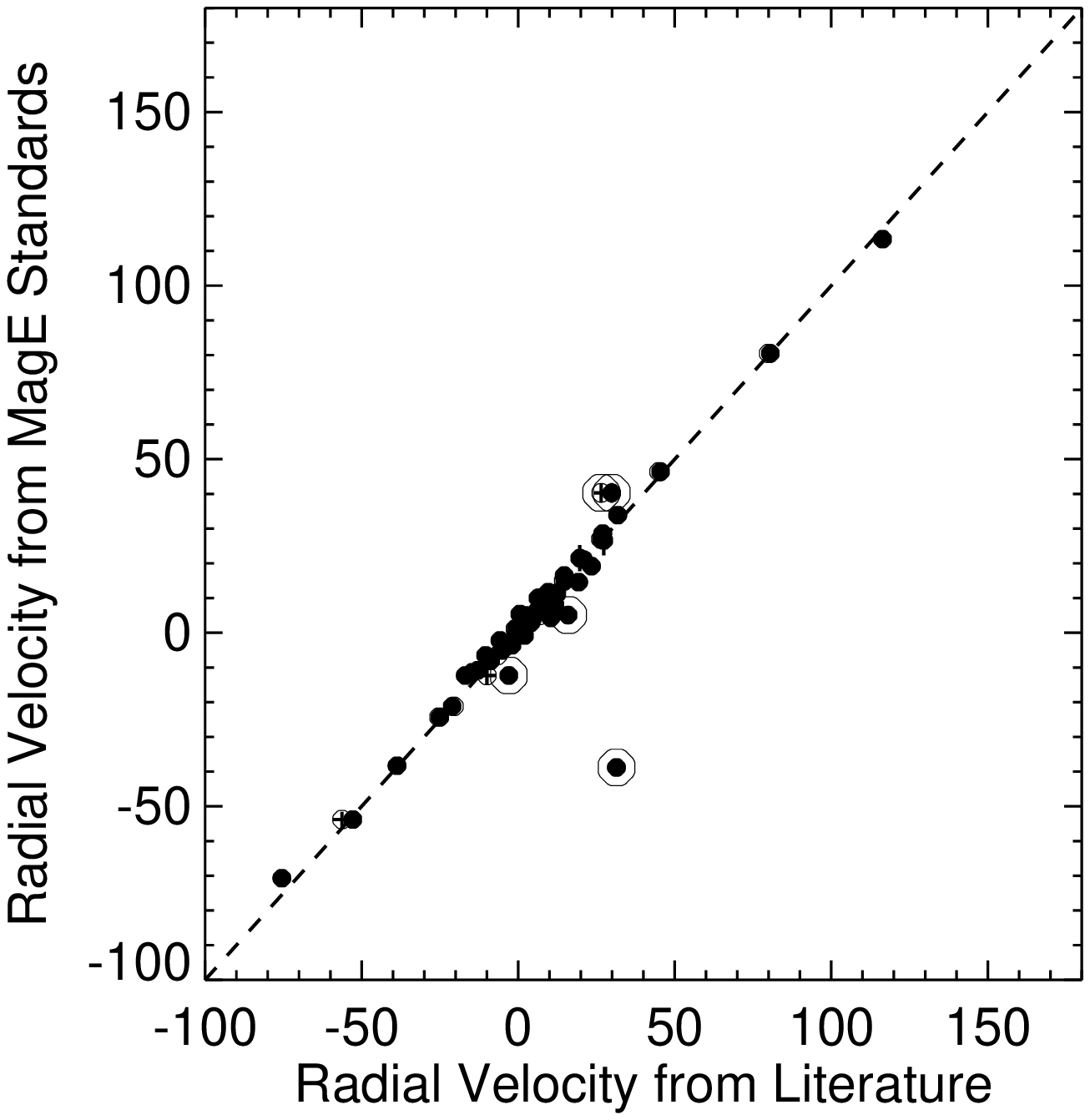}
\caption{Comparison of RVs measured from line centers (upper left panel), cross-correlation with SDSS templates (upper right panel) and cross-correlation with MagE RV standards (lower panel) to previously published high precision measurements ($\leq$3~{\kms}; Table~\ref{table:sample}).  The consistent deviant, J0517$-$3349, is highlighted by an open circle; this appears to be a sign reversal in the RV reported by \citet{2008ApJ...684.1390R}.  The other { deviants in the lower panel, J0500+0330, J0835+0819 and J2351$-$2537, are discussed further in Section~3.4}.
\label{figure:litcomp}}
\end{figure}

%self-cross vs lit; self-cross vs lines
\begin{figure}
\centering
\includegraphics[scale=0.45]{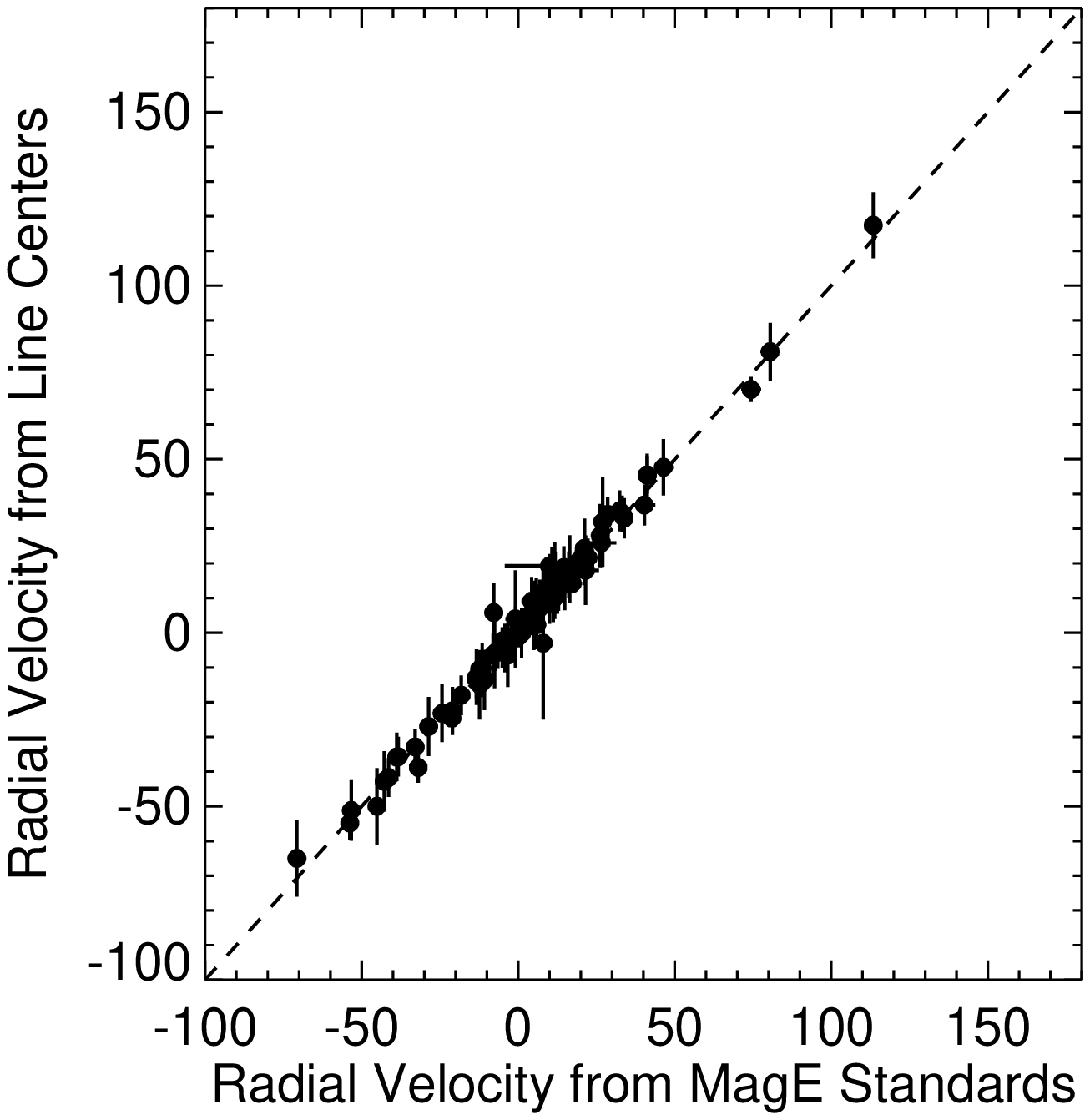}
\includegraphics[scale=0.45]{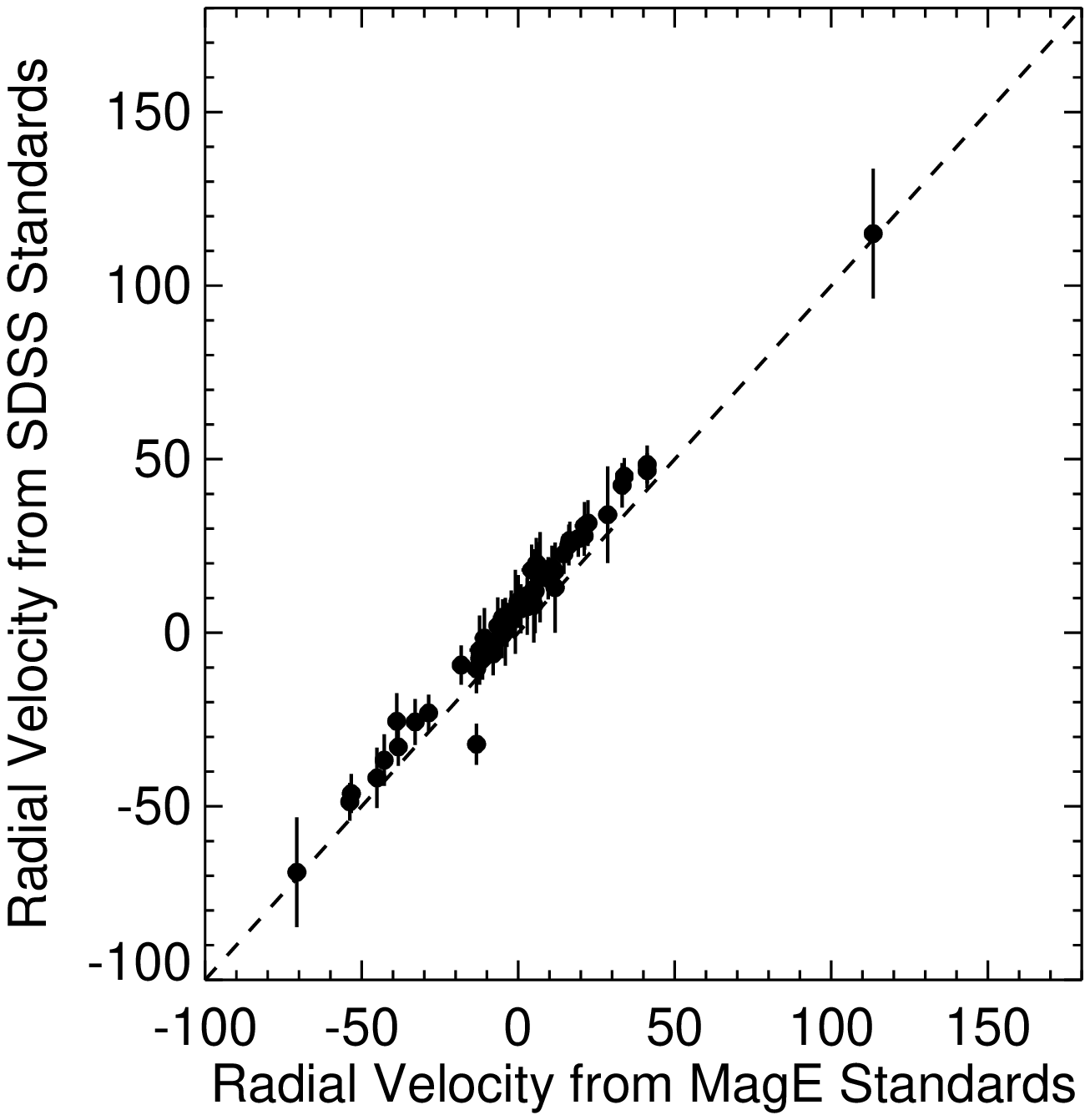}
\caption{Comparison of RV measurements between line centers, SDSS templates, and MagE RV standards. 
\label{figure:methodcomp}}
\end{figure}

%UVW plots
\begin{figure}
%\centering
\includegraphics[scale=0.4]{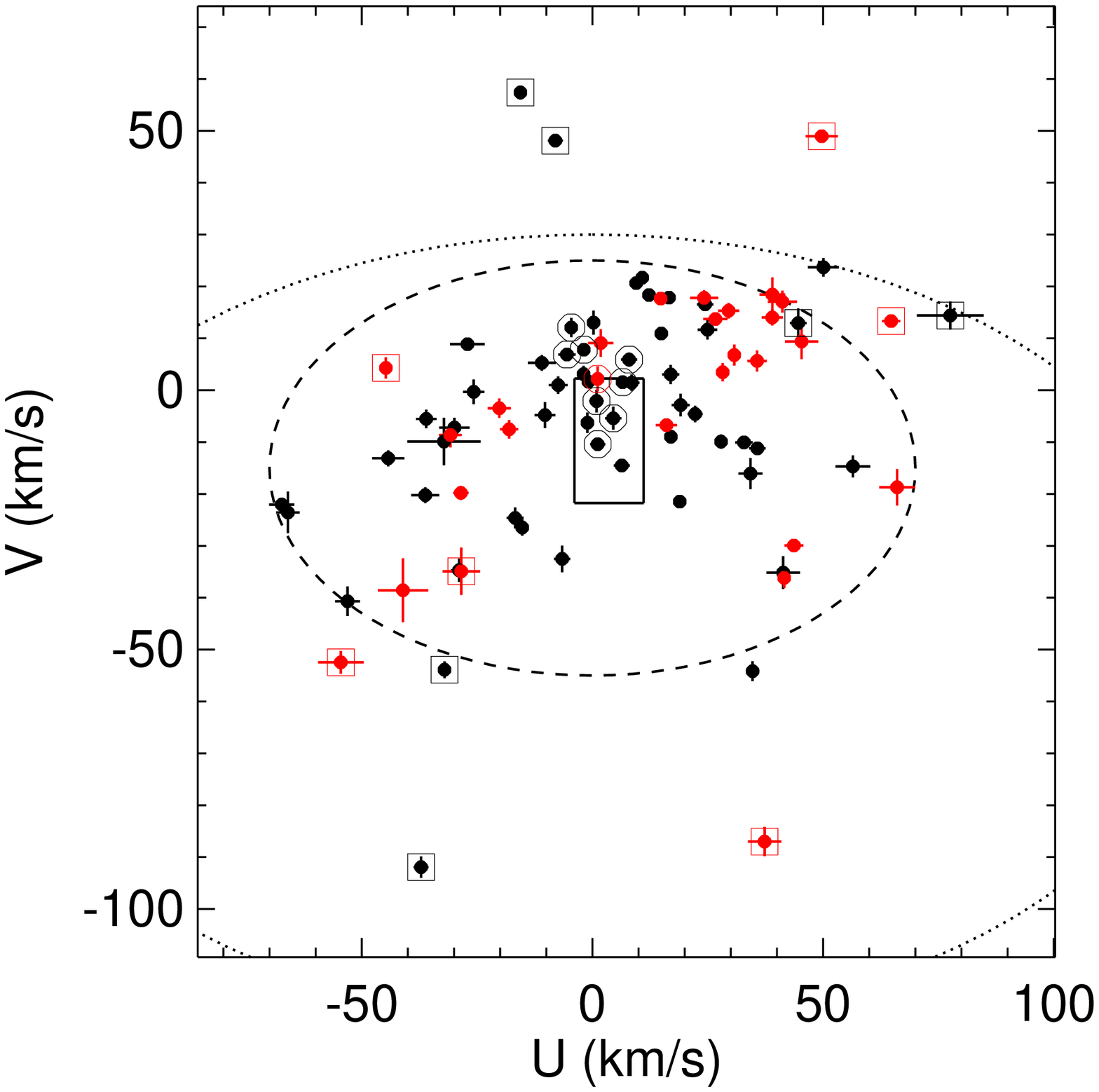} 
\includegraphics[scale=0.4]{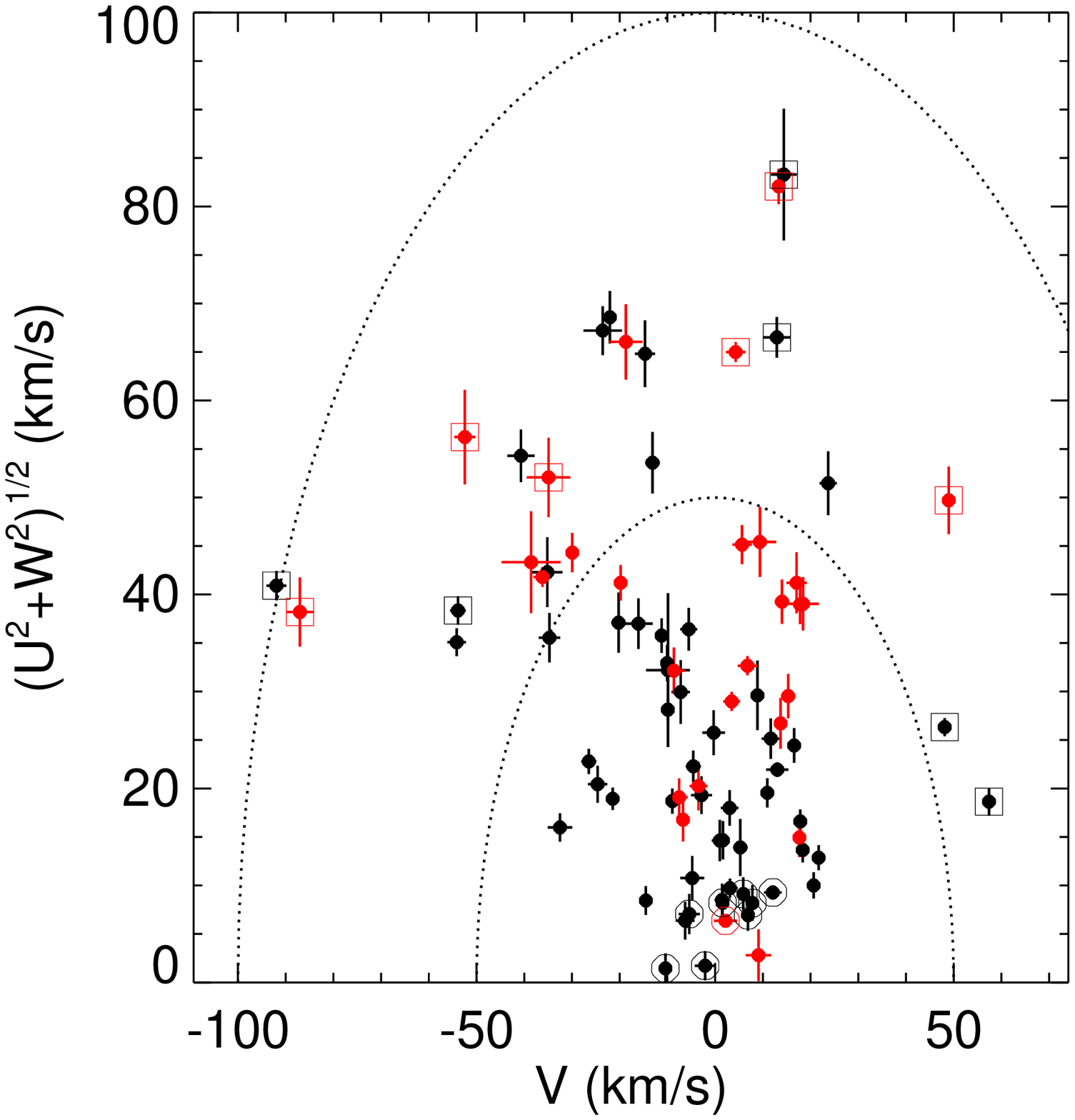}  \\
\includegraphics[scale=0.4]{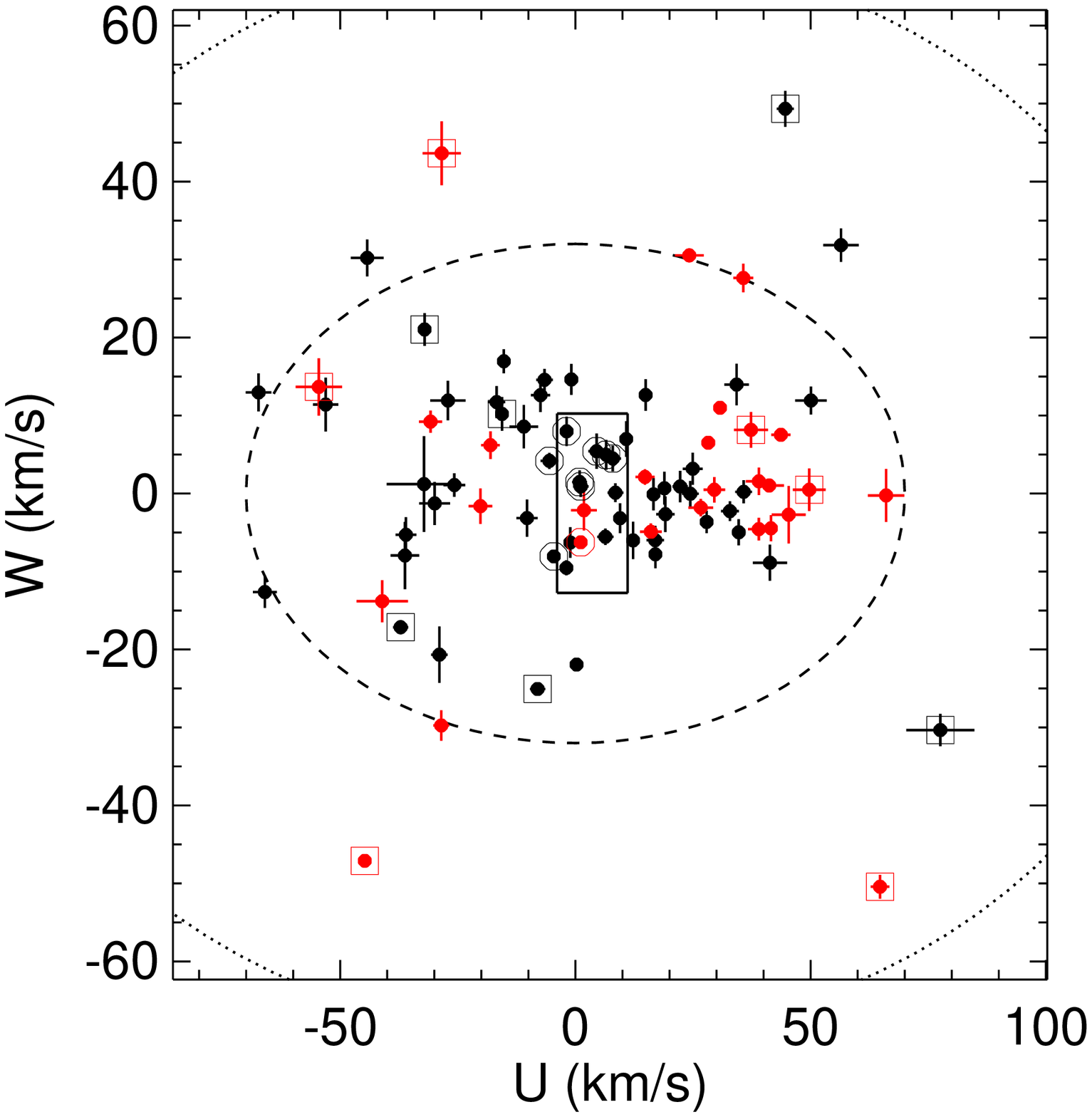}
\includegraphics[scale=0.4]{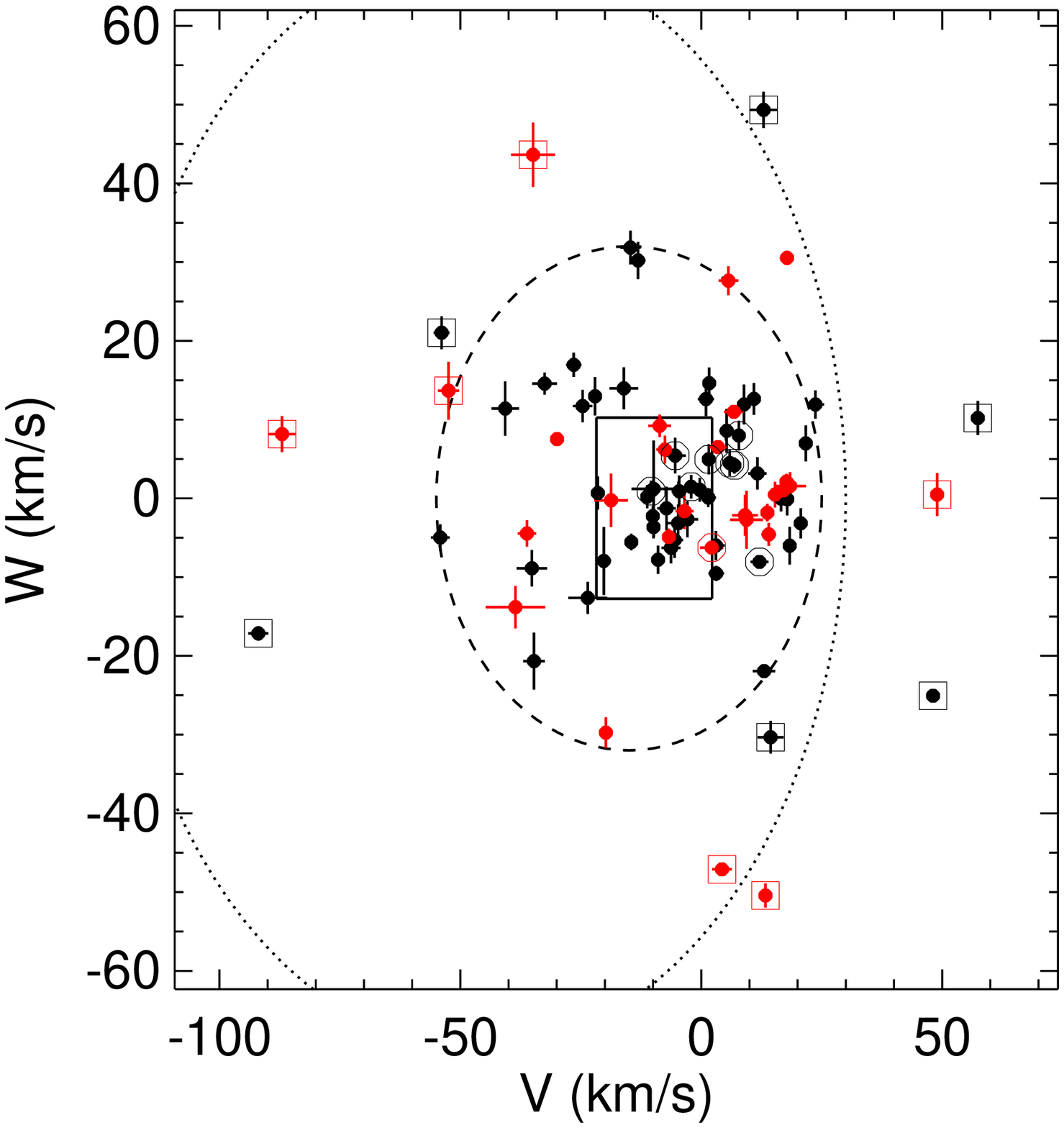} \\
\caption{$UVW$ space motions for our sample in the Local Standard of Rest. Upper left, lower left and lower right plots compare velocity components to the 2$\sigma$ velocity means and dispersions of Galactic thin disk (dashed lines) and thick disk (dotted lines) populations from \citet{2003A&A...410..527B}. Also shown is the ``good box'' of \citet{2004ARA&A..42..685Z}, a rough locus of nearby YMG members.  The upper right panel shows a Toomre plot, with dotted lines delineating 50~{\kms} steps of constant $v_{tot}$ =  $(U^2+V^2+W^2)^{1/2}$.
Late-M dwarfs are indicated by black symbols, L dwarfs by red symbols.  Sources identified as  intermediate thin/thick disk stars are highlighted by open squares; sources exhibiting Li~I absorption are highlighted by open circles.
\label{figure:UVW}}
\end{figure}

\begin{figure}
\centering
\includegraphics[scale=0.32]{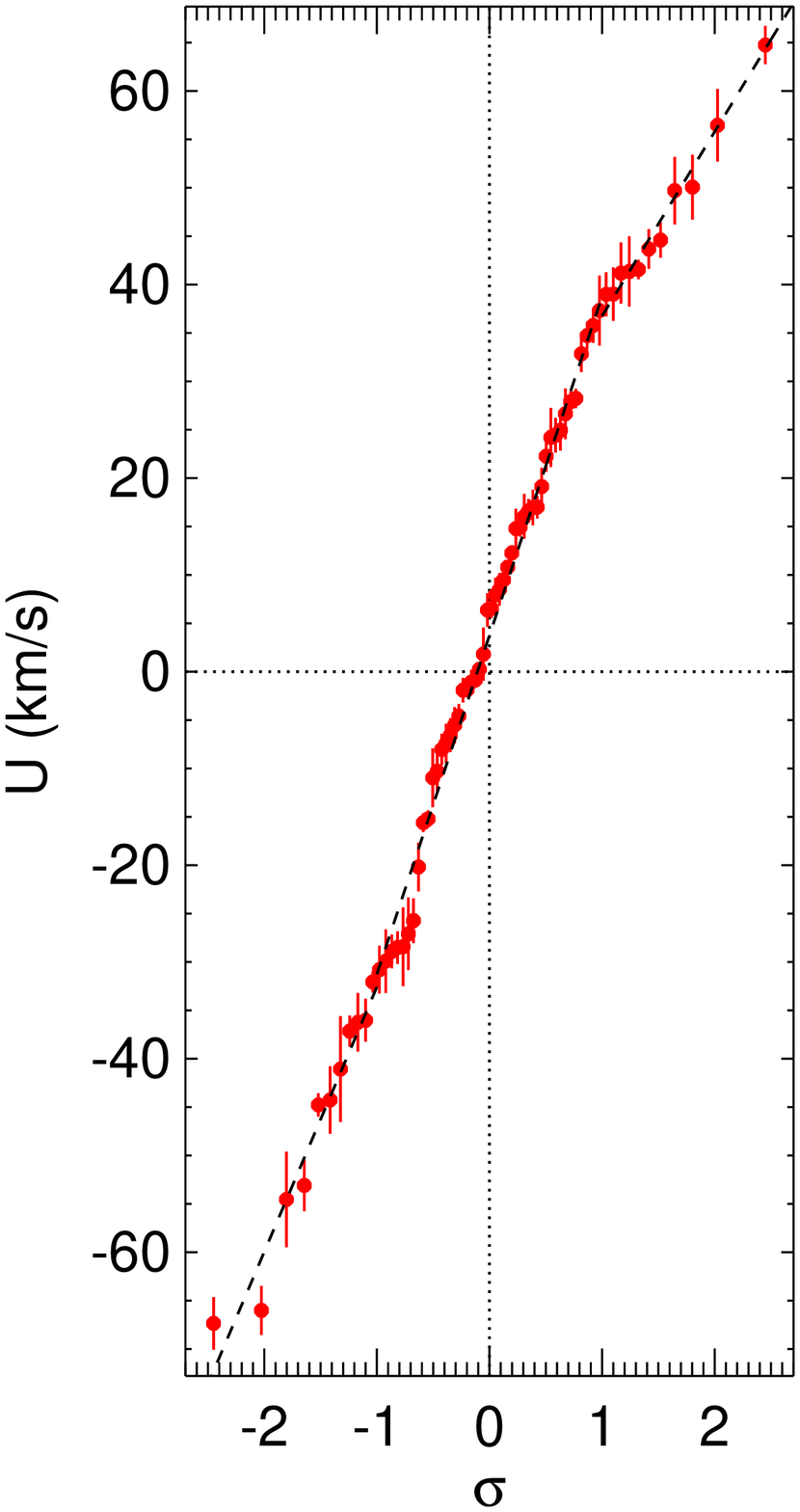}
\includegraphics[scale=0.32]{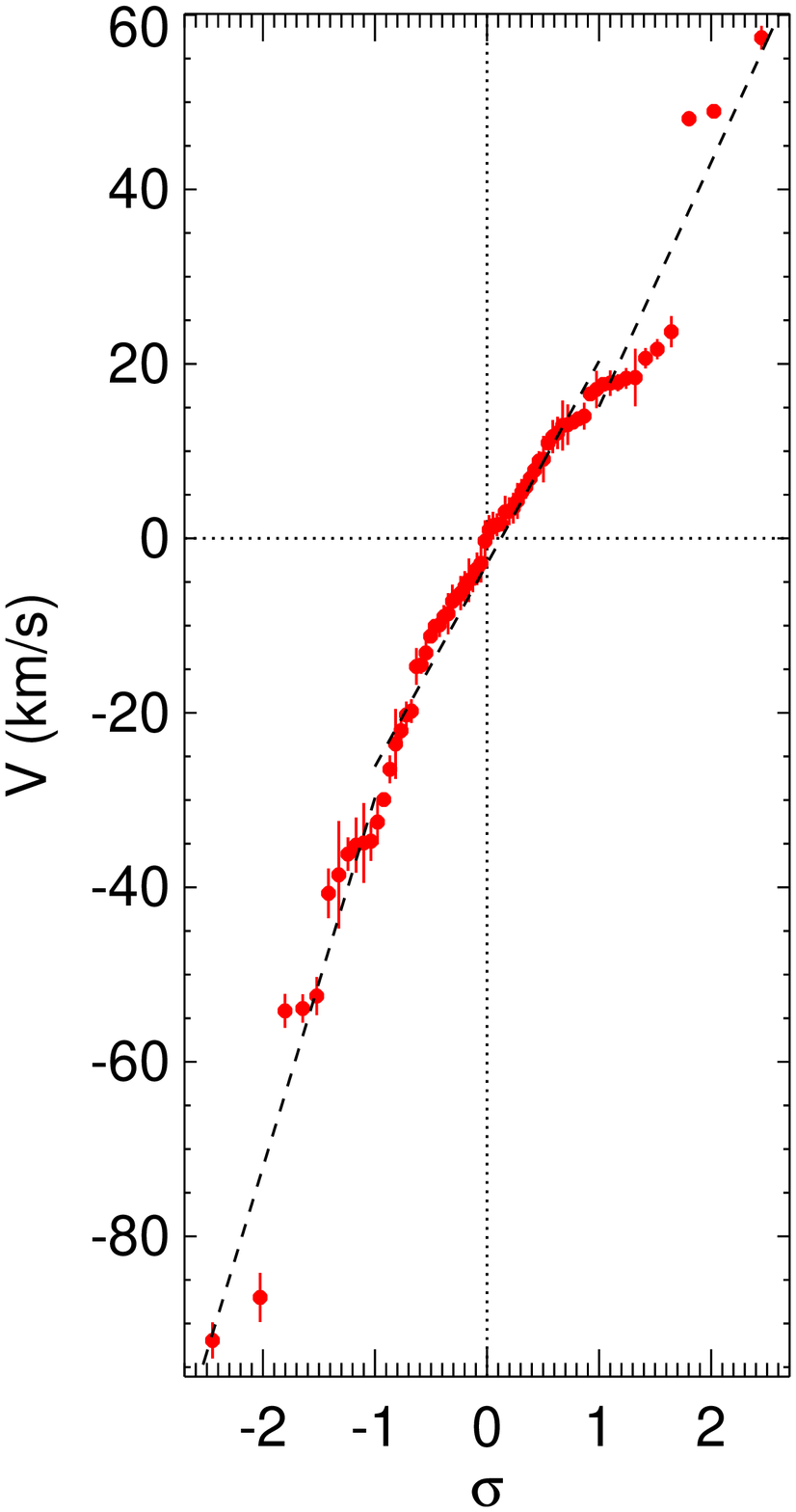}
\includegraphics[scale=0.32]{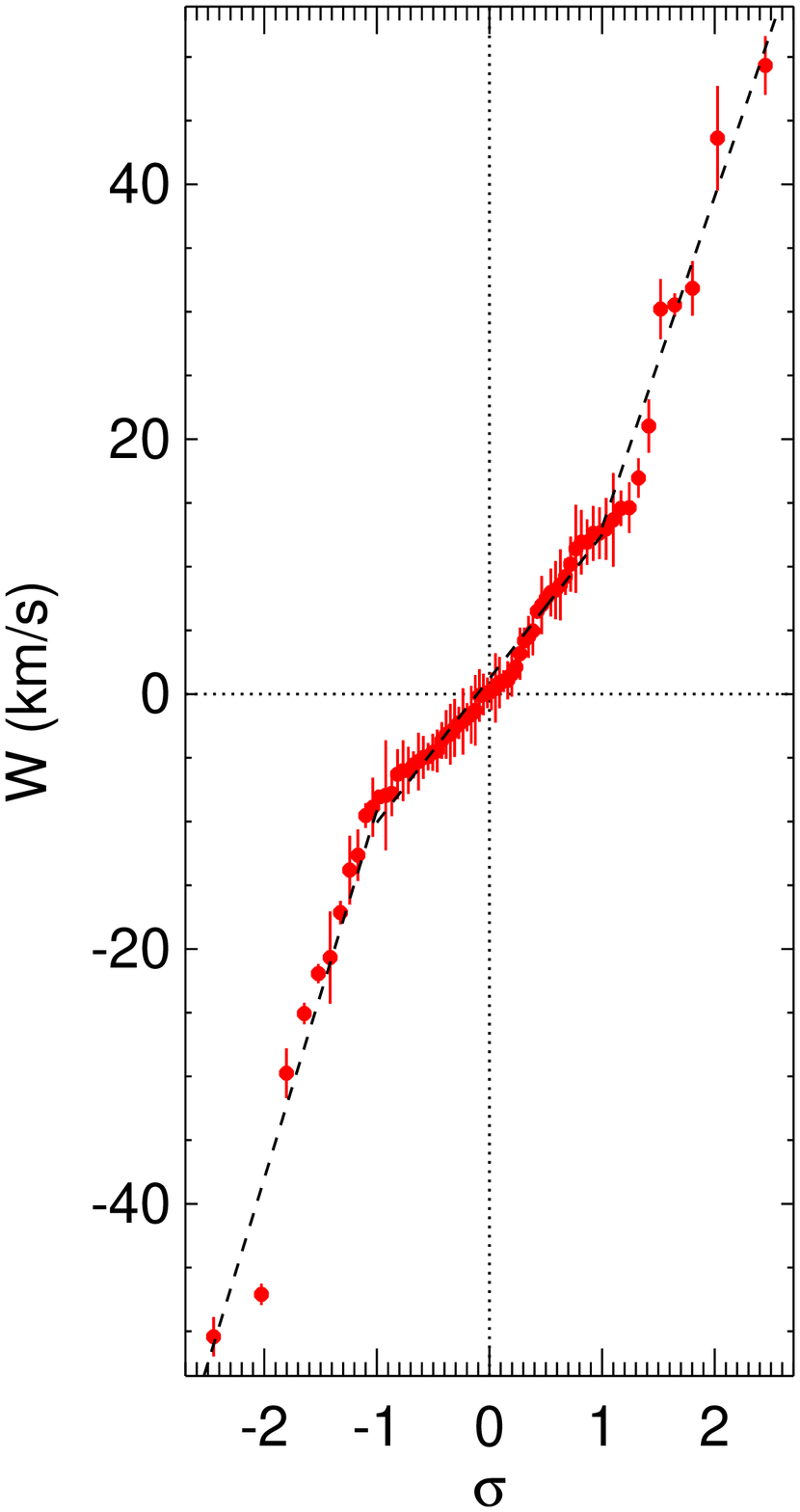}
\caption{Probit plots for $U$,  $V$, and $W$ velocity components for sources in our sample within 20 pc of the Sun. Individual velocity measurements and uncertainties are indicated in red.  Separate linear fits are shown for core ($|\sigma| < 1$) and warm ($|\sigma| > 1$) populations.  Dotted lines mark the mean of the velocity distributions ($\sigma$ = 0) and zero velocity.
\label{figure:disperse_20pc}}
\end{figure}

\begin{figure}
\centering
\includegraphics[scale=0.32]{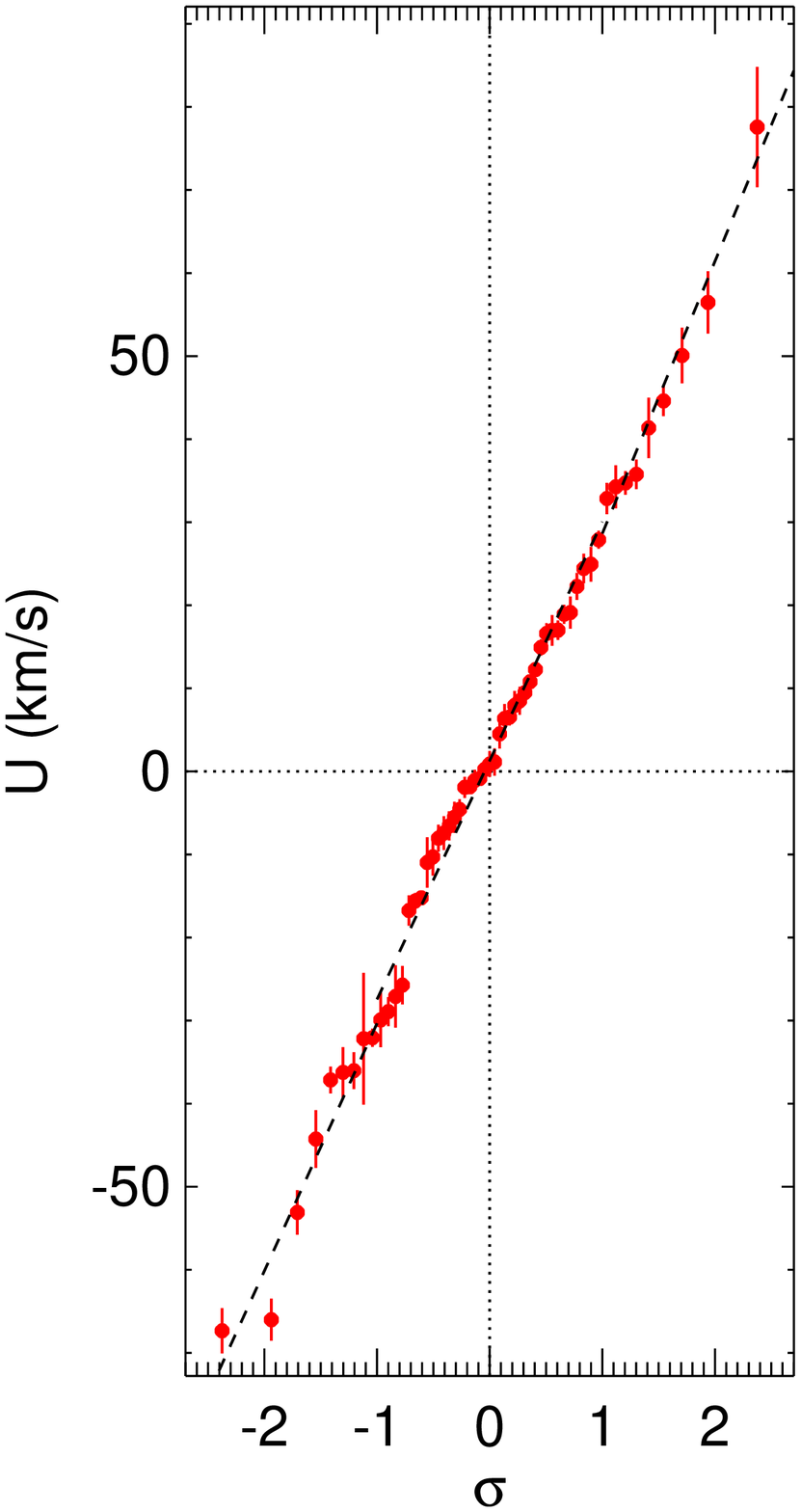}
\includegraphics[scale=0.32]{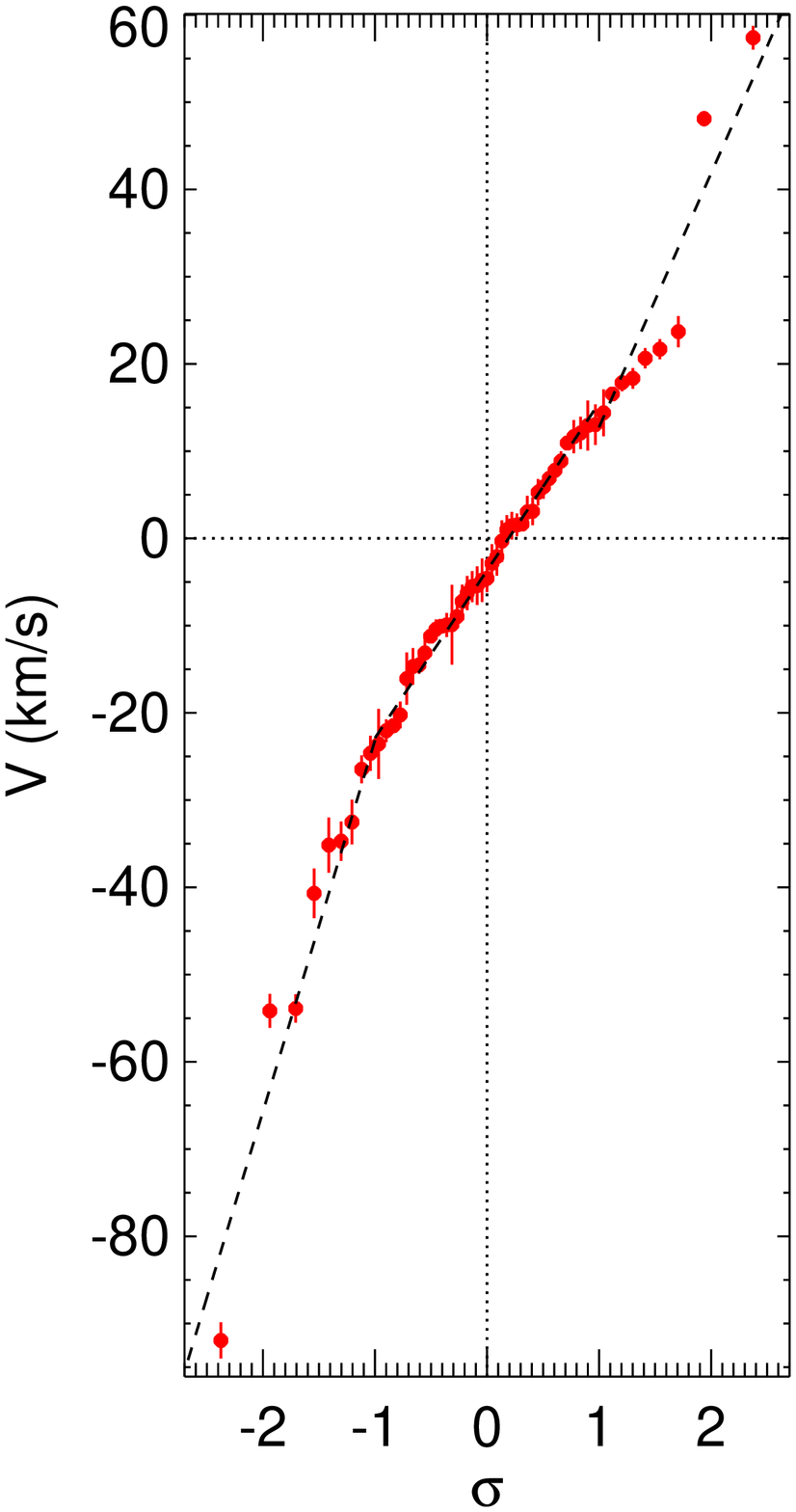}
\includegraphics[scale=0.32]{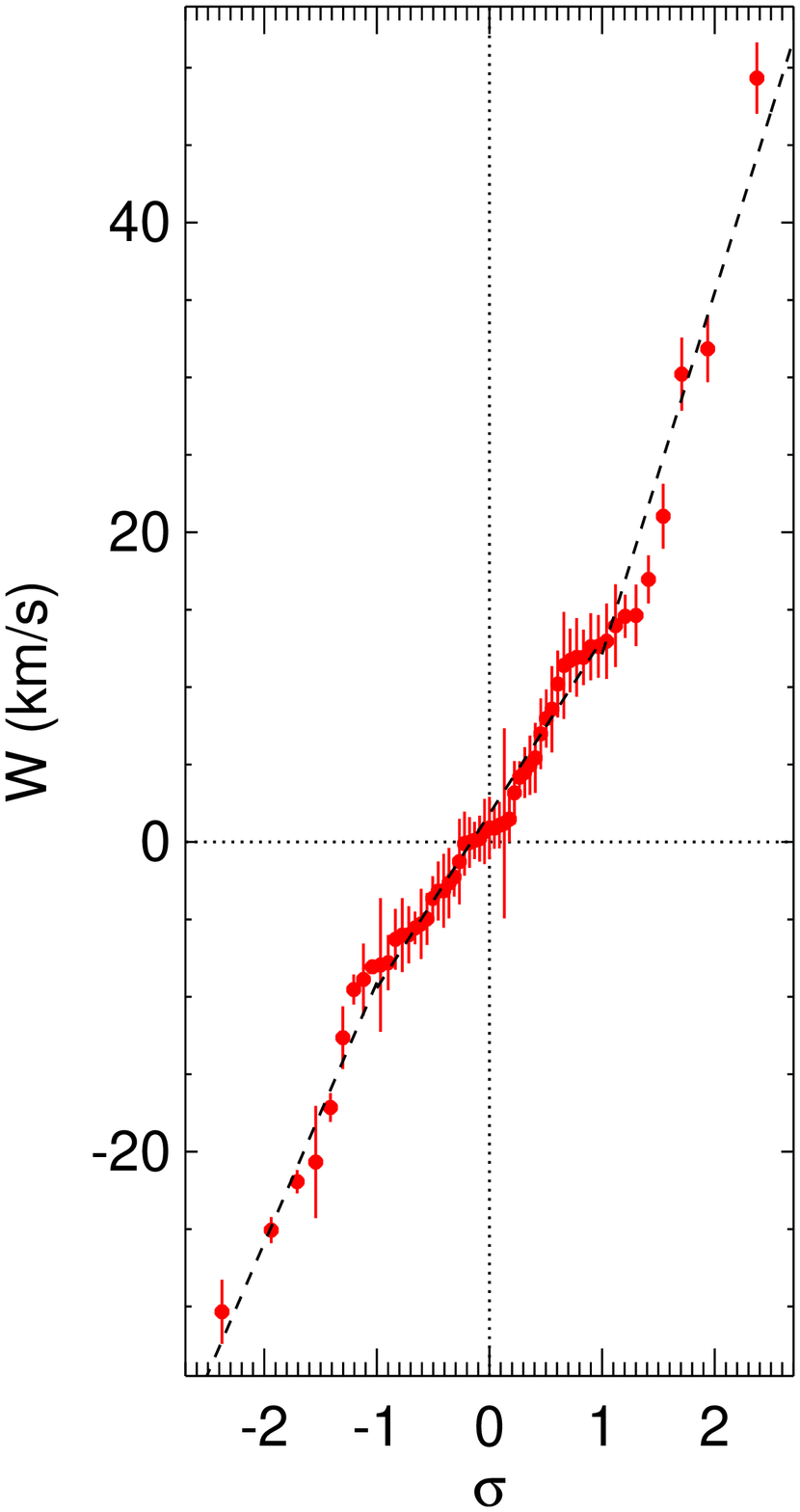}
\caption{Same as Figure~\ref{figure:disperse_20pc} for the late-M dwarfs in our sample. 
\label{figure:disperse_mdwarf}}
\end{figure}

\begin{figure}
\centering
\includegraphics[scale=0.32]{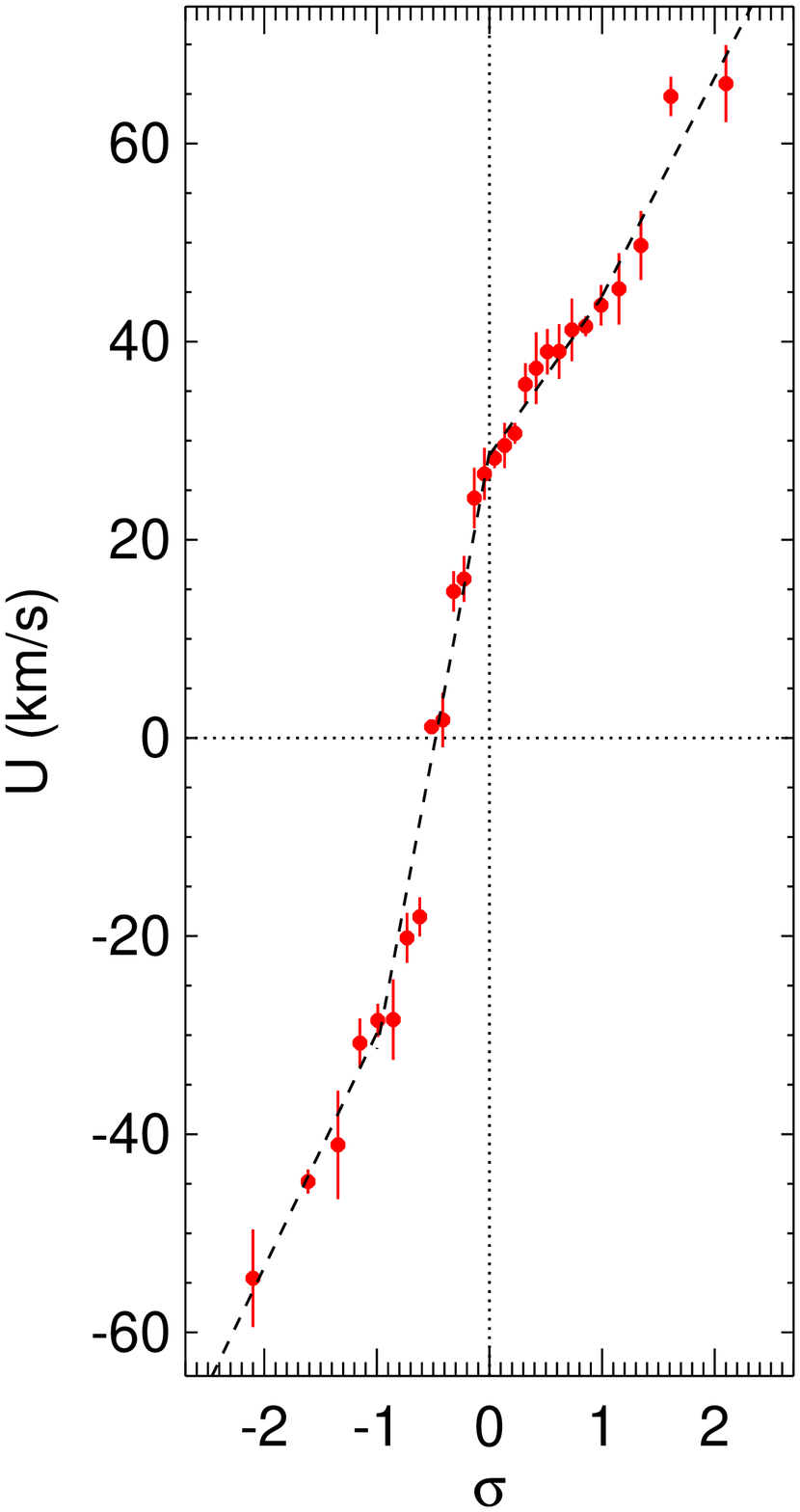}
\includegraphics[scale=0.32]{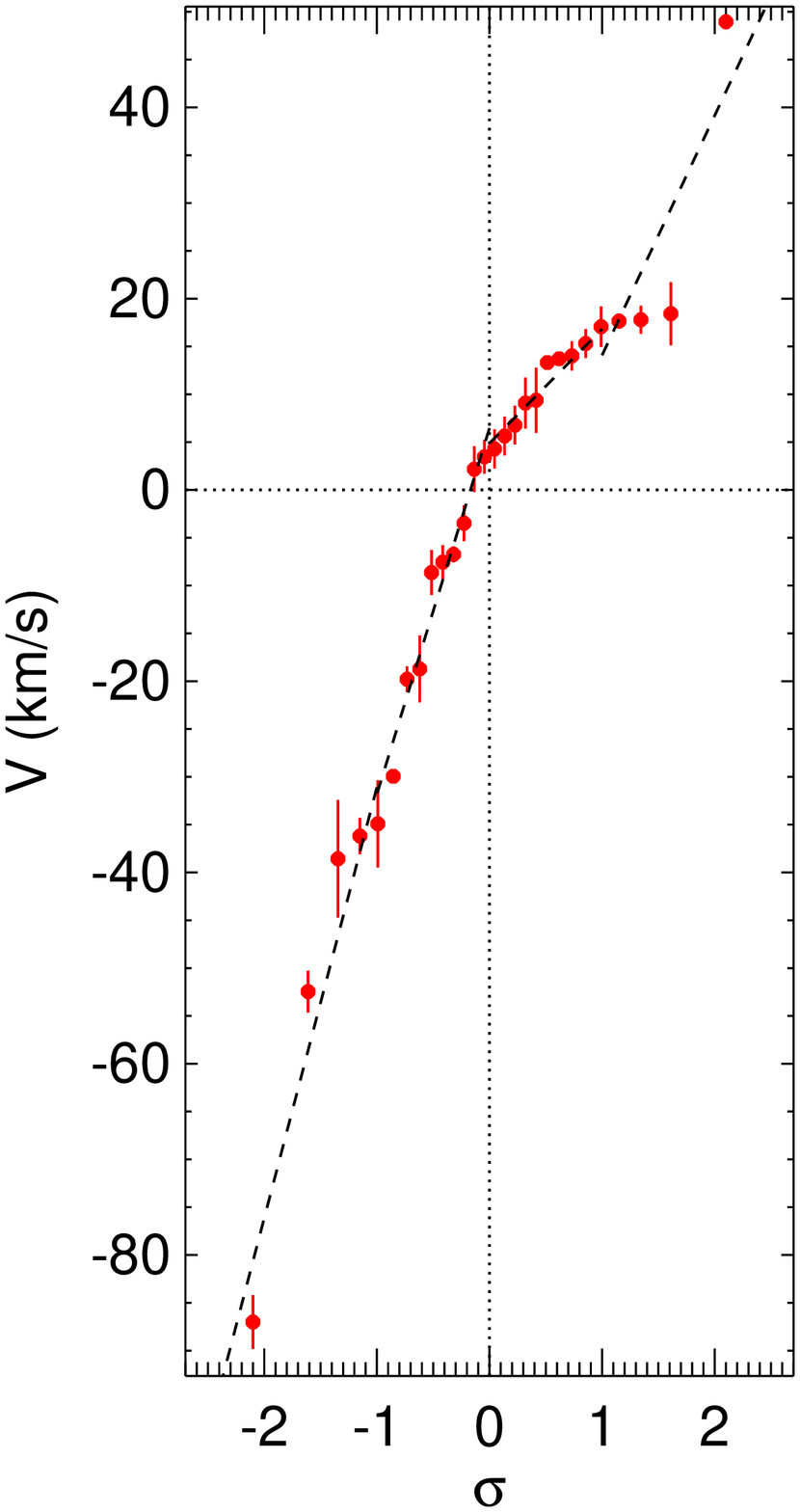}
\includegraphics[scale=0.32]{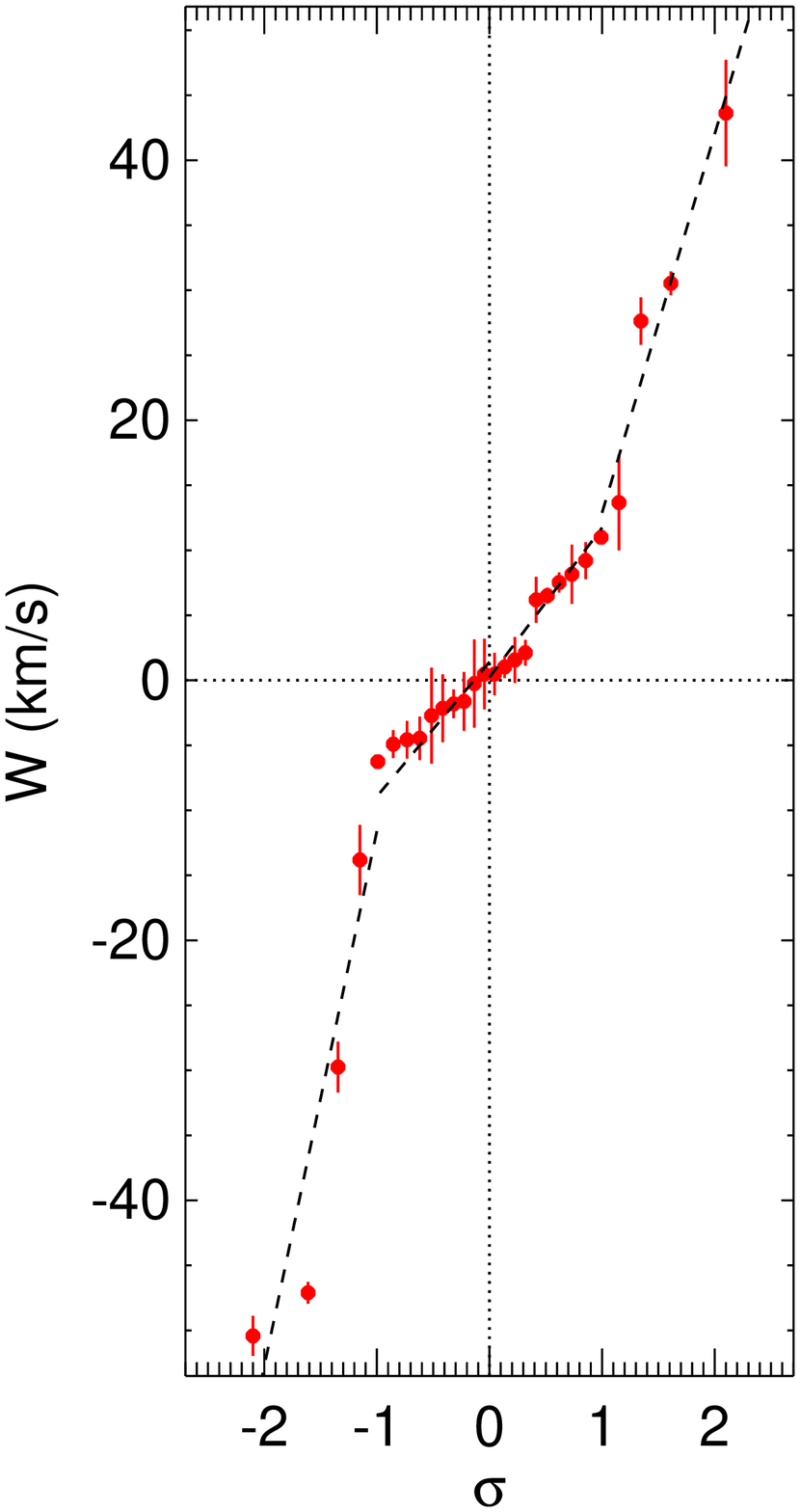}
\caption{Same as Figure~\ref{figure:disperse_20pc} for the L dwarfs in our sample.  In this case, separate fits are made within the core population for $\sigma > 0$ and $\sigma < 0$, which have significantly different linear slopes.
\label{figure:disperse_ldwarf}}
\end{figure}

\clearpage
%UVW distribution for 20 pc sample
\begin{figure}
\centering
\includegraphics[scale=0.32]{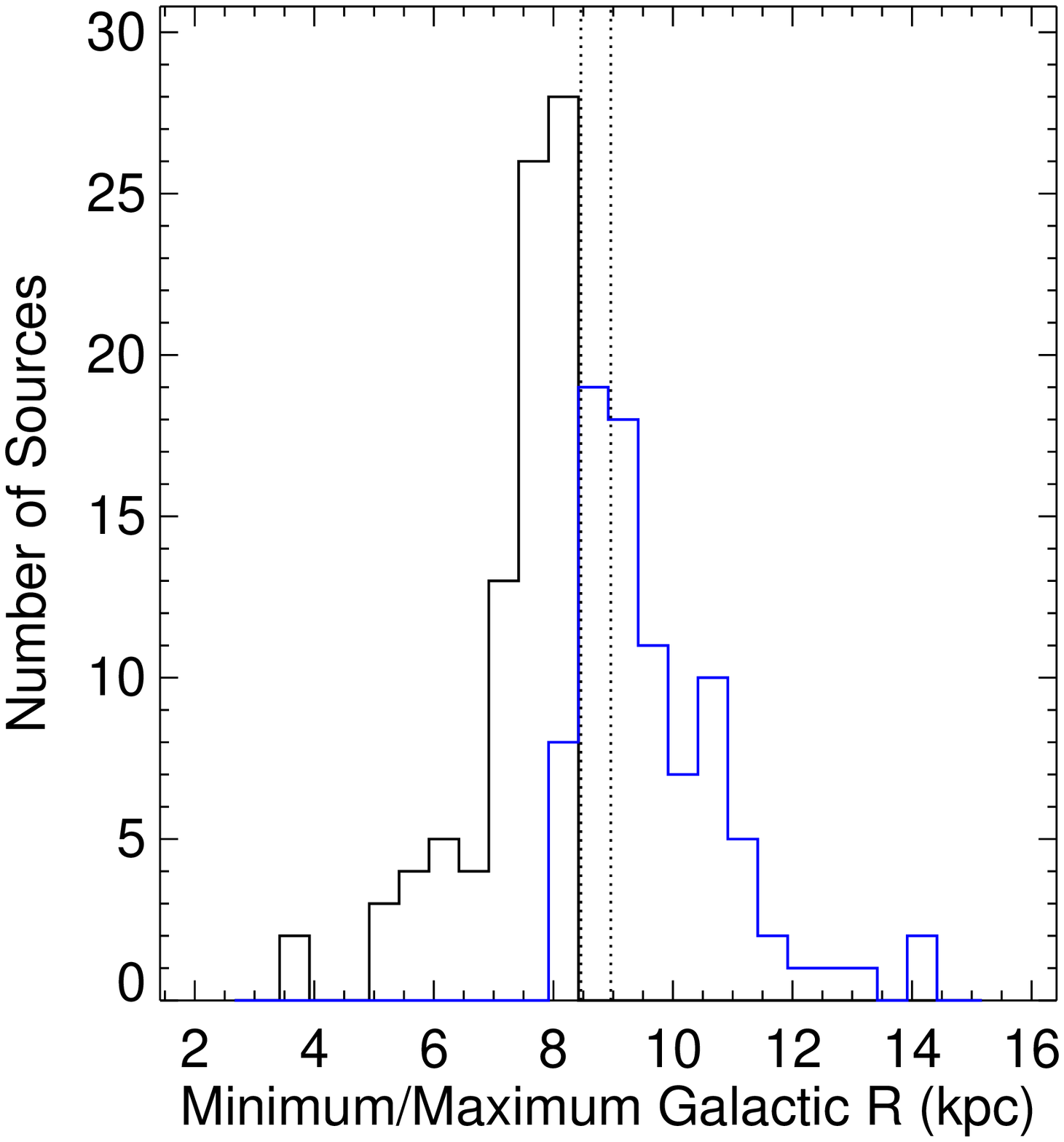}
\includegraphics[scale=0.32]{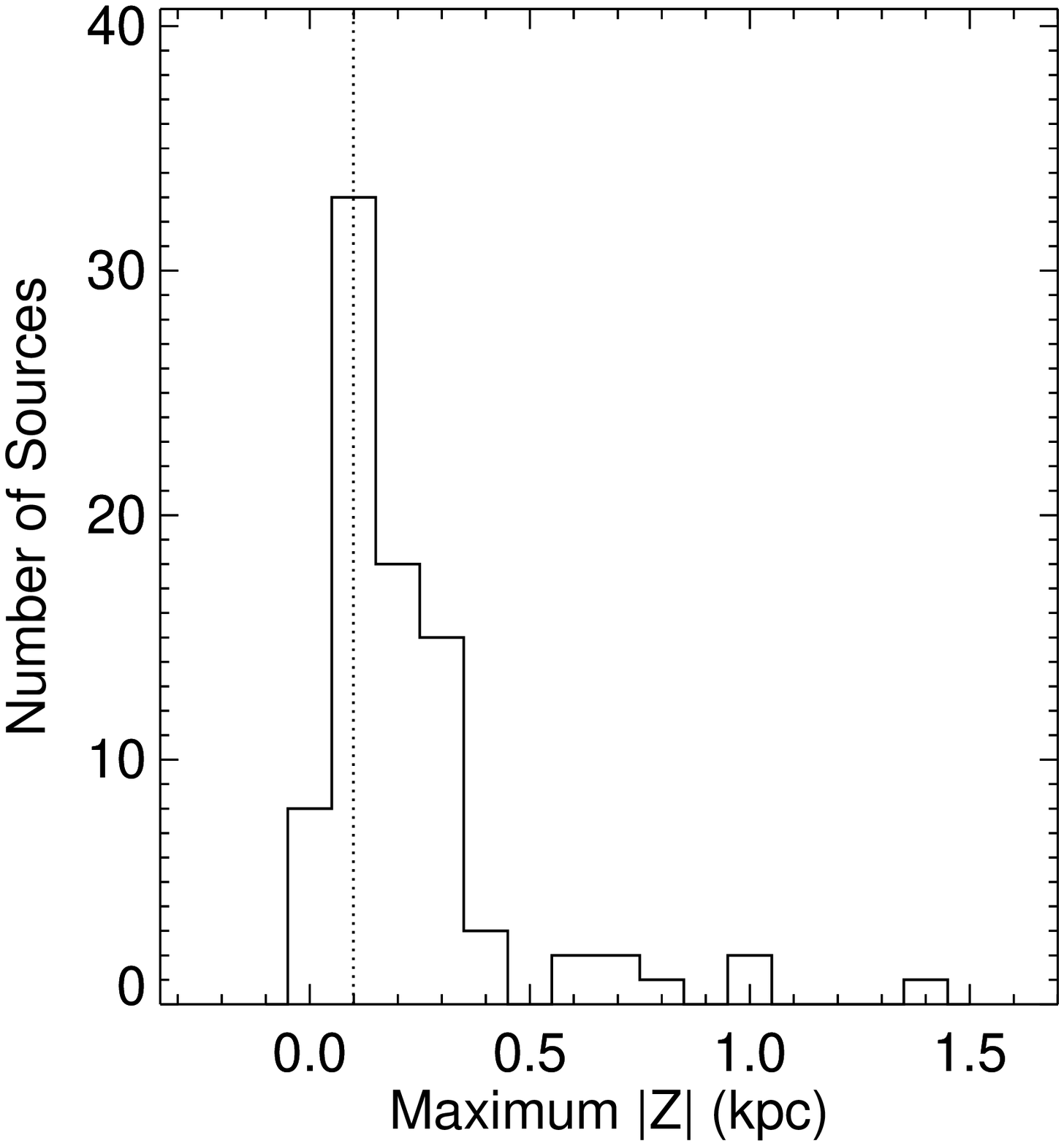}
\includegraphics[scale=0.32]{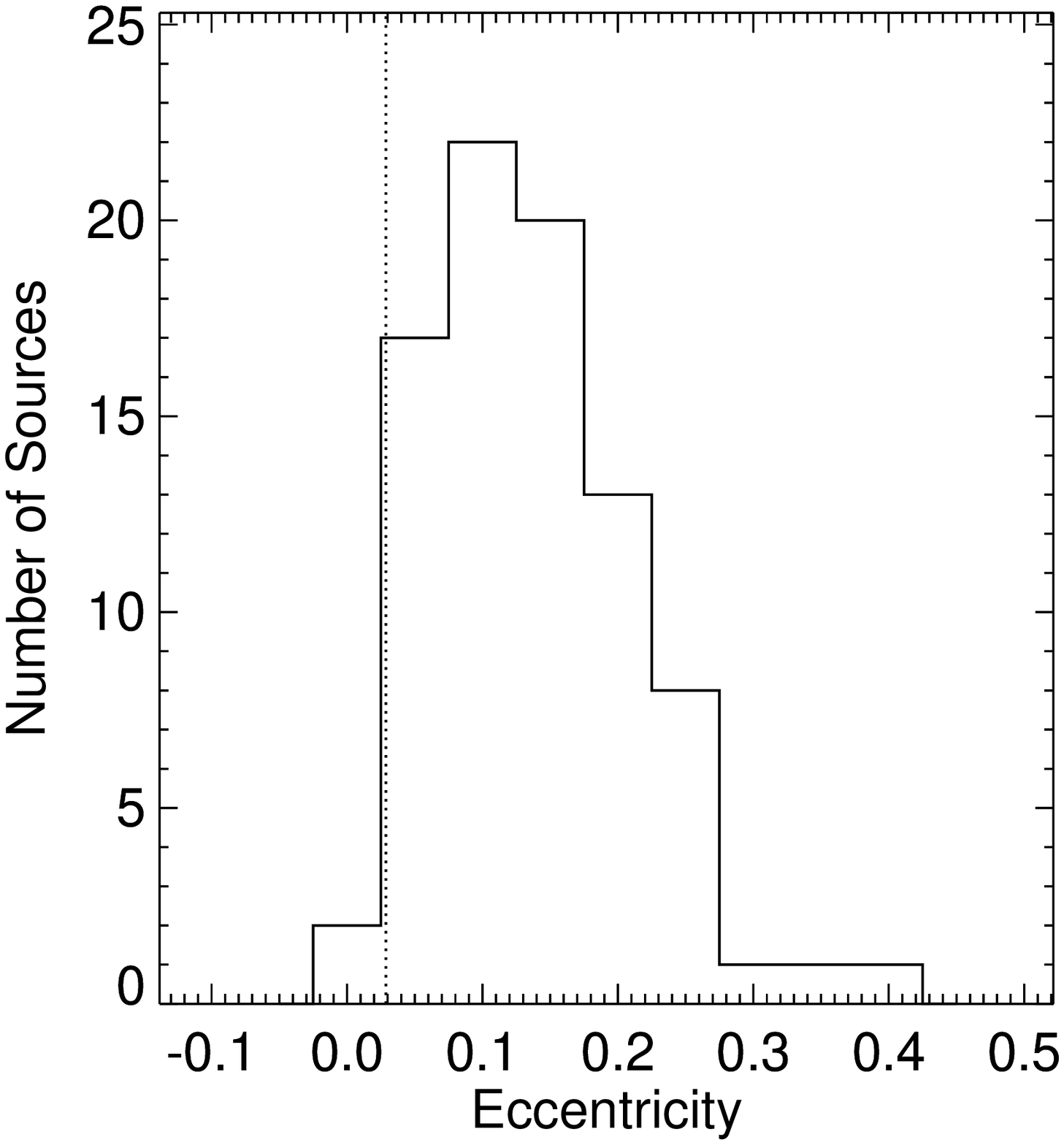}
\includegraphics[scale=0.32]{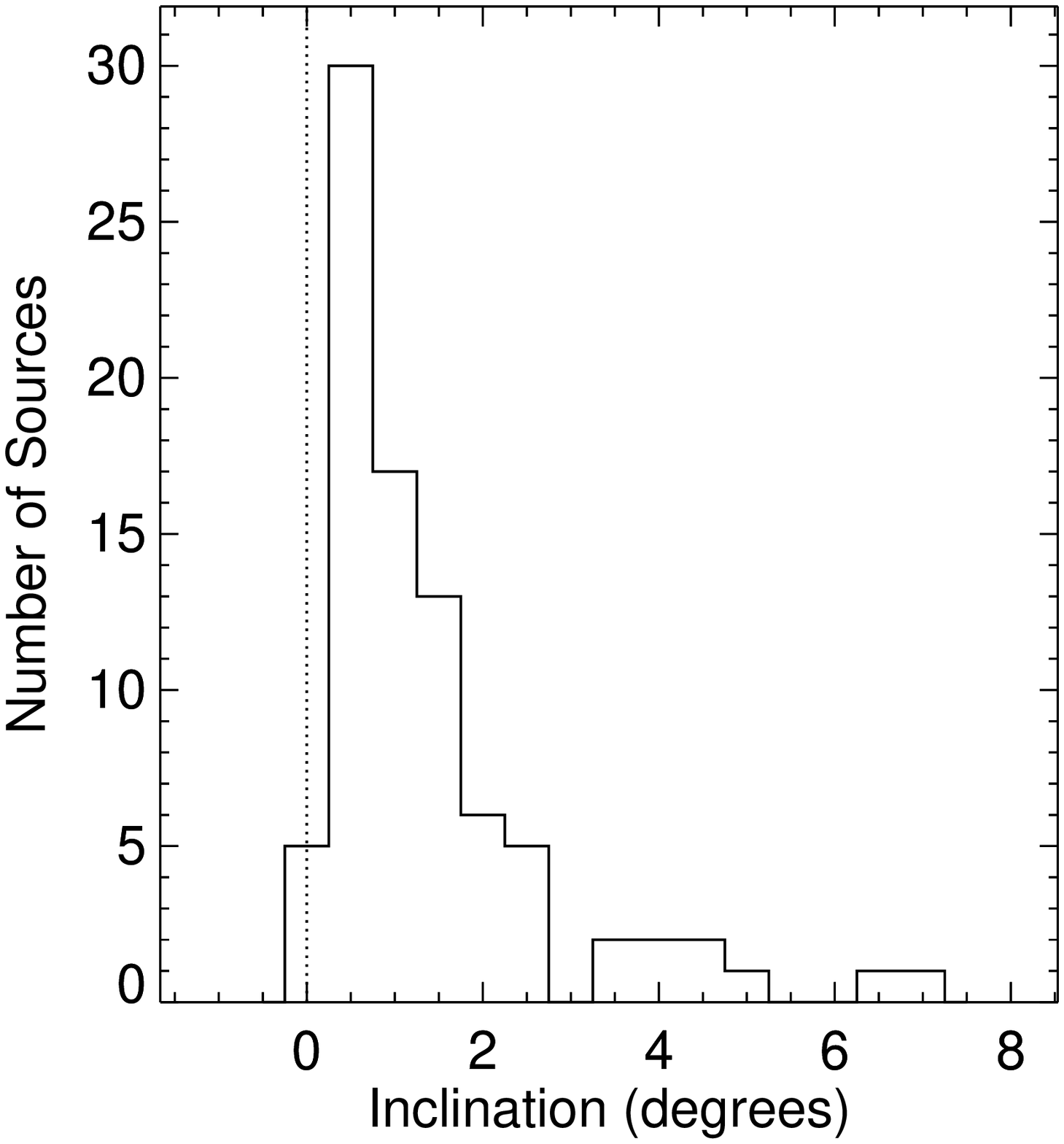}
\caption{Distribution of Galactic orbital parameters for sources in our sample: 
(upper left) minimum (black) and maximum (blue) Galactic radius, 
(upper right) maximum absolute vertical displacement, 
(lower left) orbital eccentricity, and
(lower right) orbital inclination.
%, and (bottom left) specific angular momentum ($\vec{v} \times \vec{r}$).
Solar values based on the same orbit calculations and assuming the current position 
($R_{\odot}$ = 8.43~kpc, $|Z_{\odot}|$ = 0.027~kpc) and
motion of the Sun relative to the LSR are indicated by the dashed lines.
\label{figure:orbits}}
\end{figure}

\begin{figure}
\centering
\plotone{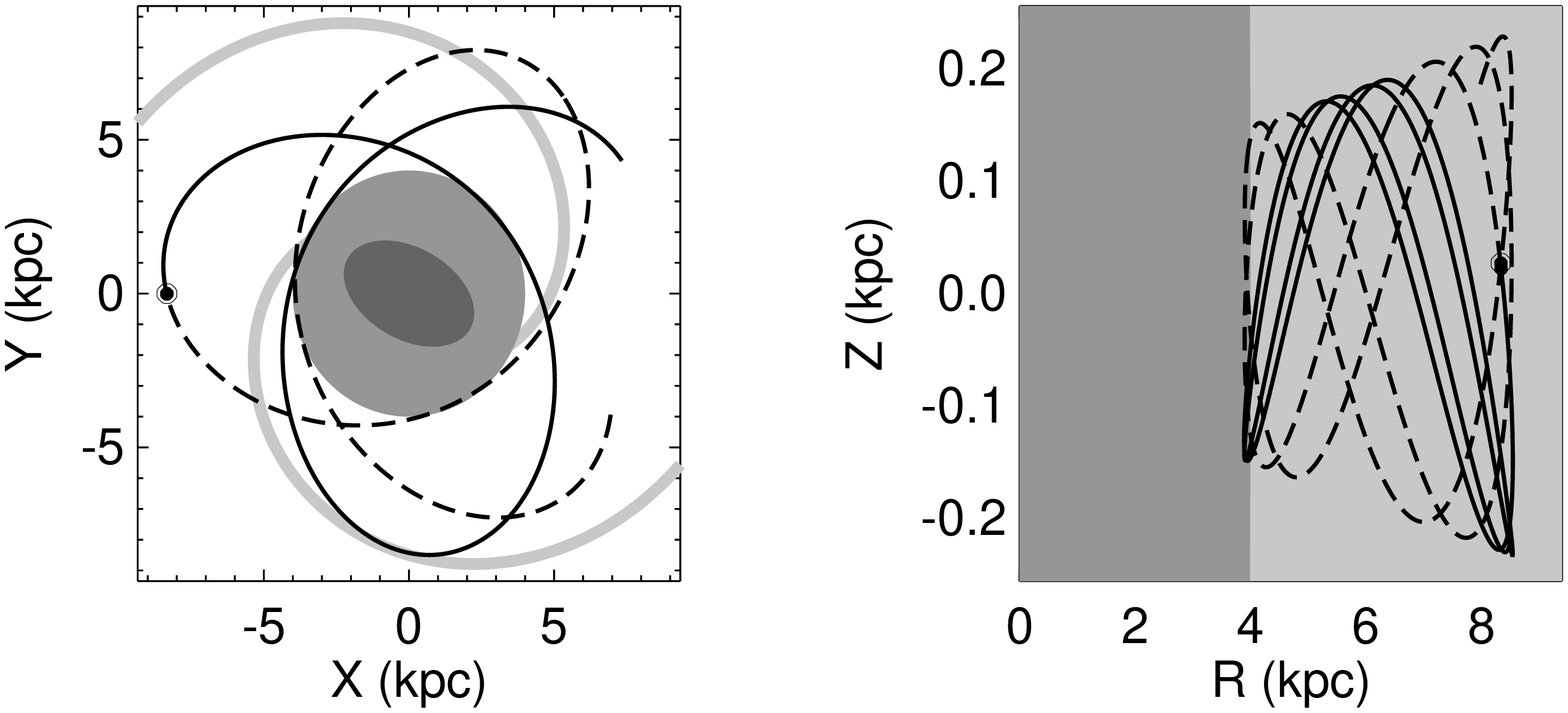}
\plotone{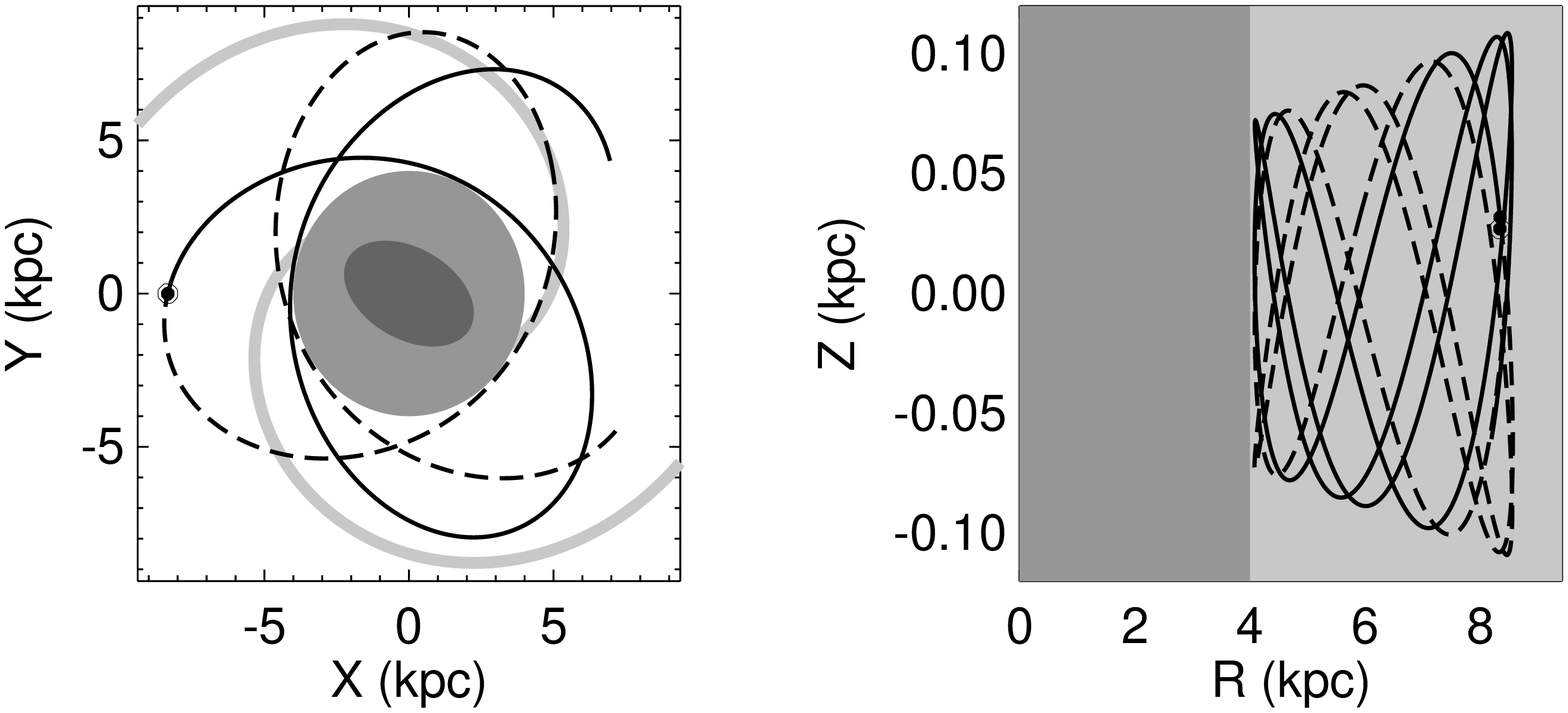}
\caption{Galactic orbits of the thick disk M8.5 J0707$-$4900 (top) and the intermediate thin/thick disk L1 J0921$-$2104 (bottom). Panels show the orbit over the past (dashed line) and future (solid line) 250~Myr about the current epoch (solid point, near the current position of the Sun), projected onto the Galactic plane (left) and in cylindrical coordinates (right).  %Also shown are the cylindrical radius (lower left) and vertical displacement from the Galactic plane (lower right) over time.
Representations of the Galactic bar (darkest gray), bulge (gray), and thin disk/major spiral arms (lightest gray) based on \citet{2008ASPC..387..375B,2008gady.book.....B} and \citet{2013MNRAS.435.1874W} are also shown.
\label{figure:orb0707}}
\end{figure}

\clearpage
%UVW distribution for 20 pc sample

\begin{figure}
\centering
\includegraphics[scale=0.4]{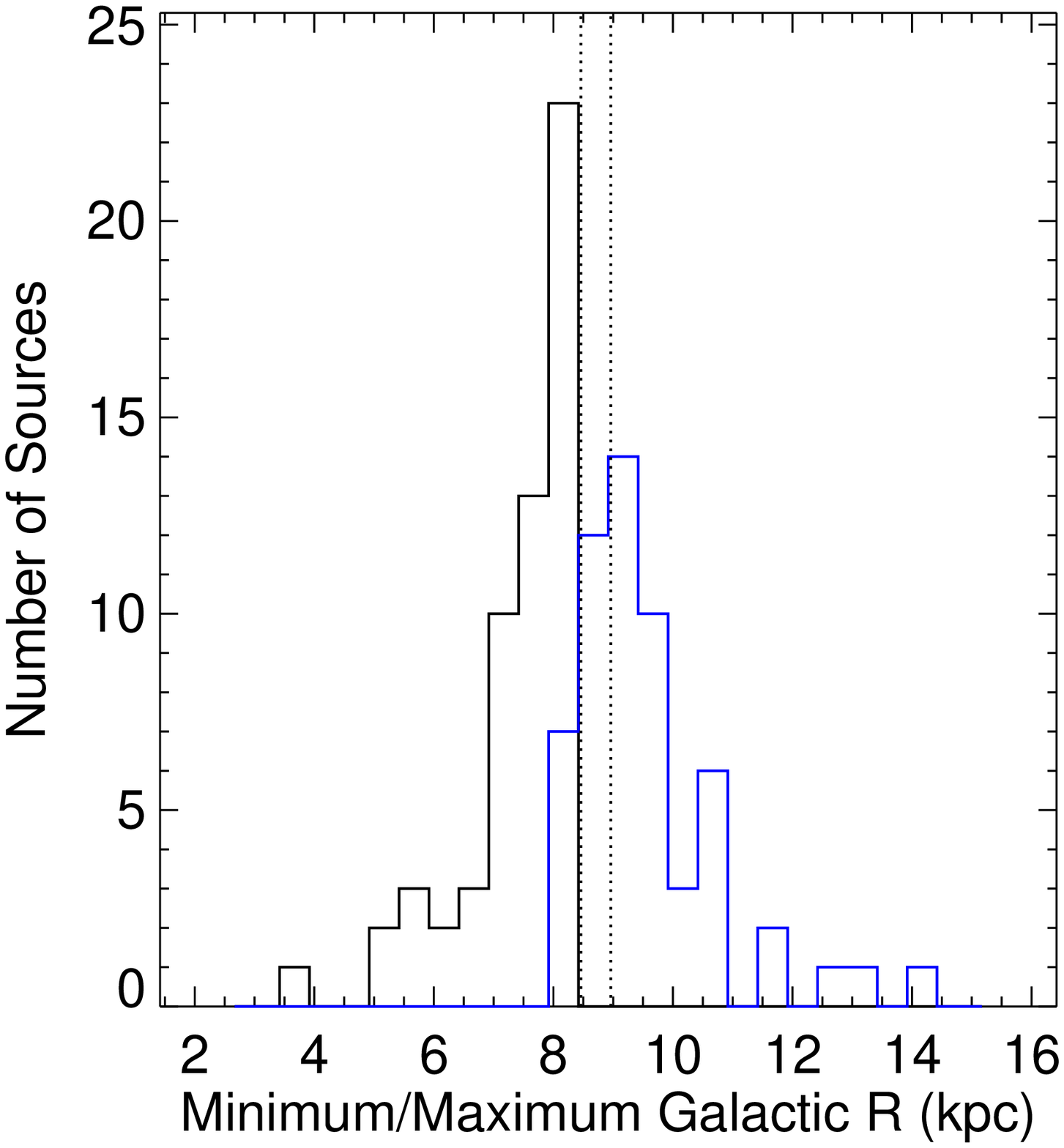}
\includegraphics[scale=0.4]{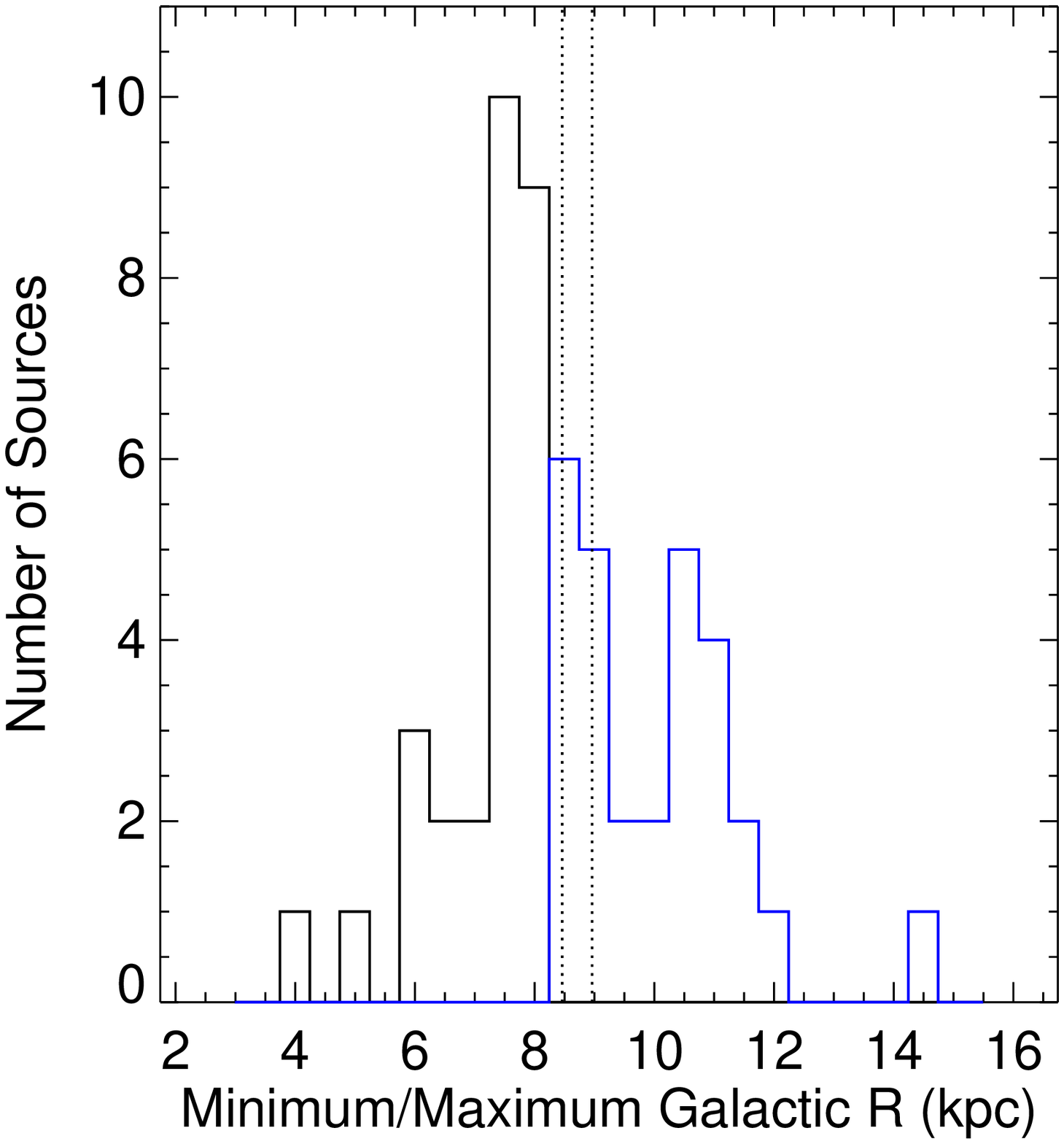}
\caption{Comparison of $R_{min}$ (black) and $R_{max}$ (blue) distributions for the late-M dwarfs (left)
and L dwarfs (right) in our sample.  Solar values are indicated by vertical dashed lines.
\label{fig:orbml}}
\end{figure}

\begin{figure}
\centering
\includegraphics[scale=0.35]{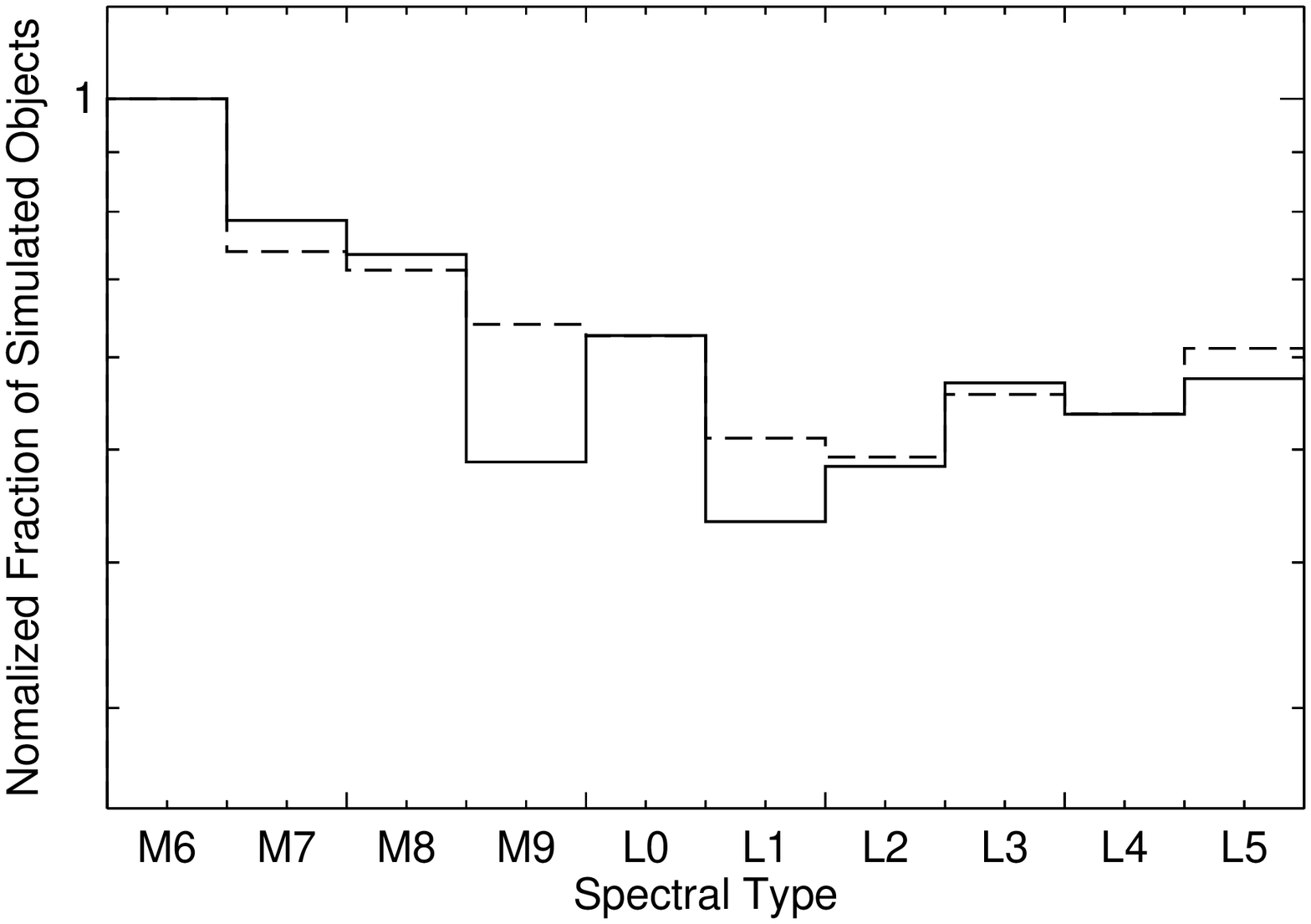}
\includegraphics[scale=0.35]{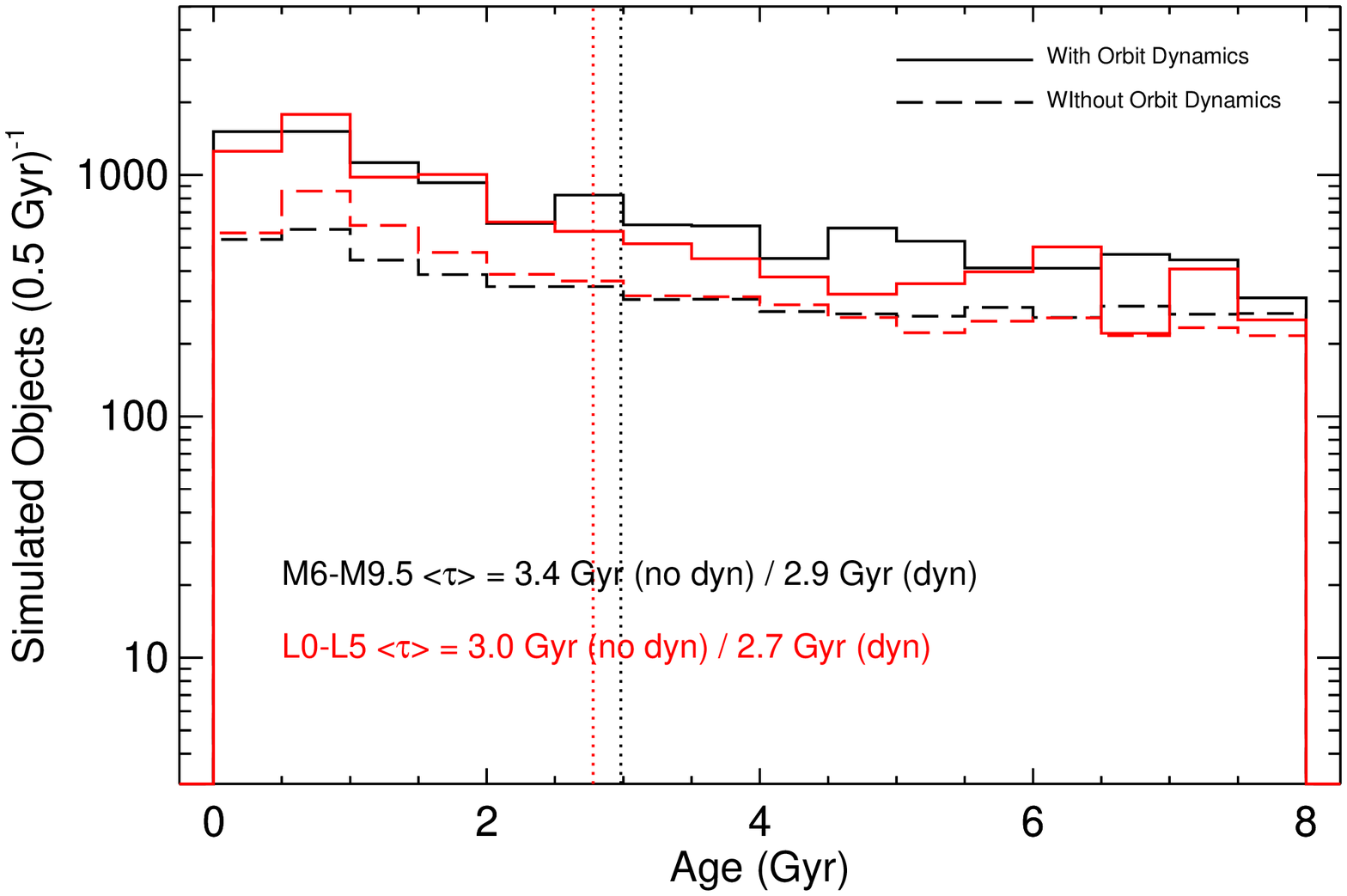} \\
\includegraphics[scale=0.35]{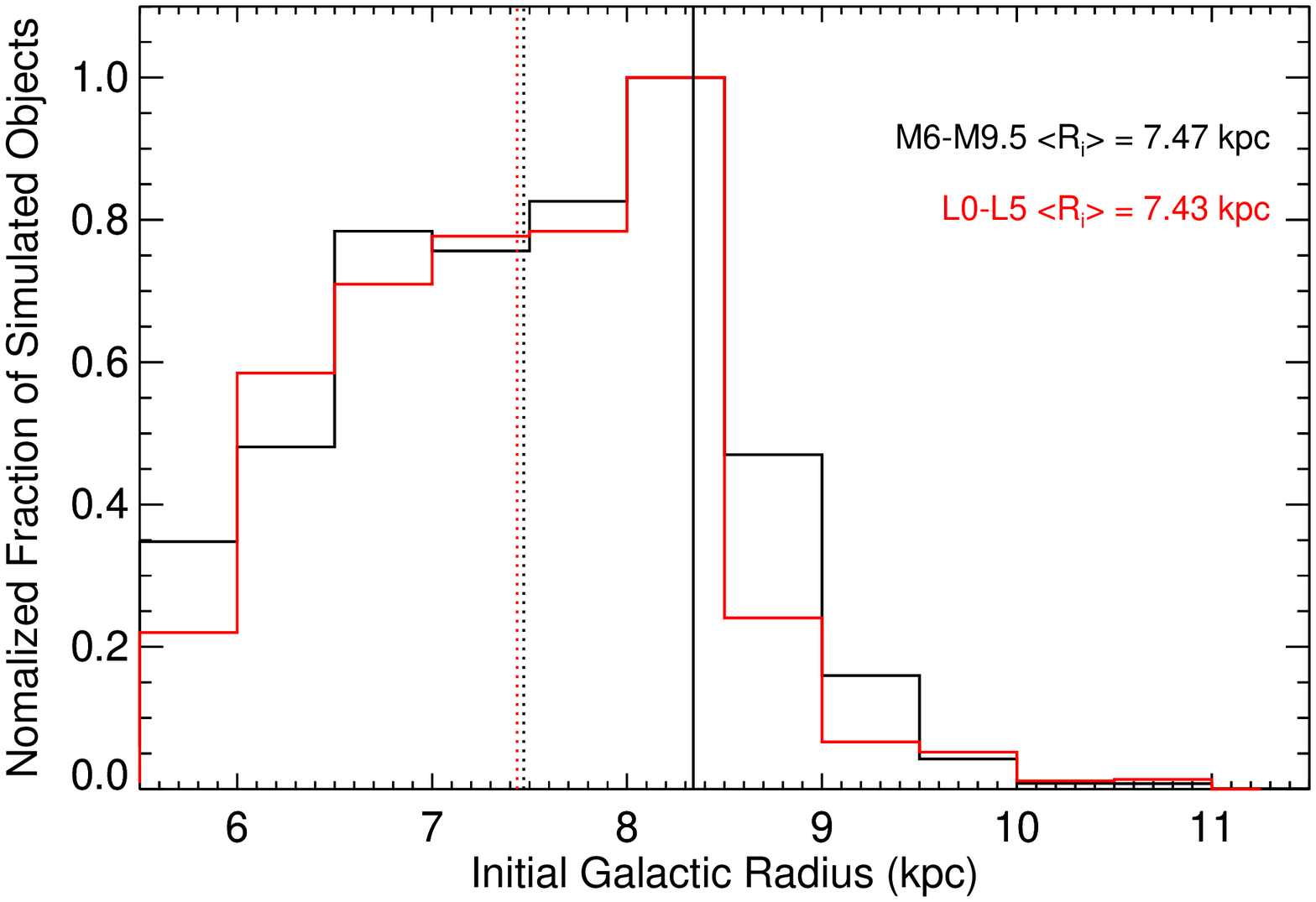}
\includegraphics[scale=0.35]{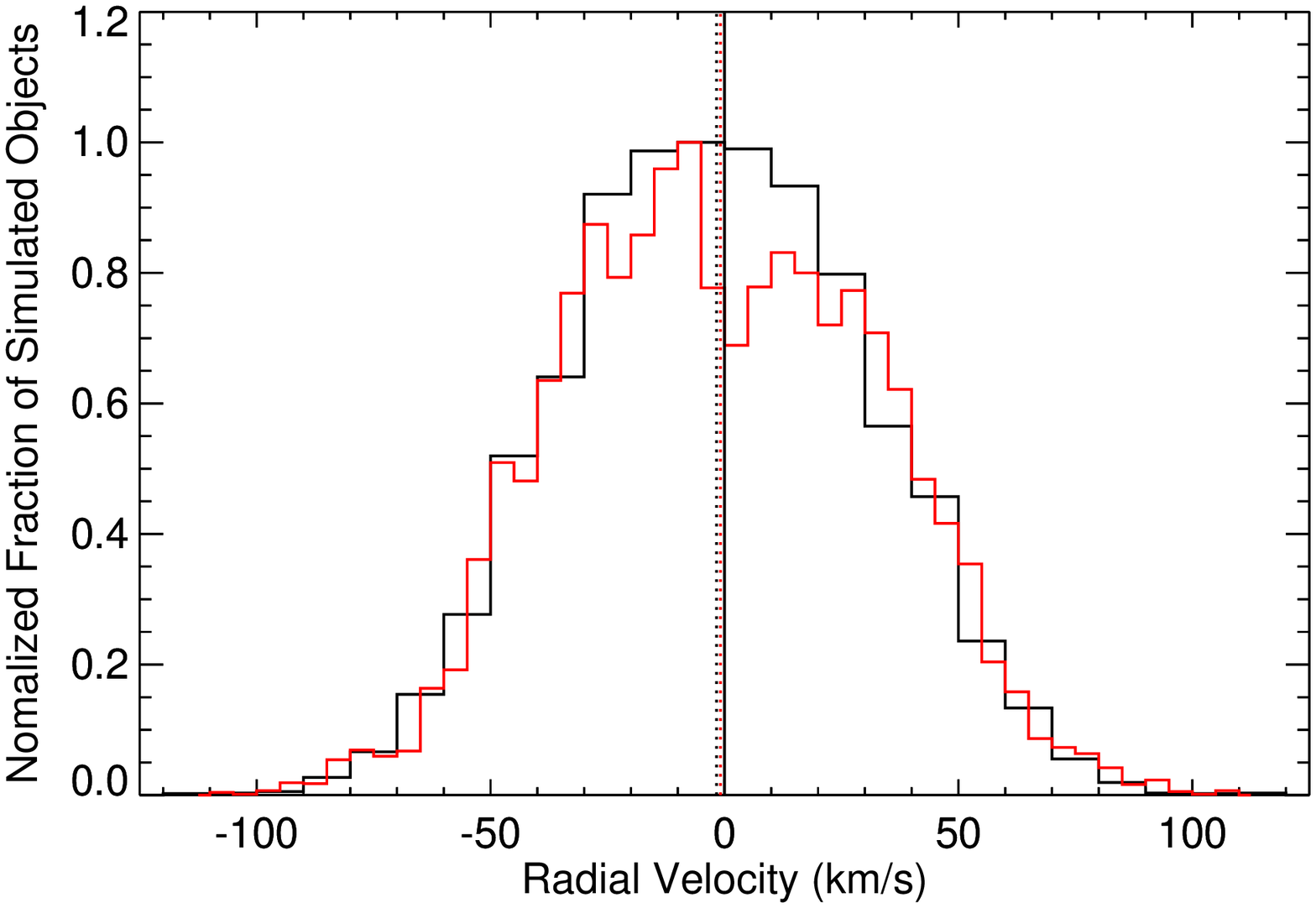} \\
\includegraphics[scale=0.35]{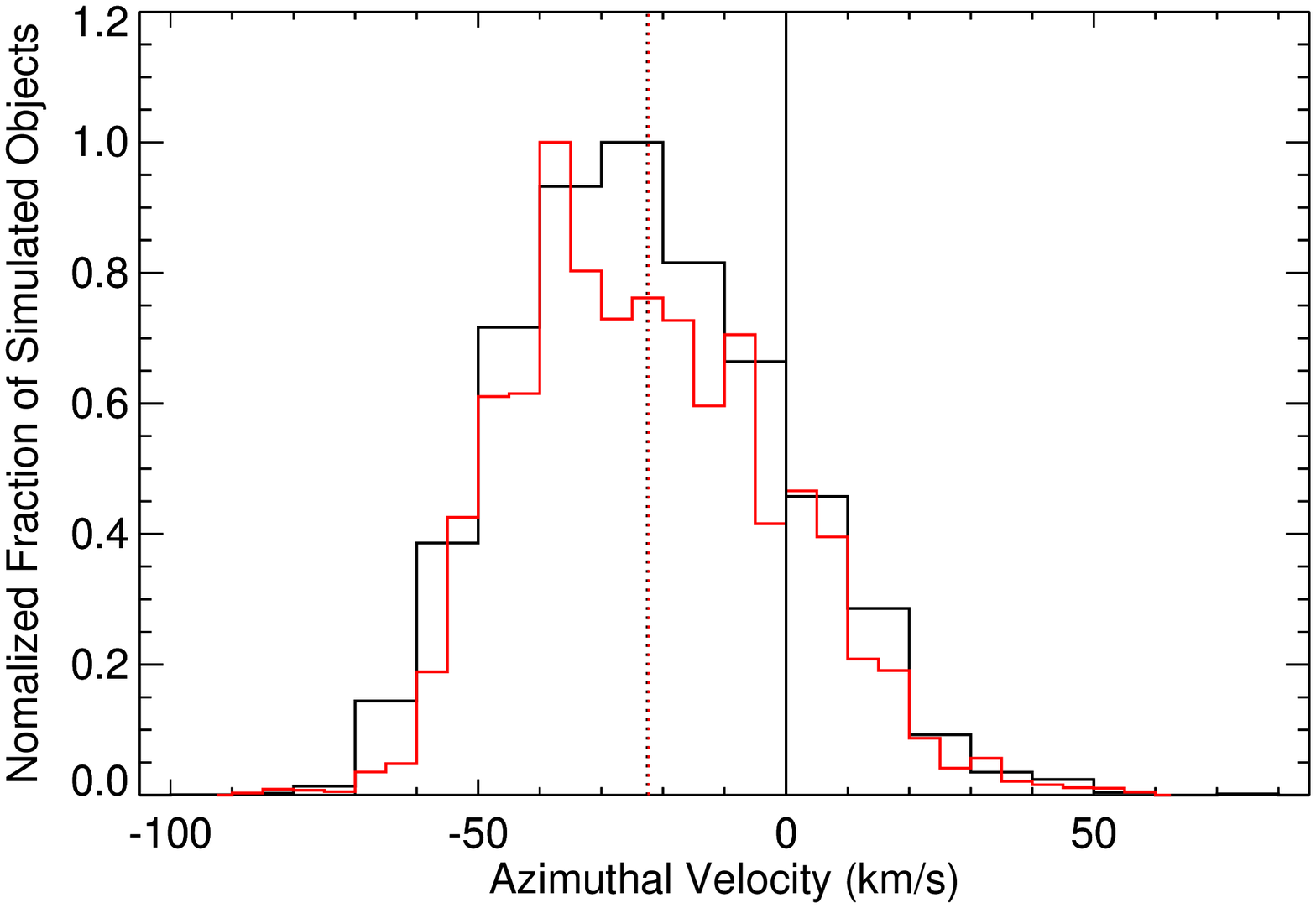}
\includegraphics[scale=0.35]{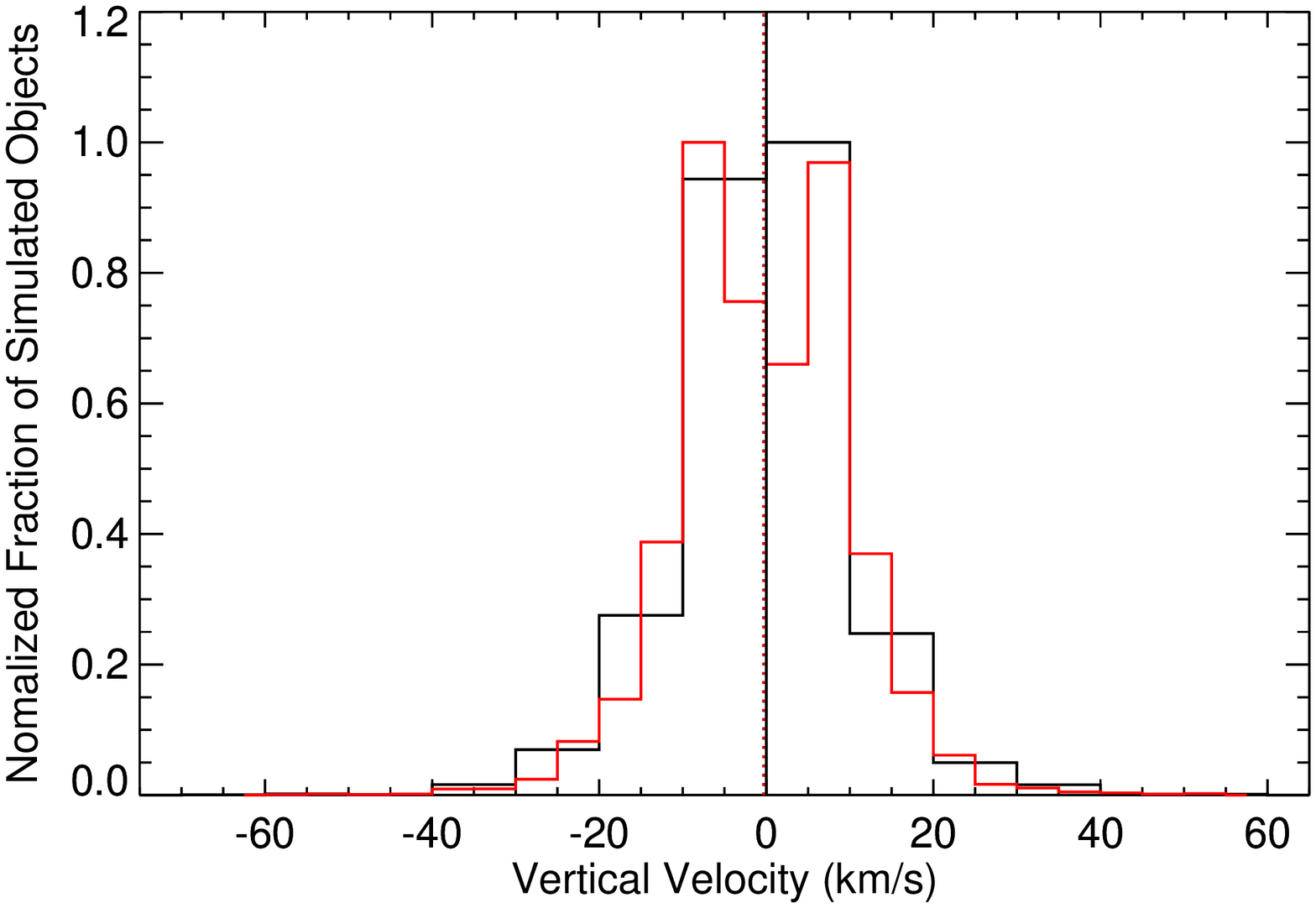}
\caption{Summary of population synthesis, dynamical evolution and local selection simulation for $\beta$ = 0.0 and $\alpha$ = 0.5. Distributions for
M6--M9.5 dwarfs are indicated in black,  L0--L5 dwarfs in red.
(Upper left): Spectral type distribution before (dashed) and after (solid) dynamical selection.
(Upper right): Age distribution before (dashed) and after (solid) dynamical selection; vertical dotted lines indicate mean ages.
(Middle left): Initial Galactic radii after dynamical selection. Dotted vertical lines indicate mean radii, solid vertical line indicates the Solar radius (8.43~kpc).
(Middle right, lower left, lower right): Galactic azimuthal, radial and vertical velocities in the Local Standard of Rest after dynamical evolution and local selection. Vertical solid lines indicate zero mean velocity, vertical dotted lines indicate mean velocities of the spectral class subsamples.
\label{fig:mfsimsummary}}
\end{figure}

\begin{figure}
\centering
\includegraphics[scale=0.35]{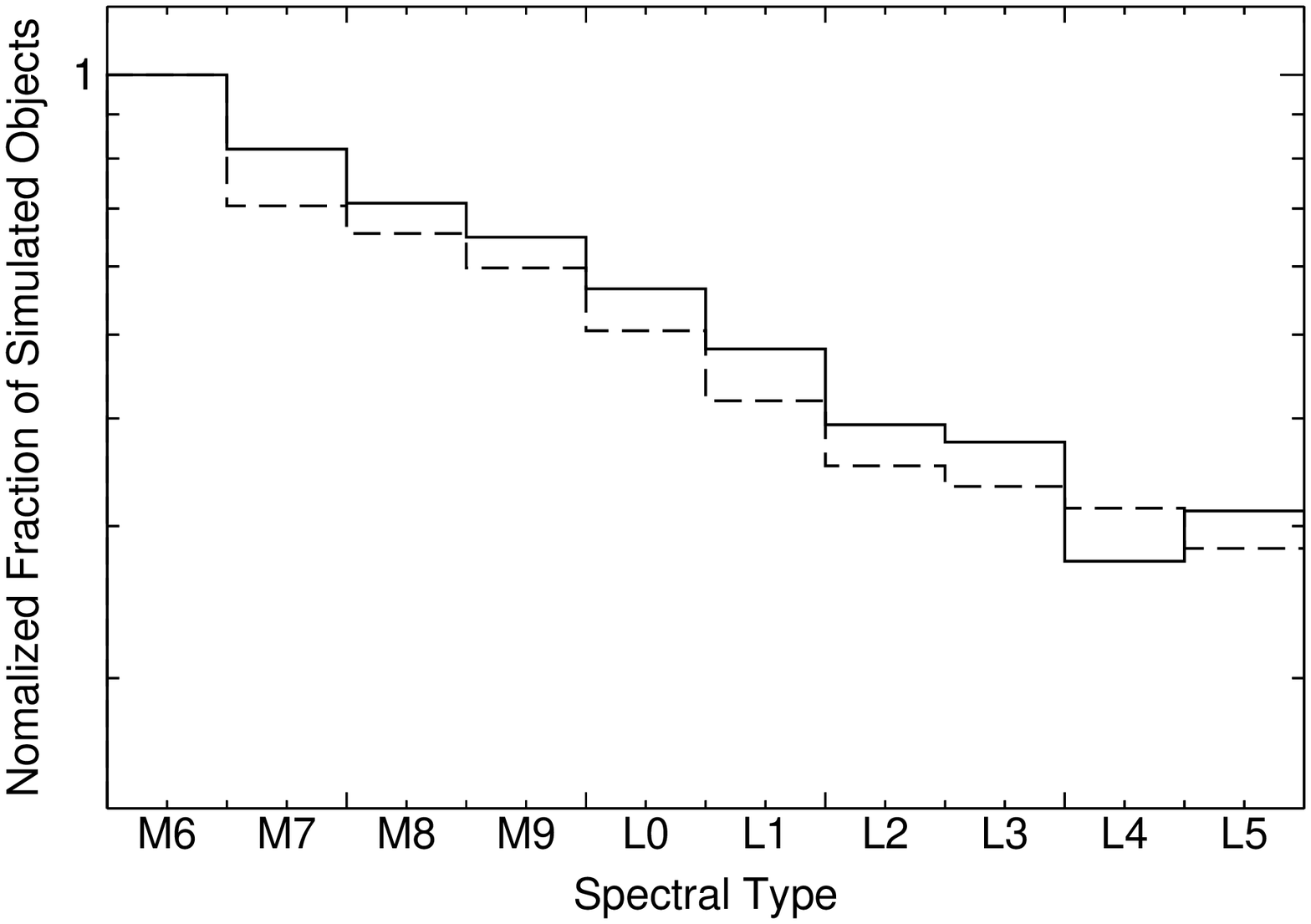}
\includegraphics[scale=0.35]{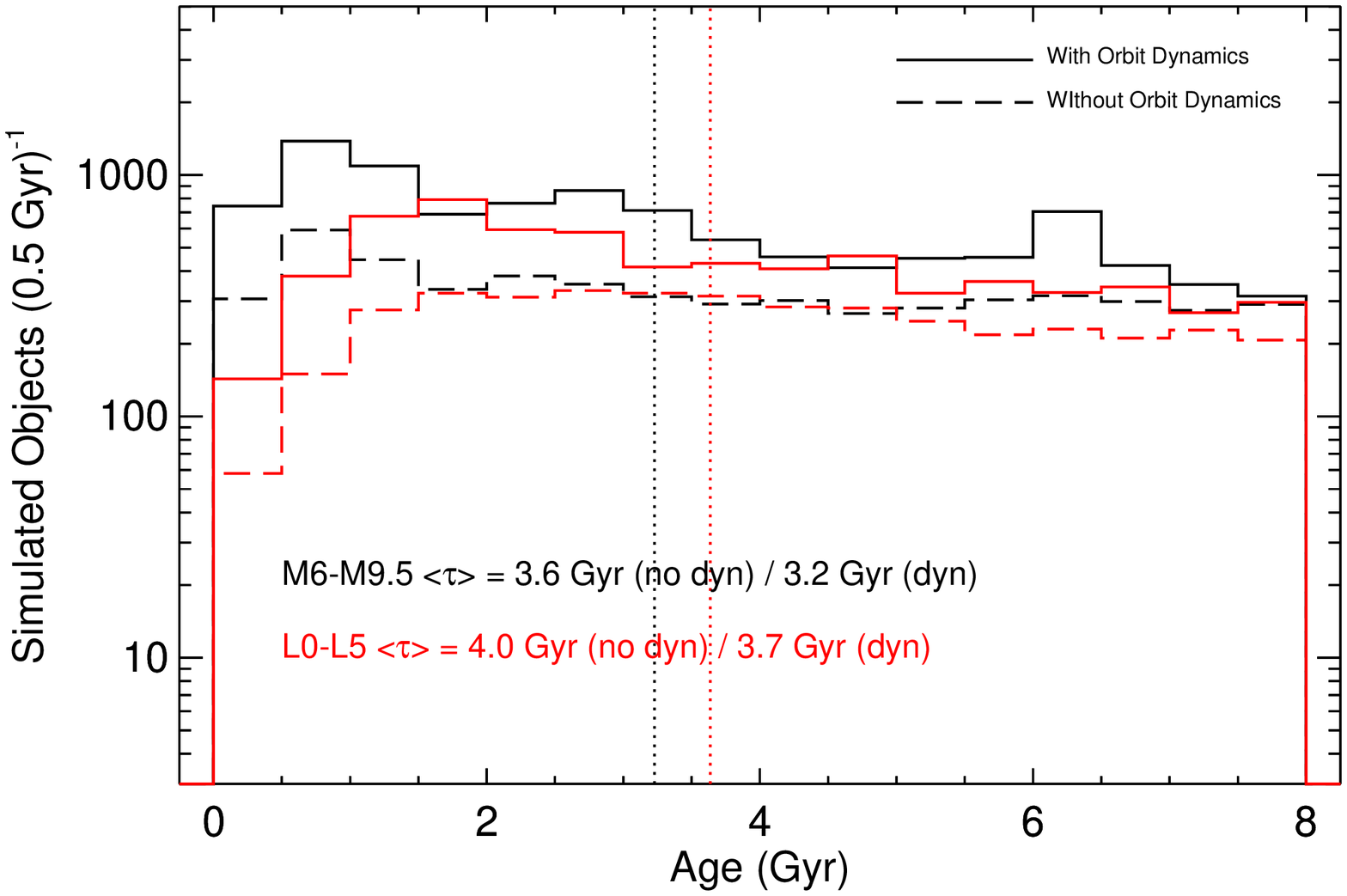} \\
\includegraphics[scale=0.35]{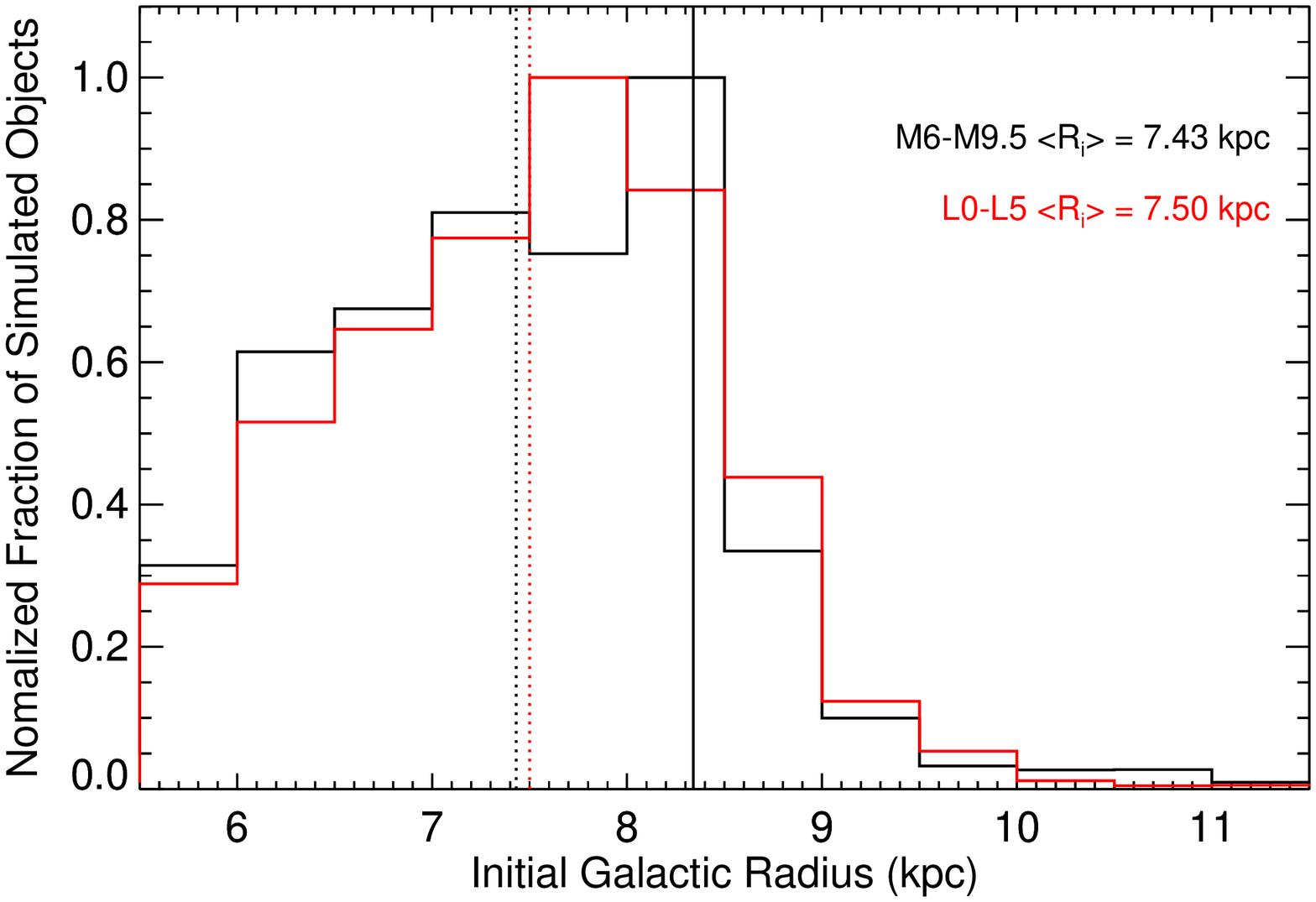}
\includegraphics[scale=0.35]{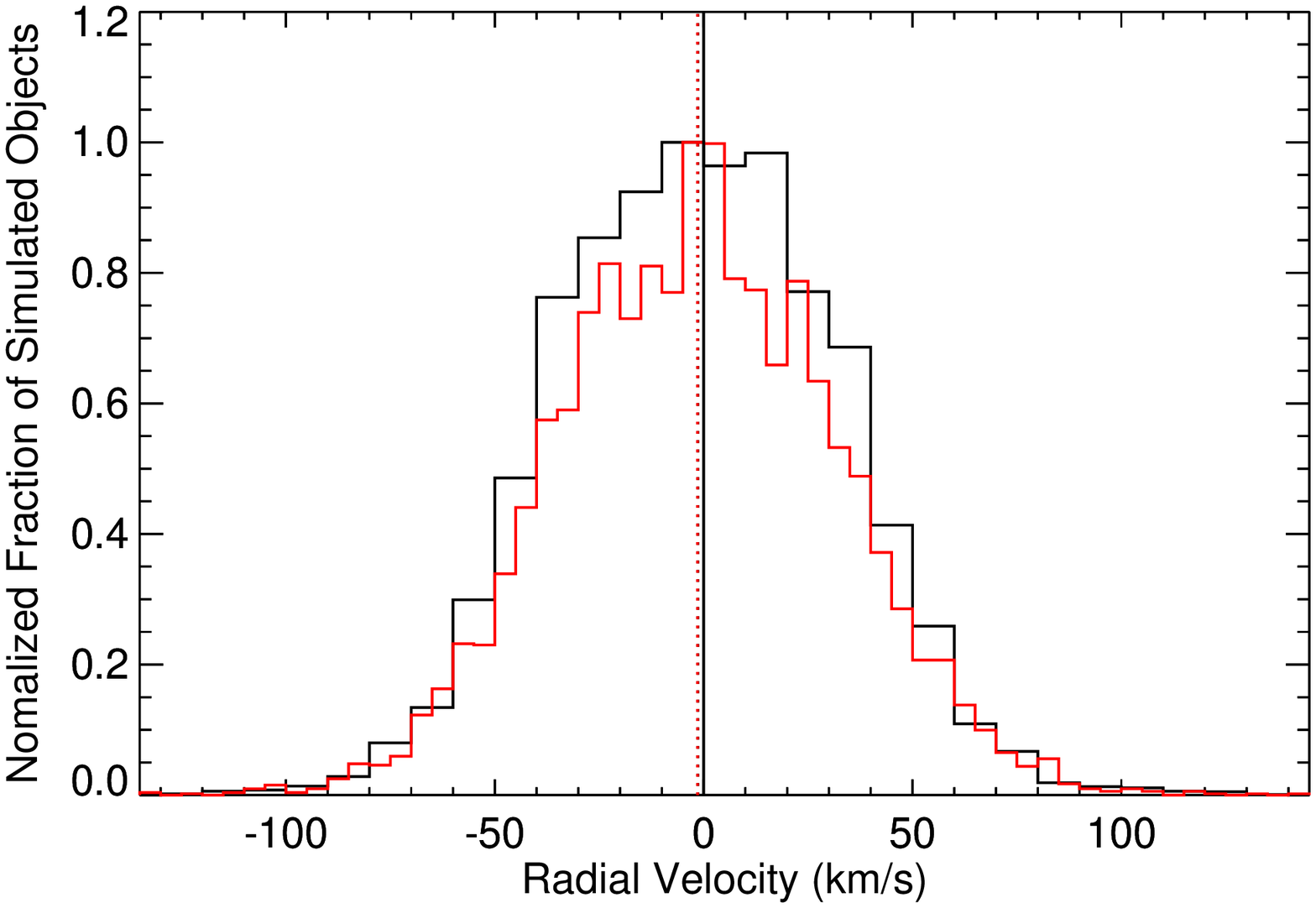} \\
\includegraphics[scale=0.35]{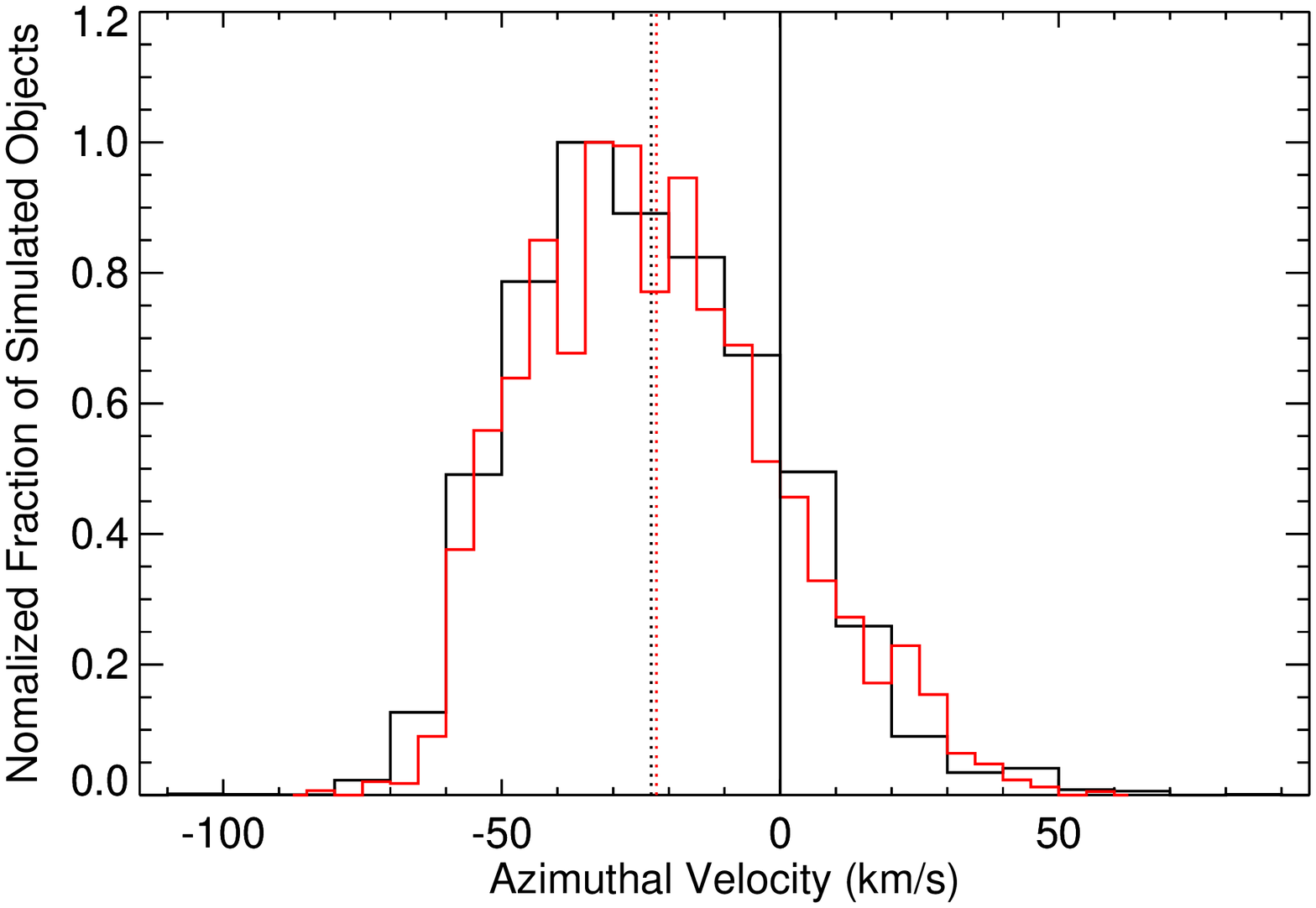}
\includegraphics[scale=0.35]{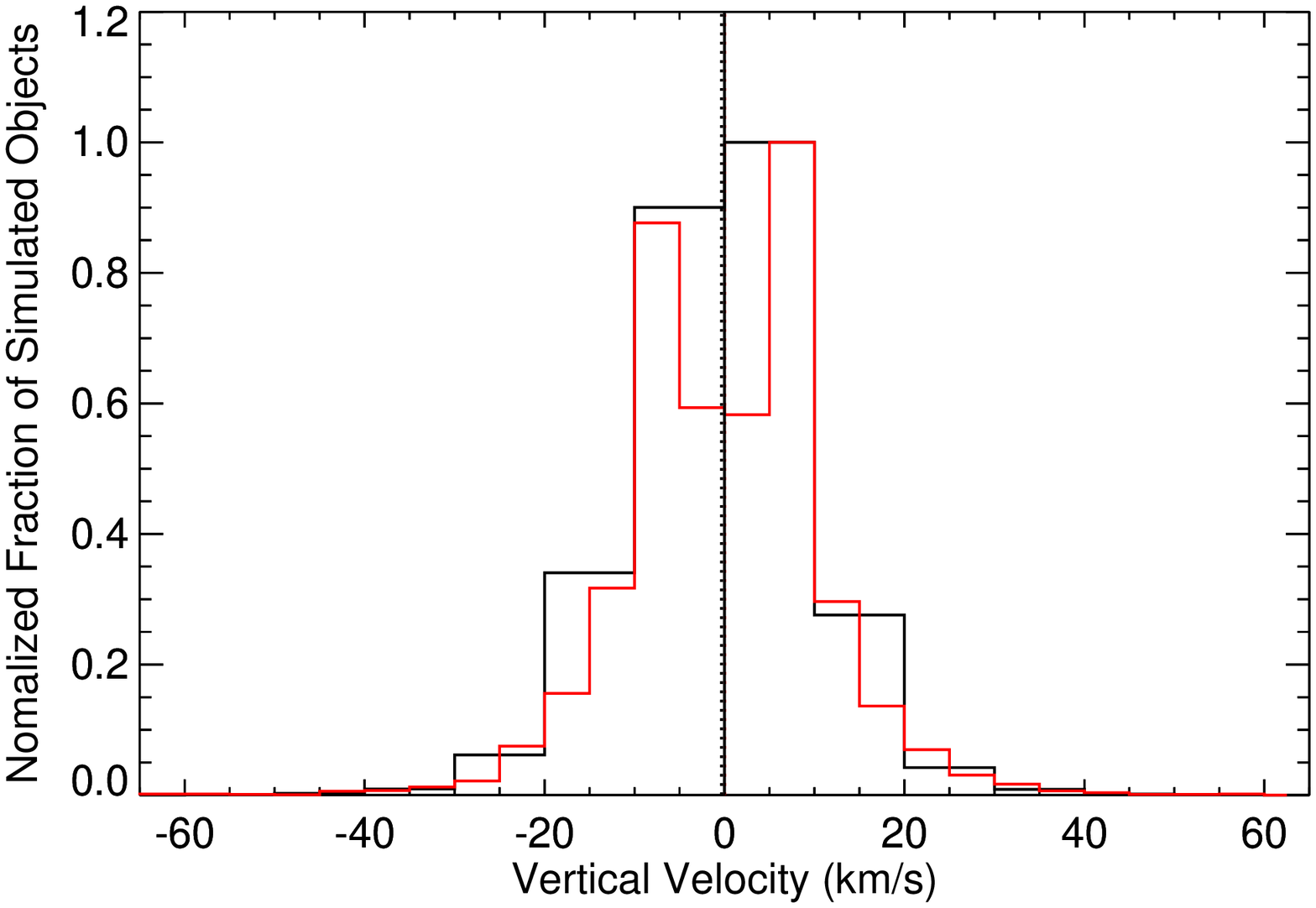}
\caption{Same as Figure~\ref{fig:mfsimsummary} but for $\beta$ = 0.5 for M $<$ 0.07~M$_{\odot}$ and 
$\beta$ = 0.0 for M $>$ 0.07~M$_{\odot}$. Note the loss of young L (brown) dwarfs in this sample, resulting in an older L dwarf
population on average.
\label{fig:mfsimsummary2}}
\end{figure}

\begin{figure}
\centering
\includegraphics[scale=0.37]{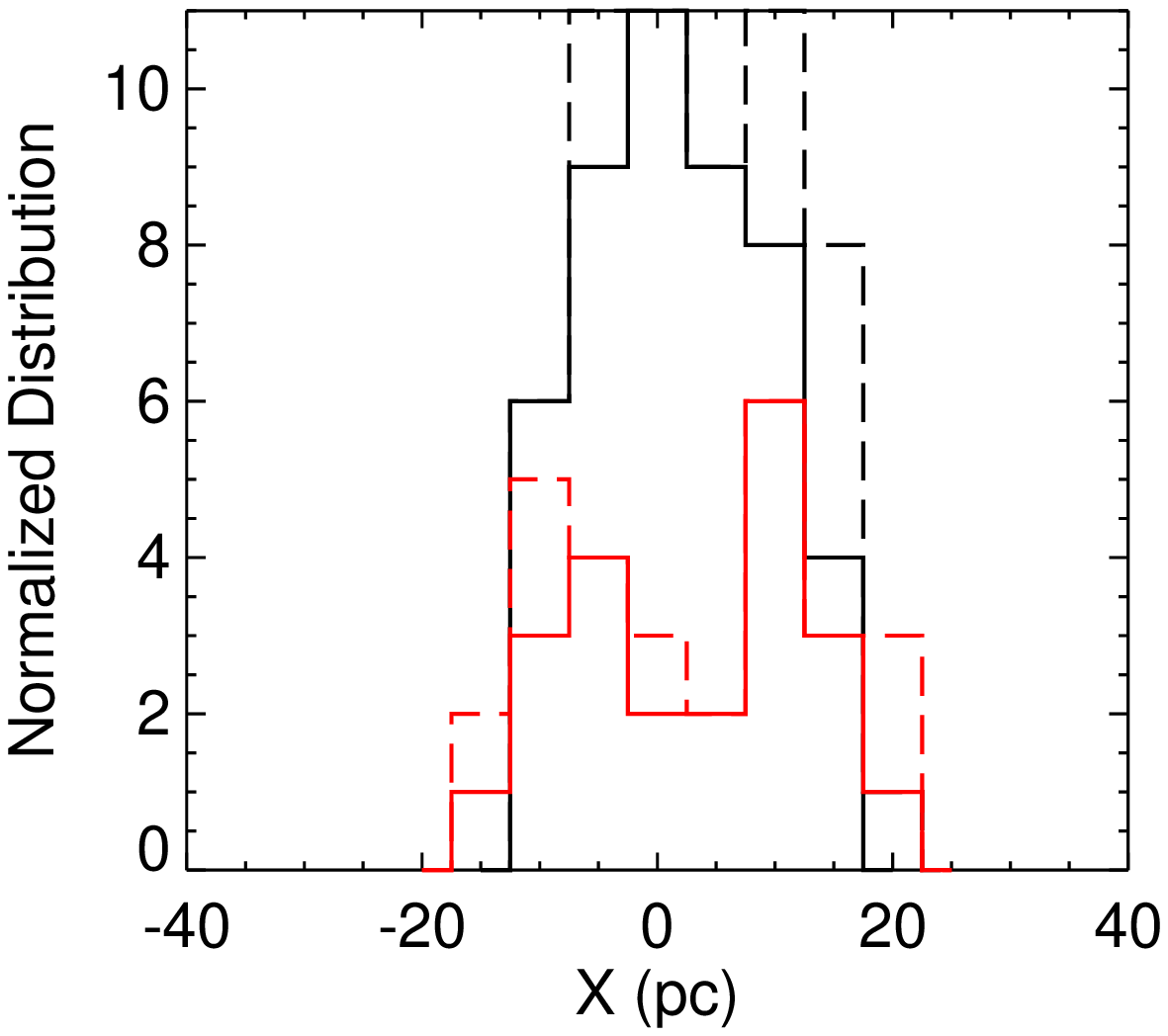} 
\includegraphics[scale=0.37]{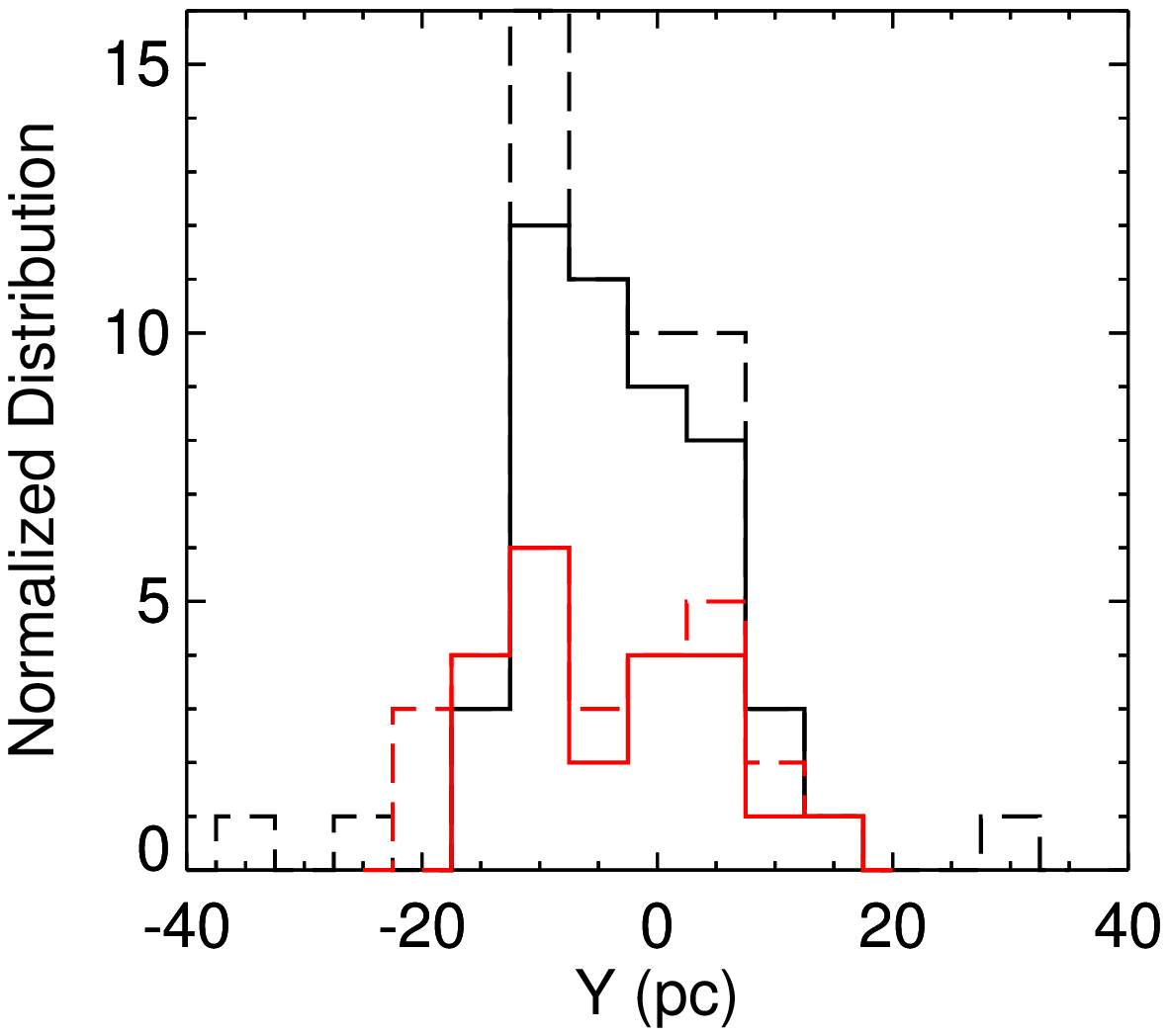} 
\includegraphics[scale=0.37]{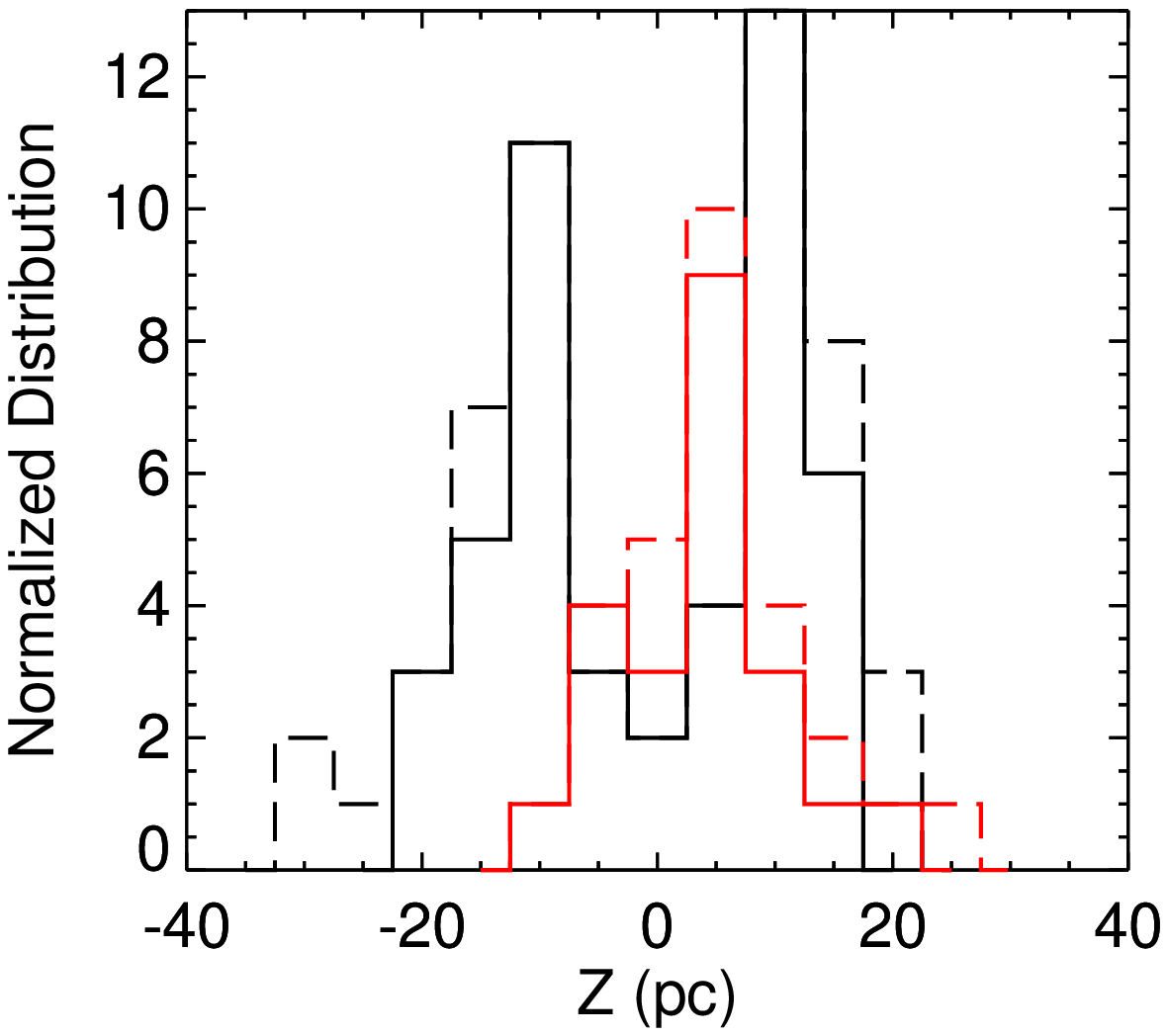} 
\caption{Distribution of { M dwarfs (black) and L dwarfs (red)} in $XYZ$ spatial coordinates. Dashed-line histograms show the distributions for all sources, solid-line histograms show the distributions for those sources within 20~pc of the Sun.
\label{fig:srcskydist}}
\end{figure}

\begin{figure}
\centering
\includegraphics[scale=0.37]{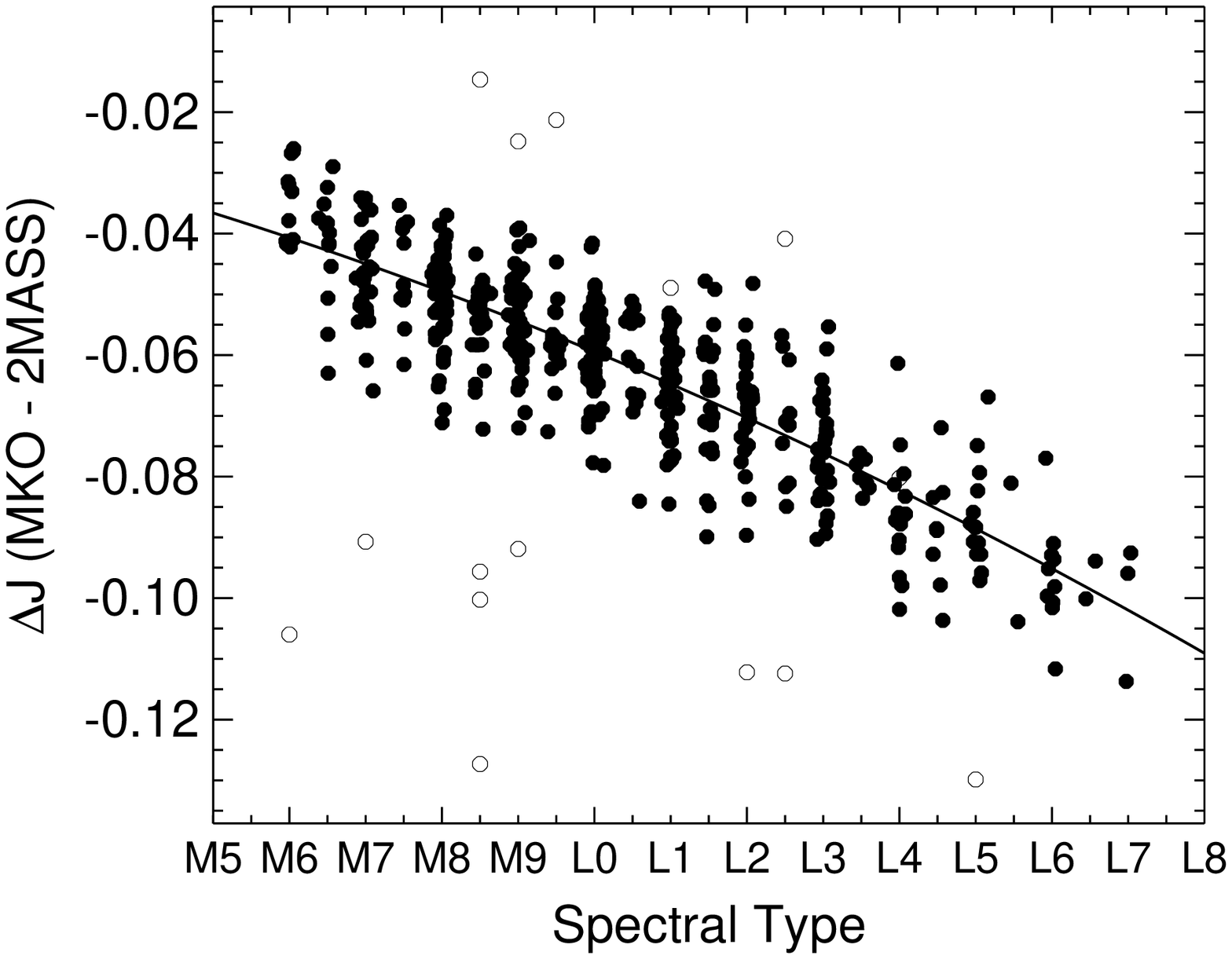} 
\includegraphics[scale=0.37]{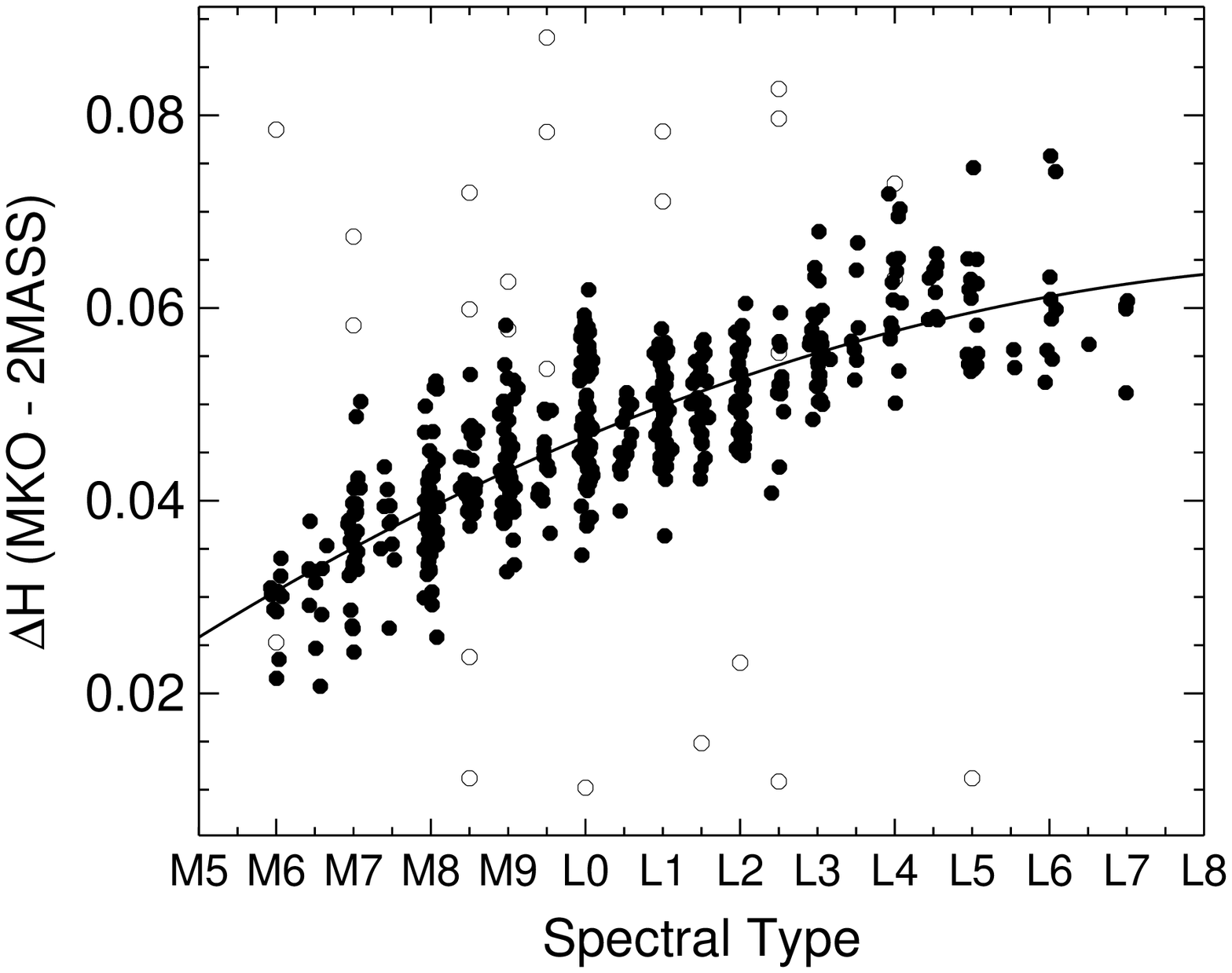} 
\includegraphics[scale=0.37]{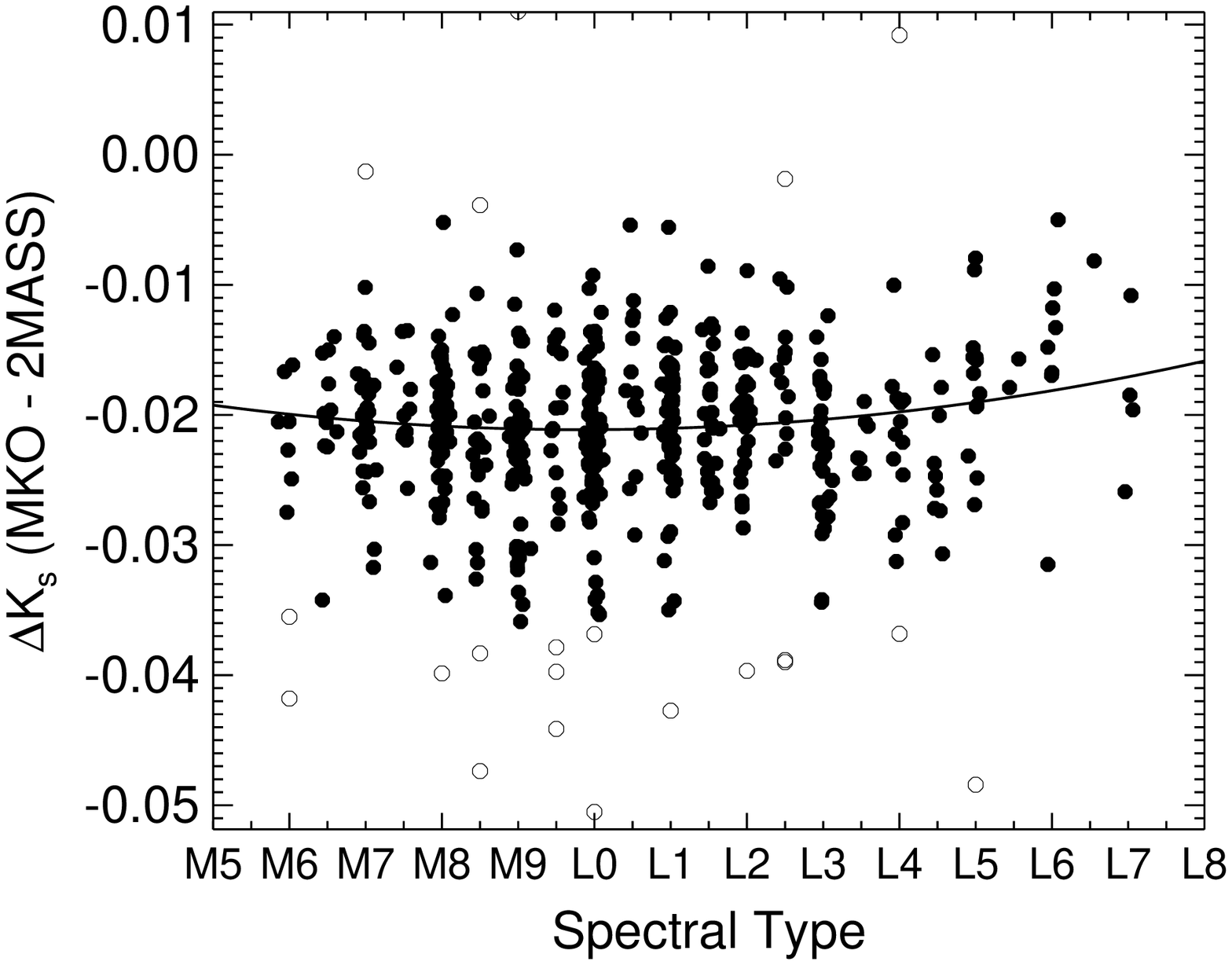} 
\caption{ Magnitude differences between 2MASS and MKO $JHK_s$ filter systems for M6-L7 dwarfs, based on spectrophotometric measurements from 533 optically-classified, high S/N near-infrared spectra in the SpeX Prism Library. Each point has a random offset in spectral type added to distinguish sources. Open circles are outliers rejected from the second-order polynomial fits.
\label{fig:mko2m}}
\end{figure}

\begin{deluxetable}{lclccccccll}
\tablecaption{Observational Sample\label{table:sample}}
\tabletypesize{\tiny}
\tablewidth{0pt}
\rotate
\tablehead{
 & & & & & & & & \colhead{ Published} \\
\colhead{Source} &
%\colhead{Other} &
\colhead{Coordinates}&
\colhead{Optical}&
\colhead{2MASS $J$}&
\colhead{$J-K_s$}&
\colhead{$\mu_{\alpha}$}&
\colhead{$\mu_{\delta}$}&
\colhead{$d$\tablenotemark{a}}&
%\colhead{\vtan}&
\colhead{$RV$}&
\colhead{Flag\tablenotemark{b}}&
\colhead{References\tablenotemark{c}}\\
\colhead{Name} &
%\colhead{Name} &
\colhead{(J2000)} &
\colhead{SpT} &
\colhead{(mag)} &
\colhead{(mag)} & 
\colhead{(\masyr)} &
\colhead{(\masyr)} &
\colhead{(pc)} &
%\colhead{(\kms)} &
\colhead{(\kms)}
 \\
}

\startdata
J0004$-$2058	&	00 04 41.44	$-$20 58 29.8	&	M9	&	12.40$\pm$0.02	&	1.01$\pm$0.03	&	756$\pm$8	&	95$\pm$10	&		(15.5$\pm$1.0)		&		\nodata		&		&			28							\\
J0024$-$0158	&	00 24 24.63	$-$01 58 20.1	&	M9.5	&	11.99$\pm$0.04	&	1.45$\pm$0.04	&	$-$80$\pm$4	&	132$\pm$6	&		11.6$\pm$0.5		&	10$\pm$3	&	UR	&			48; 51			\\
J0041$-$5621AB	&	00 41 35.38	$-$56 21 12.7	&	M8	&	11.96$\pm$0.02	&	1.10$\pm$0.03	&	98$\pm$6	&	$-$50$\pm$8	&		(20.7$\pm$1.8)		&	2.4$\pm$1	&	Li, VB, THA	&			45; 55, 51	\\
J0102$-$3737	&	01 02 51.00	$-$37 37 43.8	&	M9	&	11.13$\pm$0.02	&	1.06$\pm$0.03	&	1456$\pm$3	&	262$\pm$4	&		12.2$\pm$0.4		&		\nodata		&		&			34							\\
J0109$-$0343	&	01 09 51.17	$-$03 43 26.4	&	M8	&	11.69$\pm$0.02	&	1.27$\pm$0.03	&	360$\pm$2	&	18$\pm$2	&		9.6$\pm$0.2		&	-10.4$\pm$0.6	&	UR	&			34; 55, 51	\\
J0123$-$6921	&	01 23 11.25	$-$69 21 37.9	&	M9	&	12.32$\pm$0.02	&	1.00$\pm$0.04	&	77$\pm$2	&	$-$25$\pm$9	&		42$\pm$5		&	7.5$\pm$0.3	&	Li, THA	&			39; 55, 51	\\
J0148$-$3024	&	01 48 38.64	$-$30 24 39.6	&	M7.5	&	12.30$\pm$0.02	&	1.08$\pm$0.03	&	$-$86$\pm$8	&	51$\pm$8	&		(18.8$\pm$2.0)		&	10$\pm$3	&		&			28; 51			\\
J0248$-$1651	&	02 48 41.00	$-$16 51 21.6	&	M8	&	12.55$\pm$0.02	&	1.13$\pm$0.03	&	20$\pm$9	&	$-$273$\pm$12	&		16.1$\pm$1.3		&	4$\pm$3	&		&			31; 51			\\
\enddata
\end{deluxetable}

\begin{deluxetable}{llccll}
\tablecaption{Observations\label{table:log}}
\tabletypesize{\scriptsize}
\tablewidth{0pt} 
\tablehead{
\colhead{} &
\colhead{Observation} &
\colhead{Integration} & 
\colhead{} &
\colhead{Flux} &
\colhead{}  \\
\colhead{Source} &
\colhead{Date (UT)} &
\colhead{Time (s)} & 
\colhead{Airmass} &
\colhead{Calibrator} &
\colhead{Conditions/Seeing}
}
\startdata
%Name	&	Date	&	Integration	&	Airmass	&	Flux Callibrator	&	Conditions	\\
J0004-2058	&	2009 Jan 11	&	800	&	1.54	&	Hiltner 600	&	seeing $\sim$0$\farcs$6	\\
J0024-0158	&	2009 Aug 27	&	450	&	1.179	&	Feige 110	&	dry, clear, windy, seeing $\sim$1	\\
J0041-5621AB	&	2009 Aug 28	&	600	&	1.329	&	Feige 110	&	seeing $\sim$0$\farcs$7-0$\farcs$8	\\
J0102-3737	&	2008 Nov 26	&	600	&	1.18	&	HR 3454	&	seeing$\sim$ 0$\farcs$8-0$\farcs$9	\\
J0109-0343	&	2009 Aug 27	&	450	&	1.12	&	Feige 110	&	dry, clear, windy, seeing $\sim$1	\\
J0123-6921	&	2009 Aug 28	&	600	&	1.427	&	Feige 110	&	seeing $\sim$0$\farcs$7	\\
J0148-3024	&	2009 Aug 29	&	600	&	1.005	&	Feige 110	&	clear; seeing ² 0$\farcs$5	\\
J0248-1651	&	2009 Aug 27	&	600	&	1.055	&	Feige 110	&	dry, clear, windy, seeing $\sim$0$\farcs$7	\\
\enddata
\end{deluxetable}

\begin{deluxetable}{llcccccccccc}
\tablecaption{Equivalent Widths and Line Strengths of Absorption and Emission Lines\label{table:lines}}
\tabletypesize{\scriptsize}
\tablewidth{0pt}
\rotate
\tablehead{
\colhead{Source} & \colhead{SpT} & \colhead{Li~I} & \colhead{Rb~I} & \colhead{Rb~I} & \colhead{Na~I} & \colhead{Na~I} & \colhead{Cs~I} & \colhead{Cs~I} & \colhead{H$\alpha$} &  \colhead{{\lha}} & \colhead{$\zeta$} \\
 & & \colhead{6710\,{\AA}} & \colhead{7802\,{\AA}} & \colhead{7950\,{\AA}} & \colhead{8186\,{\AA}} & \colhead{8197\,{\AA}} & \colhead{8523\,{\AA}} & \colhead{8946\,{\AA}} & \colhead{6583\,{\AA}}  }
\startdata
J0004-2058 & M9 & $<$1.3 & 1.54$\pm$0.17 & 1.24$\pm$0.15 & 1.7$\pm$0.2 & 2.1$\pm$0.3 & 0.8$\pm$0.2 & $<$0.7 & -4.2$\pm$0.4 & -5.04$\pm$0.06 & 0.973$\pm$0.018 \\
J0024-0158 & M9.5 & $<$0.4 & 1.62$\pm$0.16 & 1.39$\pm$0.18 & 1.49$\pm$0.18 & 1.59$\pm$0.18 & 1.01$\pm$0.14 & 0.8$\pm$0.2 & -0.13$\pm$0.12 & -6.57$\pm$0.13 & 0.56$\pm$0.03 \\
J0041-5621AB & M8 & 0.7$\pm$0.2 & 0.94$\pm$0.09 & 0.70$\pm$0.18 & 1.59$\pm$0.12 & 1.8$\pm$0.2 & 0.55$\pm$0.08 & $<$0.7 & -22$\pm$3 & -4.01$\pm$0.06 & 1.208$\pm$0.016 \\
J0102-3737 & M9 & $<$0.2 & 1.33$\pm$0.11 & 0.98$\pm$0.16 & 1.83$\pm$0.14 & 2.2$\pm$0.2 & 0.58$\pm$0.13 & $<$1.1 & -7.24$\pm$0.16 & -4.76$\pm$0.06 & 1.030$\pm$0.004 \\
J0109-0343 & M8 & $<$0.4 & 1.03$\pm$0.12 & 0.59$\pm$0.16 & 1.47$\pm$0.17 & 1.7$\pm$0.2 & 0.56$\pm$0.11 & $<$0.9 & -3.58$\pm$0.12 & -4.82$\pm$0.06 & 1.20$\pm$0.04 \\
J0123-6921 & M9 & 0.56$\pm$0.17 & 0.93$\pm$0.09 & 0.62$\pm$0.13 & 1.40$\pm$0.10 & 1.66$\pm$0.18 & $<$0.5 & $<$0.8 & -11.9$\pm$1.4 & -4.51$\pm$0.06 & 1.28$\pm$0.03 \\
J0148-3024 & M7.5 & $<$0.4 & 1.33$\pm$0.14 & 1.09$\pm$0.10 & 1.7$\pm$0.2 & 2.1$\pm$0.2 & 0.37$\pm$0.04 & $<$0.6 & -7.8$\pm$0.8 & -4.40$\pm$0.06 & 1.052$\pm$0.008 \\
J0248-1651 & M8 & $<$0.5 & 1.15$\pm$0.15 & 0.7$\pm$0.2 & 1.85$\pm$0.17 & 2.0$\pm$0.2 & 0.49$\pm$0.13 & $<$0.9 & -11.1$\pm$1.8 & -4.20$\pm$0.06 & 1.14$\pm$0.02 \\
\enddata
\end{deluxetable}

\begin{deluxetable}{lccccc}
\tablecaption{Mean Values of Gravity-Sensitive Indices\label{table:gravindices}}
\tabletypesize{\scriptsize}
\tablewidth{0pt} 
%\rotate
\tablehead{ \colhead{SpT} & \colhead{\#} & \colhead{Na-a + Na-b} & \colhead{VO7900} & \colhead{TiO-b} & \colhead{CrH-a + FeH-a} }
\startdata
M7	&	12	&	2.84$\pm$0.16	&	1.23$\pm$0.11	&	1.87$\pm$0.11	&	2.17$\pm$0.10	\\
M8	&	20	&	2.83$\pm$0.15	&	1.24$\pm$0.08	&	2.06$\pm$0.10	&	2.25$\pm$0.13	\\
M9	&	13	&	2.68$\pm$0.14	&	1.10$\pm$0.06	&	2.04$\pm$0.11	&	2.18$\pm$0.12	\\
L0	&	4	&	2.47$\pm$0.09	&	1.04$\pm$0.14	&	1.5$\pm$0.2	&	2.7$\pm$0.3	\\
L1	&	8	&	2.63$\pm$0.16	&	1.08$\pm$0.09	&	1.46$\pm$0.11	&	2.79$\pm$0.13	\\
L2	&	3	&	2.48$\pm$0.12	&	1.13$\pm$0.06	&	1.33$\pm$0.04	&	2.81$\pm$0.08	\\
L3	&	6	&	2.43$\pm$0.08	&	1.28$\pm$0.13	&	1.21$\pm$0.09	&	3.02$\pm$0.17	\\
L4	&	3	&	2.441$\pm$0.018	&	1.37$\pm$0.09	&	1.20$\pm$0.08	&	3.40$\pm$0.08	\\
L5	&	2	&	2.34$\pm$0.02	&	1.86$\pm$0.04	&	1.063$\pm$0.005	&	3.4$\pm$0.2	\\
\enddata
\end{deluxetable}

\begin{deluxetable}{llccccccccccl}
\tablecaption{Low Surface Gravity Dwarfs\label{table:lowg}}
\tabletypesize{\scriptsize}
\tablewidth{0pt} 
\rotate
\tablehead{ &  & \multicolumn{2}{c}{Na-a + Na-b} & \multicolumn{2}{c}{VO7900} & \multicolumn{2}{c}{TiO-b} & \multicolumn{2}{c}{CrH-a + FeH-a} & & \colhead{Prior} \\
\colhead{Source} & \colhead{SpT} & \colhead{Value} & \colhead{Low g?} & \colhead{Value} & \colhead{Low g?} & \colhead{Value} & \colhead{Low g?} & \colhead{Value} & \colhead{Low g?} & \colhead{Li~I?} & \colhead{YMG} & \colhead{Ref}  }
\startdata
\hline
\multicolumn{12}{c}{Strong Signatures of Low Surface Gravity} \\
\hline
J0909-0658	&	L1	&	2.28$\pm$0.03	&	Y?	&	0.86$\pm$0.010	&	Y?	&	1.52$\pm$0.02	&	N	&	2.50$\pm$0.02	&	Y?	 & 	N	&	 \\
J1139-3159	&	M9$\gamma$	&	2.151$\pm$0.012	&	Y	&	0.94$\pm$0.004	&	Y?	&	2.316$\pm$0.017	&	Y?	&	1.845$\pm$0.007	&	Y	 & 	Y	&	TWA & 1 \\
J1510-2818	&	M9	&	2.367$\pm$0.014	&	Y?	&	1.002$\pm$0.004	&	Y?	&	2.358$\pm$0.016	&	Y	&	2.021$\pm$0.007	&	Y?	 & 	Y	&	 \\
J2000-7523	&	M9	&	2.183$\pm$0.013	&	Y	&	0.94$\pm$0.005	&	Y?	&	2.42$\pm$0.02	&	Y	&	1.920$\pm$0.008	&	Y?	 & 	N	&	$\beta$PMG & 2, 3 \\
J2045-6332	&	M9	&	2.222$\pm$0.009	&	Y	&	1.044$\pm$0.004	&	Y?	&	2.274$\pm$0.012	&	Y?	&	2.073$\pm$0.005	&	Y?	 & 	Y	&	$\beta$PMG & 3  \\
\hline
\multicolumn{12}{c}{Weak Signatures of Low Surface Gravity} \\
\hline
J0041-5621AB	&	M8	&	2.385$\pm$0.011	&	Y?	&	1.282$\pm$0.005	&	N	&	2.216$\pm$0.012	&	Y?	&	2.075$\pm$0.006	&	N	 & 	Y	&	THA & 2, 3 \\
\enddata
\tablerefs{
(1) \citet{2002ApJ...575..484G}; 
(2) \citet{2010MNRAS.409..552G}; 
(3) \citet{2014ApJ...783..121G}; 
(4) \citet{2003A&A...400..297R}.}
\end{deluxetable}

%do we want to include 1635 here?
\begin{deluxetable}{llcccccl}
\tablecaption{Radial Velocity Measurements\label{table:rvs}}
\tabletypesize{\scriptsize}
\tablewidth{0pt} 
%\rotate
\tablehead{
\colhead{} &
\colhead{} &
\colhead{Line} &
\colhead{SDSS }&
\colhead{MagE} &
\colhead{Adopted} &
\colhead{Previous} &
\colhead{} \\
\colhead{Source} &
\colhead{SpT} &
\colhead{Centers} &
\colhead{ Templates}&
\colhead{ Standards} &
\colhead{ Value} &
\colhead{Measures} &
\colhead{Ref.} \\
& & 
\colhead{(\kms)} &
\colhead{(\kms)} &
\colhead{(\kms)} &
\colhead{(\kms)} &
\colhead{(\kms)} 
}

\startdata
J0004$-$2058	&	M9	&	$-$33$\pm$5	&	$-$26$\pm$4	&	$-$32.9$\pm$0.9	&	$-$32.9$\pm$2.2	&		\nodata						&				\\		
J0024$-$0158	&	M9.5	&	9$\pm$7	&	18$\pm$5	&	4.3$\pm$1.6	&	4.3$\pm$2.6	&	10$\pm$3					&	5			\\				
J0041$-$5621AB	&	M8	&	7$\pm$4	&	13$\pm$5	&	5.0$\pm$0.8	&	5.0$\pm$2.2	&	2.4$\pm$1.0, 7$\pm$3	&	3, 5	\\										
J0102$-$3737	&	M9	&	$-$2$\pm$2	&	4.5$\pm$1.2	&	$-$4.9$\pm$1.4	&	$-$4.9$\pm$2.4	&		\nodata						&				\\		
J0109$-$0343	&	M8	&	$-$6$\pm$5	&	2$\pm$7	&	$-$6.5$\pm$0.9	&	$-$6.5$\pm$2.2	&	$-$10.4$\pm$0.6, $-$7$\pm$3	&	3, 5	\\										
J0123$-$6921	&	M9	&	13$\pm$5	&	16$\pm$3	&	9.6$\pm$1.1	&	9.6$\pm$2.3	&	7.5$\pm$0.3, 11$\pm$3	&	3, 5	\\										
J0148$-$3024	&	M7.5	&	15$\pm$11	&	18$\pm$4	&	11.7$\pm$1.4	&	11.7$\pm$2.4	&	10$\pm$3					&	5			\\				
J0248$-$1651	&	M8	&	5$\pm$5	&	8$\pm$6	&	2.9$\pm$0.8	&	2.9$\pm$2.2	&	4$\pm$3					&	5			\\
\enddata
\tablenotetext{a}{Measured value differs by more than 3$\sigma$ from previously published value(s); see Section~3.1 and 3.4.}
\tablecomments{ 
(1) \citet{2010ApJ...723..684B};
(2) \citet{2005AJ....130.1871B};
(3) \citet{2010MNRAS.409..552G};
(4) \citet{2003ApJ...583..451M};
(5) \citet{2009ApJ...705.1416R};  
(6) \citet{2010AJ....139.1808S};
(7) \citet{2010A&A...512A..37S};
(8) \citet{1998MNRAS.301.1031T};
(9) \citet{2008AJ....135..785W};
(10) \citet{2012ApJS..203...10T}}
\end{deluxetable}

\begin{deluxetable}{llcccccc}
\tablecaption{Final Radial Velocities and Heliocentric Space Motions\label{table:uvw}}
\tabletypesize{\scriptsize}
\tablewidth{0pt} 
%\rotate
\tablehead{
\colhead{Source} & \colhead{SpT} & \colhead{Adopted RV} & \colhead{U} & \colhead{V} & \colhead{W} &\colhead{$P$(TD)/$P$(D)} & \colhead{Population\tablenotemark{a}} \\
 & & \colhead{(\kms)} & \colhead{(\kms)} & \colhead{(\kms)} & \colhead{(\kms)} & }
\startdata
J0004$-$2058 & M9 & $-$33$\pm$2 & $-$44$\pm$3 & $-$13.1$\pm$1.5 & 30$\pm$2 & 0.04 & D \\
J0024$-$0158 & M9.5 & 4$\pm$3 & 10.8$\pm$0.4 & 21.7$\pm$1.2 & 7$\pm$2 & 0.01 & D \\
J0041$-$5621AB & M8 & 5$\pm$2 & 6.5$\pm$1.0 & 1.5$\pm$1.3 & 5$\pm$2 & 0.01 & D \\
J0102$-$3737 & M9 & $-$5$\pm$2 & $-$67$\pm$3 & $-$22.0$\pm$1.3 & 13$\pm$2 & 0.03 & D \\
J0109$-$0343 & M8 & $-$7$\pm$2 & $-$0.9$\pm$0.7 & 1.6$\pm$0.7 & 15$\pm$2 & 0.01 & D \\
J0123$-$6921 & M9 & 1$\pm$2 & 4.5$\pm$1.8 & $-$5$\pm$2 & 5$\pm$2 & 0.01 & D \\
J0148$-$3024 & M7.5 & 12$\pm$2 & 12.3$\pm$0.8 & 18.3$\pm$1.2 & $-$6$\pm$2 & 0.01 & D \\
J0248$-$1651 & M8 & 3$\pm$2 & 22.3$\pm$1.6 & $-$4.6$\pm$1.6 & $\pm$2 & 0.01 & D \\
\enddata
\tablenotetext{a}{Stars assigned to thin disk (D), thick disk (TD) and intermediate between thin and thick disk (D/TD)  kinematic populations based on relative membership probabilities $P$(TD)/$P$(D) $<$ 0.1,  $P$(TD)/$P$(D) $>$10 and 0.1 $<$ $P$(TD)/$P$(D) $<$ 10, respectively; see Section~4.1 and \citet{2003A&A...410..527B}.}
\end{deluxetable}

\clearpage

\begin{deluxetable}{lcccccccccl}
\tablecaption{Velocity Dispersions and Group Kinematic Ages\label{table:dispersions}}
\tabletypesize{\scriptsize}
\tablewidth{0pt}
\rotate
\tablehead{
\colhead{Sample} & \colhead{N} & \colhead{$\langle{U}\rangle$} & \colhead{$\langle{V}\rangle$} & \colhead{$\langle{W}\rangle$} & \colhead{$\sigma_U$} & \colhead{$\sigma_V$} & \colhead{$\sigma_W$} & \colhead{$\sigma_v$} & \colhead{Age} & \colhead{Note} \\
 & & \colhead{(\kms)} & \colhead{(\kms)} & \colhead{(\kms)} & \colhead{(\kms)} & \colhead{(\kms)} & \colhead{(\kms)} & \colhead{(\kms)} & \colhead{(Gyr)} }
\startdata
All Sources & 85 & 5.4$\pm$0.3 & -4.9$\pm$0.2 & 1.4$\pm$0.3 & 31.9$\pm$0.4 & 25.0$\pm$0.3 & 15.8$\pm$0.3 & 43.5$\pm$0.3 & 4.8$\pm$0.2 & Unweighted \\
 & 85 &  &   &  & 38.7$\pm$0.6 & 27.1$\pm$0.4 & 19.6$\pm$0.4 & 51.1$\pm$0.6 & 4.8$\pm$0.1 & $|W|$ Weighted  \\
 & 65 & & &  & 34.4$\pm$0.4 & 21.2$\pm$0.4 & 10.7$\pm$0.4 & 41.8$\pm$0.5 & 4.3$\pm$0.2 & Core \\
 & 36 & & &  & 27$\pm$2 & 34$\pm$8 & 26$\pm$2 & 52$\pm$6 & 8$\pm$3 & Wing \\
$d$ $<$ 20~pc & 70 & 2.7$\pm$0.3 & -5.1$\pm$0.3 & 1.1$\pm$0.3 & 31.6$\pm$0.4 & 27.0$\pm$0.3 & 16.5$\pm$0.3 & 44.8$\pm$0.4 & 5.2$\pm$0.2 & Unweighted \\
 & 70 &  &   &  & 37.7$\pm$0.5 & 28.6$\pm$0.4 & 20.3$\pm$0.3 & 51.5$\pm$0.4 & 4.9$\pm$0.1 & $|W|$ Weighted  \\
 & 54 & & &  & 35.0$\pm$0.5 & 23.2$\pm$0.4 & 11.3$\pm$0.4 & 43.5$\pm$0.6 & 4.8$\pm$0.3 & Core \\
 & 30 & & &  & 23$\pm$4 & 35$\pm$7 & 27$\pm$2 & 51$\pm$5 & 8$\pm$2 & Wing \\
\enddata
\tablecomments{Ages for Unweighted, Core and Warm populations are based on Equation~\ref{equation:age2} using the full set of parameters given in \citet{2009MNRAS.397.1286A}.  Ages for the $|W|$-Weighted populations are based on Equation~\ref{equation:age1} using parameters $\sigma_{v,0}$ = 10~{\kms}, $\gamma_{v,p}$ = 1.1$\times$10$^4$~({\kms})$^3$~Gyr$^{-1}$ and $T_{\gamma}$ = 5~Gyr \citep{1977A&A....60..263W}.}
\end{deluxetable}

\clearpage 

\begin{deluxetable}{ll|cc|cc|cc|cc|cc|cc|cc|cc|cc|cc|cc|cc}
\tablecaption{Average Ages of Simulated Populations\label{table:mfsimage}}
\tabletypesize{\scriptsize}
\tablewidth{0pt} 
\rotate
\tablehead{
%& \multicolumn{1}{c}{} &	\multicolumn{24}{c}{$\beta$} \\ 
& \multicolumn{1}{c}{$\beta$} 	&	\multicolumn{4}{|c}{-0.5}	&	\multicolumn{4}{|c}{0.0}	&	\multicolumn{4}{|c}{0.5}	&	\multicolumn{4}{|c}{1.0}		&	\multicolumn{4}{|c}{0.0/0.5\tablenotemark{a}}		&	\multicolumn{4}{|c}{0.5/0.0\tablenotemark{a}}	\\ 
\tableline
	\multicolumn{2}{c}{Dynamic Selection\tablenotemark{b}}	&	\multicolumn{2}{|c}{No}	&	\multicolumn{2}{|c}{Yes}	&	\multicolumn{2}{|c}{No}	&	\multicolumn{2}{|c}{Yes}	&	\multicolumn{2}{|c}{No}	&	\multicolumn{2}{|c}{Yes}	&	\multicolumn{2}{|c}{No}	&	\multicolumn{2}{|c}{Yes}	&	\multicolumn{2}{|c}{No}	&	\multicolumn{2}{|c}{Yes}	&	\multicolumn{2}{|c}{No}	&	\multicolumn{2}{|c}{Yes}	\\ 
\tableline
\colhead{$\alpha$}	&	\colhead{SpT}	&	\multicolumn{1}{|c}{$\bar{\tau}$}	&	\colhead{$\sigma_{\tau}$}	&	\multicolumn{1}{|c}{$\bar{\tau}$}	&	\colhead{$\sigma_{\tau}$}	&	\multicolumn{1}{|c}{$\bar{\tau}$}	&	\colhead{$\sigma_{\tau}$}	&	\multicolumn{1}{|c}{$\bar{\tau}$}	&	\colhead{$\sigma_{\tau}$}	&	\multicolumn{1}{|c}{$\bar{\tau}$}	&	\colhead{$\sigma_{\tau}$}	&	\multicolumn{1}{|c}{$\bar{\tau}$}	&	\colhead{$\sigma_{\tau}$}	&	\multicolumn{1}{|c}{$\bar{\tau}$}	&	\colhead{$\sigma_{\tau}$}	&	\multicolumn{1}{|c}{$\bar{\tau}$}	&	\colhead{$\sigma_{\tau}$}	&	\multicolumn{1}{|c}{$\bar{\tau}$}	&	\colhead{$\sigma_{\tau}$}	&	\multicolumn{1}{|c}{$\bar{\tau}$}	&	\colhead{$\sigma_{\tau}$}	&	\multicolumn{1}{|c}{$\bar{\tau}$}	&	\colhead{$\sigma_{\tau}$}	&	\multicolumn{1}{|c}{$\bar{\tau}$}	&	\colhead{$\sigma_{\tau}$}	
}
\startdata
-0.5	&	M6--M9.5	&	1.5	&	1.4	&	1.4	&	1.3	&	3.5	&	2.3	&	3.0	&	2.3	&	6.0	&	1.8	&	5.5	&	2.1	&	6.9	&	0.9	&	6.9	&	1.0	&	5.8	&	2.1	&	5.5	&	2.3	&	3.6	&	2.3	&	3.3	&	2.3	\\
	&	L0--L5.5	&	1.5	&	1.3	&	1.4	&	1.2	&	3.2	&	2.2	&	2.8	&	2.2	&	5.8	&	1.9	&	5.7	&	1.8	&	6.9	&	1.0	&	6.8	&	1.0	&	4.5	&	2.7	&	4.1	&	2.8	&	4.1	&	2.0	&	4.0	&	2.0	\\
0.0	&	M6--M9.5	&	1.5	&	1.4	&	1.2	&	1.1	&	3.4	&	2.3	&	2.9	&	2.2	&	6.0	&	1.7	&	5.7	&	1.9	&	6.9	&	1.0	&	6.9	&	0.9	&	5.6	&	2.1	&	5.5	&	2.2	&	3.6	&	2.3	&	3.2	&	2.1	\\
	&	L0--L5.5	&	1.4	&	1.3	&	1.3	&	1.2	&	3.1	&	2.2	&	2.6	&	2.1	&	5.7	&	1.9	&	5.5	&	1.9	&	6.8	&	1.1	&	6.8	&	1.0	&	4.4	&	2.7	&	3.9	&	2.8	&	4.0	&	2.0	&	3.6	&	2.0	\\
0.5	&	M6--M9.5	&	1.5	&	1.4	&	1.4	&	1.3	&	3.4	&	2.3	&	2.8	&	2.2	&	6.0	&	1.8	&	5.7	&	2.1	&	6.9	&	1.0	&	6.9	&	1.1	&	5.6	&	2.2	&	5.2	&	2.4	&	3.5	&	2.3	&	2.9	&	2.2	\\
	&	L0--L5.5	&	1.4	&	1.2	&	1.2	&	1.1	&	3.1	&	2.3	&	2.6	&	2.1	&	5.7	&	1.9	&	5.4	&	2.1	&	6.9	&	1.1	&	6.8	&	1.2	&	4.1	&	2.8	&	3.7	&	2.7	&	4.0	&	2.0	&	3.4	&	2.0	\\
1.0	&	M6--M9.5	&	1.4	&	1.4	&	1.4	&	1.3	&	3.4	&	2.3	&	3.0	&	2.2	&	5.9	&	1.8	&	5.7	&	2.0	&	6.9	&	1.0	&	6.9	&	1.2	&	5.5	&	2.2	&	5.2	&	2.5	&	3.5	&	2.3	&	2.9	&	2.2	\\
	&	L0--L5.5	&	1.3	&	1.2	&	1.1	&	1.1	&	2.8	&	2.2	&	2.3	&	2.0	&	5.6	&	2.0	&	5.2	&	2.3	&	6.8	&	1.1	&	6.9	&	1.1	&	3.9	&	2.8	&	3.1	&	2.7	&	3.9	&	2.1	&	3.6	&	1.9	\\
\enddata
\tablenotetext{a}{Different values of $\beta$ used for brown dwarfs (first number) and stars (second number).}
\tablenotetext{b}{Restricted to those sources whose orbits pass within 50~pc of the Sun after given initial Galactic radius and velocity components based on age-dependent dispersion functions; see Section~5.}
\end{deluxetable}

\clearpage

\begin{deluxetable}{lllcccccllll}
\tablecaption{Candidate Members of Young Moving Groups and Associations Based on BANYAN II Analysis\label{table:ymg}}
\tabletypesize{\scriptsize}
\tablewidth{0pt} 
%\rotate
\tablehead{
\colhead{Source}&
\colhead{SpT}&
\colhead{Moving}&
\colhead{$P_M$}&
\colhead{$P_C$}&
\colhead{$\Delta{D}$} &
\colhead{$\Delta{V}$} &
\colhead{H$\alpha$} &
\colhead{Li~I} &
\colhead{Low $g$} &
\colhead{Ref.} \\
 & & \colhead{Group} & & & 
\colhead{(pc)}&
\colhead{(\kms)} \\
}
\startdata
J0041$-$5621AB	&	M6.5+M9	&	THA	&	100	&	$<$0.1	&	7	&	1.4	& Y  &  Y  & N  & 1,2 \\
J0123$-$6921	&	M7.5	&	THA	&	99.9	&	$<$0.1	&	11	&	2 & Y  & Y  & Y 	& 2 \\
J0339$-$3525	&	M9	&	ARG	&	17	&	0.3	&	25	&	13	& Y  & (Y)\tablenotemark{a}  &  Y & 2 \\
J0823$-$4912	&	L1.5	&	$\beta$PMG	&	30	&	0.4	&	23	&	7	& N  &  Y & Y & 3 \\
J1139$-$3159	&	M9$\gamma$	&	TWA	&	99.6	&	0.1	&	21	&	5	& Y  & Y  & Y   & 4 \\
%J1456$-$2809	&	M7	&	ABD	&	94	&	4	&	21	&	3 & Y  &  N  & N &  Probable interloper	\\
J1510$-$2818	&	M9	&	ARG	&	60	&	34	&	24	&	4	& Y  & N  & Y  & 5 \\
J2000$-$7523	&	M9	&	$\beta$PMG	&	99	&	4	&	20	&	2	& Y  &  N & Y & 1,2   \\
J2045$-$6332	&	M9	&	$\beta$PMG	&	87	&	0.2	&	11	&	8	& Y  &  Y & Y  & 1,2 \\
\enddata
\tablenotetext{a}{Marginal detection in MagE data; previously reported detection in \citet{1998MNRAS.296L..42T} and \citet{2009ApJ...705.1416R}}.
\tablecomments{Moving groups and associations: ABD = AB Doradus Moving Group, ARG = Argus Association, $\beta$PMG = $\beta$ Pictoris Moving Group, THA = Tucana Horologium. $P_M$ gives the probability of membership, $P_C$ gives the probability of field contamination.}
\tablerefs{
(1) \citet{2010MNRAS.409..552G}; 
(2) \citet{2002ApJ...575..484G}; 
(3) \citet{sahlmann2015};
(4) \citet{2014ApJ...783..121G}; 
(5) This paper.}
\end{deluxetable}

\begin{deluxetable}{ccccc}
\tablecaption{ Polynomial Coefficients for Magnitude Difference Relations Between MKO and 2MASS Systems\label{table:mko2m}}
\tabletypesize{\scriptsize}
\tablewidth{0pt} 
%\rotate
\tablehead{
\colhead{Filter}&
\colhead{$c_0$}&
\colhead{$c_1$}&
\colhead{$c_2$}&
\colhead{$\sigma$} \\
}
\startdata
$J$ & $-$5.97605e-3 & $-$1.49741e-4 & $-$1.26157e-4 & 0.0078 \\
$H$ &  $-$8.41293e-2 &  9.70498e-3 & $-$1.58315e-4 &  0.0052 \\
$K_s$ &   1.02933e-2 &  $-$3.16442e-3 &  7.96614e-5 &  0.0053 \\
\enddata
\tablecomments{Coefficients are given for the polynomial relation MKO - 2MASS = $\sum_{i=0}^2c_i[SpT]^i$ where SpT(M0) = 10, SpT(M5) = 15, SpT(L0) = 20, etc.}
\end{deluxetable}

\end{document}